\documentclass[a4paper,11pt]{article}
\usepackage{lineno}
\usepackage{amsthm}
\usepackage{amssymb}
\usepackage{bbm}
\usepackage[T1]{fontenc}
\usepackage[utf8]{inputenc}
\DeclareUnicodeCharacter{2010}{-}
\usepackage{graphicx}
\usepackage{subcaption}
\usepackage{tikz-cd}
\usetikzlibrary{calc}
\usepackage{amsmath}
\usepackage{cite}

\newcommand{\dd}{\mathrm{d}}

\newcommand{\cK}{\mathcal{K}}
\newcommand{\cI}{\mathcal{I}}
\newcommand{\cJ}{\mathcal{J}}
\newcommand{\cA}{\mathcal{A}}

\usepackage{color}
\definecolor{dark-gray}{gray}{0.20}
\definecolor{gray}{gray}{0.30}
\definecolor{light-gray}{gray}{0.80}
\definecolor{dark-red}{rgb}{0.7,0,0}
\definecolor{dark-green}{rgb}{0.1,0.4,0}
\definecolor{dark-blue}{rgb}{0.3,0.3,0.7}
\definecolor{dark-blue2}{RGB}{0, 114,178}
\definecolor{light-blue}{rgb}{0.8,0.8,1}
\definecolor{swamp}{RGB}{240, 199, 197}
\definecolor{antiquefuchsia}{rgb}{0.57, 0.36, 0.51}

\usepackage{hyperref}
\usepackage{cleveref}

\newcommand{\bbR}{{\mathbb{R}}}
\newcommand{\bbQ}{{\mathbb{Q}}}
\newcommand{\bbZ}{{\mathbb{Z}}}

\usepackage{multirow}

\hypersetup{
	colorlinks=true,
	linkcolor=dark-blue2,
	citecolor=dark-blue2,
	urlcolor=dark-blue2,
	linktoc=section
}

\usepackage{tocloft}

\begin{document}

\baselineskip=18pt  
\numberwithin{equation}{section}  
\allowdisplaybreaks  

\thispagestyle{empty}

\vspace*{-2cm}
\begin{flushright}
	\texttt{CERN-TH-2026-228\\ZMP-HH/26-31}
\end{flushright}

\vspace*{2.6cm}
\begin{center}
{\Large{\textbf{Arithmetic of Duality Defects in 4d Maxwell Theory
}}} 
\\
\vspace*{1.5cm}
Damian van de Heisteeg$^{1}$, Ling Lin$^{2,3}$, Giacomo Maggiorotti$^{4,5}$\\

\vspace{1.0cm}
{\it $^1$Theoretical Physics Department, CERN, 1211 Geneva 23, Switzerland}\\
\vspace*{.1cm}
{\it $^2$Dipartimento di Fisica e Astronomia, Universit\`{a} di Bologna, via Irnerio 46, Bologna, Italy}\\
\vspace*{.1cm}
{\it $^3$INFN, Sezione di Bologna, viale Berti Pichat 6/2, Bologna, Italy}\\
\vspace*{.1cm}
{\it $^4$II. Institut f\"ur Theoretische Physik, Universit\"at Hamburg, Notkestrasse 9,\\ 22607 Hamburg, Germany}\\
\vspace*{.1cm}
{\it $^5$Zentrum f\"ur Mathematische Physik, Universit\"at Hamburg, Bundesstrasse 55, \\ 20146 Hamburg, Germany}

\vspace*{0.8cm}
\end{center}
\vspace*{.5cm}

\noindent We study the arithmetic structure of non-invertible duality defects in 4d $U(1)$ Maxwell theory. 
First, we recast the problem of realizing an elliptic $SL(2,\mathbb{Q})$ transformation as a duality defect in terms of complex multiplication for two-tori. 
The resulting classification includes well-studied cases in the literature, but extends to infinite Diophantine families of defects at a given gauge coupling. 
Second, we describe their worldvolume theories explicitly, with the Chern--Simons couplings determined by using continued fractions and tridiagonal continuant matrices.  
Third, we work out the fusion products of these defects, extracting the condensation defects and decoupled sectors both by lattice arithmetic and from the worldvolume descriptions. 
Finally, we illustrate our findings with the familiar $N$-ality defects (with $N=2,3$), but also give examples of infinite order.

\newpage
\setcounter{tocdepth}{2}
\tableofcontents

\newpage

\section{Introduction}
Geometry provides an invaluable framework for organizing physical symmetries. Its application to ordinary global symmetries dates back as far as Noether's theorem, which yields conservation laws for, e.g., rotational and translational symmetries. The advent of generalized global symmetries \cite{Gaiotto:2014kfa} has expanded these opportunities, by equating symmetry generators with topological operators.
The interplay between geometry and topological operators are particularly rich for theories with a geometrical origin, such as those obtained by geometric engineering in string theory (as pioneered in, e.g., \cite{Witten:1998wy,Belov:2006xj,Freed:2012bs,Apruzzi:2021nmk, Apruzzi:2022rei, GarciaEtxebarria:2022vzq, Bergman:2022otk, Heckman:2022muc}) or 2d worldsheet CFTs with geometric target spaces 
\cite{Gukov:2002nw,Fuchs:2007tx, Gaberdiel:2011fg, Bachas:2012bj}.
But even a theory as elementary as pure electromagnetism carries a geometry of its own: the lattice of dyonic line operators, equipped with a Dirac pairing and a coupling-dependent metric. 

In this work we consider 4d $U(1)$ Maxwell theory, which possesses an $SL(2,\mathbb{Z})$ duality group  \cite{Montonen:1977sn}.
Depending on the complexified gauge coupling $\tau = \tfrac{\theta}{2\pi}+\tfrac{i}{g^2}$, such dualities can be combined with half-space gauging of discrete one-form symmetries to construct non-invertible defects \cite{Choi:2021kmx, Kaidi:2021xfk, Choi:2022zal}.
These constructions lift Kramers--Wannier-like dualities, long familiar in 2d CFTs \cite{Verlinde:1988sn,Vafa:1989ih, Fuchs:2007tx}, to four dimensions. 
Interestingly, when $\tau$ is a so-called \emph{imaginary quadratic number}, 4d Maxwell possesses at special values of $\tau$ an infinite set of codimension-one topological defects, whose action on the charge lattice realizes any elliptic $SL(2,\bbQ)$-matrix \cite{Niro:2022ctq, Sela:2024okz}.\footnote{More recently, topological defects realizing $SL(2,\mathbb{R})$ have also been proposed in \cite{Hasan:2024aow,Arbalestrier:2024oqg}; we will not address these defects in our approach.}
Motivated by this observation, we want to systematically explore the connection between these rational defects, their half-space gauging realization, and the arithmetic of the special coupling values.

Our approach utilizes the auxiliary two-torus $T^2$ of complex structure $\tau$ that underlies Maxwell theory at coupling $\tau$.
We argue that non-invertible defects are present precisely when this $T^2$ enjoys additional \textit{complex multiplication} (CM) symmetries.
As mentioned above, the corresponding value of the coupling is then \textit{imaginary quadratic}, $\tau \in \mathbb{Q}(i\sqrt{d})$, i.e., it satisfies a quadratic equation with integer coefficients.
The CM symmetry map is naturally represented by an elliptic $SL(2, \bbQ)$-matrix $D$---the same matrix that describes the defect action on the charge lattice.
Moreover, we provide an explicit realization of these defects as \emph{minimal duality} defects, namely, by stacking two $SL(2,\mathbb{Z})$ duality transformations with a single half-space gauging interfaces.\footnote{This improves on the construction in \cite{Niro:2022ctq}, where an $SL(2, \bbQ)$-defect is obtained from stacking several gauging interfaces.
Furthermore, we also demonstrate that the defects are independent of ambiguities exhibited generically by the underlying matrix calculus.
}
At a given CM coupling we encounter infinite Diophantine families of such duality defects. These include the ``duality'', ``triality'' defects and generalizations thereof in \cite{Choi:2021kmx, Kaidi:2021xfk, Choi:2022zal, Cordova:2023ent}\footnote{In these works the ``duality'' defect is specifically of order 2 (modulo invertible defects); in our work we shall instead refer to duality defects as any defect obtained by stacking $SL(2,\bbZ)$ transformations and a gauging interface.} as special cases, but also go beyond such $N$-ality defects: most members are of infinite order, i.e.,~their self-fusions keep producing non-invertible defects within the family.
Let us remark that the connection between complex multiplication on $T^2$ and non-invertible defects has been previously noticed by \cite{Damia:2024xju} in the context of 2d $c=2$ CFTs, but a systematic deconstruction of this connection has, to our knowledge, not been attempted before in the literature.

After reviewing the basics of non-invertible defects obtained from half-space gauging in section~\ref{sec:review}, we use, in section~\ref{sec:CMdefects}, the complex multiplication symmetry to recast the action of the defect on the Wilson--'t Hooft lines and Gukov--Witten surface operators in terms of the two-torus. Performing a Smith decomposition of the $SL(2,\mathbb{Q})$ matrix $D$ representing the CM map, we construct the defect explicitly: it decomposes into two $SL(2,\mathbb{Z})$ duality interfaces that surround a half-space gauging wall.  The kernel of the corresponding CM map then makes the non-invertibility manifest: the Gukov--Witten operators that can end on the defect, i.e., those generating the discrete dyonic one-form subgroup that is being gauged, are precisely the points of $T^2 = \mathbb{C}/(\mathbb{Z}+\tau \mathbb{Z})$ that map to zero. 

In section \ref{sec:worldvolumeaction} we describe the worldvolume theory of our duality defects explicitly as a Chern--Simons theory. Building on the S- and T-duality defects of \cite{Gaiotto:2008ak, Kapustin:2009av}, we first construct the general 3d defect theory for an arbitrary $SL(2,\mathbb{Z})$ duality interface. This construction leverages the decomposition of any $SL(2,\mathbb{Z})$ element into a word in the generators $S = \left(\begin{smallmatrix}
    0 & -1 \\
    1 & 0
\end{smallmatrix}\right)$ and $T = \left(\begin{smallmatrix}
    1 & 1 \\
    0 & 1
\end{smallmatrix}\right)$, which follows from a continued fraction expansion. The resulting Chern--Simons theories are specified by tri-diagonal $K$-matrices, so-called continuant matrices, that couple the 4d gauge fields and the 3d $U(1)$ Chern--Simons fields. Using the Smith decomposition of an $SL(2,\mathbb{Q})$ defect, we then assemble its Chern--Simons theory by combining the two $SL(2,\mathbb{Z})$ wall theories with an electric gauging without torsion.
Importantly, this description also allows us to proof that two Smith decompositions $D = U_1 \, \text{diag} \, V_1 = U_2 \, \text{diag} \, V_2$ of the same $SL(2,\bbQ)$ matrix $D$ gives the same 3d Chern--Simons theory including SPT-phases.
This is crucial to establish that at fixed coupling $\tau$ (of CM type), the construction of section \ref{sec:CMdefects} associates a unique duality defect to every CM map $D$.

As a final arithmetic structure, we describe the fusion rules of the defects in section \ref{sec:fusion}. Besides other non-invertible duality defects within the family, the fusion product generically contains condensation defects and decoupled TQFT sectors. By considering the Gukov--Witten surfaces between the defects that can end on them, we determine which one-form subgroups condense and how the decoupled TQFTs arise. These two sectors are distinguished by whether the Gukov--Witten surface wraps a two-cycle parallel to the defects---giving rise the condensation defect---or a two-chain that extends between the two defects---inducing line operators at its boundaries that gives rise to the decoupled TQFT; see figure \ref{fig:decomoposit_cond_TQFT_defe} for a detailed depiction. With our explicit Chern--Simons description at hand, we verify the orders of the decoupled sectors in examples, collected in section \ref{sec:examples}. 

\section{1-form symmetries and topological defects in 4d Maxwell theory}\label{sec:review}
In this section we review the construction of topological duality defects in 4d Maxwell theory \cite{Choi:2021kmx, Choi:2022zal, Niro:2022ctq}, fixing along the way the conventions used in the rest of the paper. We begin in subsection \ref{ssec:4dMaxwell} by recalling the basic ingredients of 4d Maxwell theory: its Wilson--'t Hooft line operators, Gukov--Witten surfaces, and electromagnetic dualities. Subsection \ref{ssec:gauging} reviews the gauging of discrete subgroups of the 1-form symmetries. Finally, subsection \ref{ssec:defectsfromgauging} describes the construction of duality defects by half-space gauging.

\subsection{Maxwell theory: 1-form symmetries and electromagnetic dualities}\label{ssec:4dMaxwell}
We begin by recalling the familiar Lagrangian of a $U(1)$ gauge theory in the presence of a non-trivial theta angle in four dimensions
\begin{equation}
    \mathcal{L}(g,\theta;A)=\frac{1}{4\pi g^2} F \wedge \ast F + \frac{i \theta}{8\pi^2} F \wedge F\,,
\end{equation}
where $F=dA$ is the field strength, while $\star$ represents the Hodge duality operator. Owing to the theta-term, the field strength magnetically dual to $F$ takes the more involved form
\begin{equation}
    \tilde{F} = d\tilde{A} = -\frac{i}{g^2} \ast F + \frac{\theta}{2\pi}F\,,
\end{equation}
where $\tilde{A}$ is the associated dual gauge field.

The theory enjoys a $U(1)_e \times U(1)_m$ global 1-form symmetry. Its symmetry operators, supported on 2-manifolds $\Sigma$, are the Gukov--Witten surface operators \cite{Gukov:2006jk, Gukov:2008sn}, which can be written in terms of the field strengths $F$ and $\tilde{F}$ as
\begin{equation}\label{eq:etaWG}
    \eta_{\mathbf{r}}(\Sigma) = \exp \left( i \int_\Sigma (r_1 F - r_2 \tilde{F}) \right) \, ,
\end{equation}
where $\mathbf{r} = (r_1, r_2) \in \bbR^2/ \bbZ^2$ parametrizes the elements of the 1-form symmetry group.  
Since both $F$ and $\tilde{F}$ are closed, these operators are topological when inserted on a closed manifold $\Sigma$. The objects charged under these two 1-form symmetries are given by the electrically charged Wilson lines and magnetically charged 't Hooft lines,
\begin{equation}
    \label{eq:wilson-thooft-lines-definition}
    W_{q_1}(\gamma) = e^{i q_1 \int_\gamma A}\,, \qquad H_{q_2}(\gamma) = e^{-i q_2 \int_{\gamma}\tilde{A}}\,.
\end{equation}
More generally, there are dyonic lines, which we denote by $WH_{\bf q}(\gamma)$,\footnote{One could imagine defining dyonic line operators as $WH_{\bf q}=e^{i \int_\gamma (q_1 A- q_2 \tilde{A})}$; this naive expression is however ill-defined, because the gauge fields $A$ and $\tilde{A}$ are not mutually local.} where ${\bf q}=(q_1,q_2)$ collects the electric and magnetic charges.
Genuine line operators carry integrally-quantized charges $q_1,\,q_2 \in \bbZ$ and span the charge lattice $\Gamma\cong\bbZ^2$. At the level of correlation functions, the symmetry action of the Gukov--Witten operators is captured by the relation
\begin{align}\label{eq:linking-maxwell_lines-surfaces}
    \langle \eta_{\bf r}(\Sigma) \, WH_{\bf q} (\gamma) \, \cdots \rangle = e^{2\pi i (-q_1 r_2 + q_2 r_1)\, \text{Link}(\Sigma, \gamma)} \langle WH_{\bf q} (\gamma) \, \cdots \rangle \, ,
\end{align}
where the ellipsis denotes possible further local operator insertions.
The linking number $\text{Link}(\Sigma,\gamma)$ counts how many times the manifolds supporting the two operators ``wrap around each other''; we henceforth set it to one, as this can always be arranged by a suitable choice of manifolds.
The remaining factor in the exponent is governed by the Dirac pairing, which we define as a symplectic (namely, non-degenerate and anti-symmetric) pairing on the vector space $\Gamma_\bbR := \Gamma \otimes \bbR = \mathbb{R}^2 \supset \Gamma$,
\begin{align}\label{eq:Dirac-pairing-4d}
\begin{split}
    &{\cal J} :\Gamma_\bbR \times \Gamma_\bbR \rightarrow \mathbb{R} \, , \qquad {\cal J}\big( {\bf r}, {\bf s} \big) = - r_1 s_2 + r_2 s_1 \equiv (r_1,r_2) \underbrace{\left( \begin{smallmatrix}
        0 & -1 \\ 1 & 0
    \end{smallmatrix} \right)}_{=:J} \left( \begin{smallmatrix}
        s_1\\s_2
    \end{smallmatrix} \right)\, .
\end{split}
\end{align}
Two line operators with (not necessarily integral) charges ${\bf r}$ and ${\bf s}$ are called mutually local if ${\cal J}({\bf r}, {\bf s}) \in \bbZ$.
The lattice $\Gamma$ of genuine line operators forms a maximal mutually local subset of $\Gamma_\bbR$,\footnote{In fact, any maximal mutually local subset $\Gamma' \subset \Gamma_\bbR$ is related to $\Gamma \cong \bbZ^2$ by a (generally non-integral) $SL(2)$-transformation, simply because $SL(2) \cong Sp(1)$ naturally preserves the symplectic Dirac pairing.
By a suitable field redefinition, $\Gamma'$ can be identified with the genuine lines of Maxwell theory at a different complexified coupling; see below in subsection \ref{ssec:gauging}.\label{footnote:maximal-local-subset}} and the canonical basis $(1,0)$ and $(0,1)$ correspond to the minimal Wilson and 't Hooft line, respectively.

Improperly quantized line operators $WH_{\bf q}(\gamma)$, with ${\bf q}=(q_1,q_2)\notin\bbZ^2$, will play an important role in what follows.
Such lines are said to be non-genuine: gauge invariance requires them to be attached to a Gukov--Witten operator of the form $\eta_{\bf q}(\Sigma)$ with $\partial\Sigma=\gamma$. 

Finally, we summarize how Maxwell theory transforms under $M = \left(\begin{smallmatrix}a & b\\
c & d \end{smallmatrix} \right) \in SL(2, \mathbb{Z})$, the non-perturbative duality symmetry \cite{Montonen:1977sn} that identifies physically equivalent theories at different values of the couplings.\footnote{We assume throughout the work that the spacetime manifold is spin, so that $SL(2,\bbZ)$ is the full duality group.}
The machinery needed to treat non-invertible duality transformations $M \in SL(2,\mathbb{Q})$ will only be set up later, but the transformation rules there take the same form. 
The couplings $\theta, g$ and field strengths $(\tilde{F}, F)$ transform in tandem
\begin{equation}\label{eq:defSL2}
    \tau \equiv \frac{\theta}{2\pi}+\frac{i}{g^2} \mapsto \tau' = M \cdot \tau = \frac{a\tau+b}{c\tau+d}\, , \qquad \begin{pmatrix}
        \tilde{F} \\
        F
    \end{pmatrix} \mapsto \begin{pmatrix} \tilde{F}'\\
    F'
    \end{pmatrix} = M \begin{pmatrix}
        \tilde{F} \\
        F
    \end{pmatrix}\, .
\end{equation}
In turn, acting on the charges of Wilson--'t Hooft lines and Gukov--Witten surfaces, we find that these must transform as\footnote{To see this, collect the field strengths into the doublet $\mathbb{F} = (\tilde{F}, F)$, and write the exponent of a Gukov--Witten operator via the Dirac pairing
$$r_1 F - r_2 \tilde{F} = - \mathcal{J}(\mathbf{r}, \mathbb{F})  = -\mathcal{J}(\mathbf{r}, M^{-1}\mathbb{F}') = - \mathcal{J}(M\mathbf{r}, \mathbb{F}')  \, ,$$
where the last step uses the $SL(2,\mathbb{R})$-invariance of $\mathcal{J}$. Thus, we read off $\mathbf{r}'=M \mathbf{r}$. The same argument applies verbatim to the Wilson--'t Hooft charges.}
\begin{equation}\label{eq:SL2WHeta}
    WH_{\mathbf{q}} \to WH_{\mathbf{q}'} = WH_{M.\mathbf{q}}\, , \qquad \eta_{\mathbf{r}} \to \eta_{\mathbf{r}'} =  \eta_{M.\mathbf{r}}\, .
\end{equation}
This also elucidates the relative signs in the definitions \eqref{eq:etaWG} and \eqref{eq:wilson-thooft-lines-definition} of the Wilson--'t Hooft lines and Gukov--Witten surfaces: had we paired the labels with the field strengths symmetrically rather than antisymmetrically, the charges would instead have transformed with $M^{-T}$.

\subsection{Discrete gauging of 1-form symmetries}\label{ssec:gauging}
We next describe the effect of gauging a discrete subgroup ${\cal Z} \subset U(1)_e \times U(1)_m$ on the charge lattice. For simplicity, we first treat the case of a purely electric subgroup ${\cal Z} = \bbZ_N \subset U(1)_e$. On general grounds, the gauging projects out states charged under ${\cal Z}$ and adds a twisted sector, so as to again produce a maximally mutually-local charge lattice $\Gamma'$ of genuine line operators. The twisted sector states are fractionally charged lines which, prior to gauging, were attached to the Gukov--Witten operators generating $\mathcal{Z}$ \cite{Gaiotto:2014kfa}. 

To make these statements precise, note first that the sector invariant under ${\cal Z} = \mathbb{Z}_N^{(e)}$ consists of the line charges
\begin{align}
    \{ {\bf q} \in \mathbb{Z}^2 \ | \ {\bf q} = (N q_1, q_2) \, , \ q_{1,2} \in \bbZ \} \equiv \Delta_{\cal Z} \, ,
\end{align}
which are precisely the charges invisible to the Gukov--Witten operators (inserted on closed surfaces $\Sigma$) that generate ${\cal Z} = \mathbb{Z}_N^{(e)}$,
\begin{align}\label{eq:gukov-witten-for-electric-Z_N}
\begin{split}
    \eta_{\mathbf{r}}(\Sigma) & = e^{i \int_\Sigma( r_1 F - r_2 \tilde{F})} \quad \text{with} \quad {\bf r} \in \Delta_{\cal Z}^\vee = \{ (a, \tfrac{b}{N}) \in \mathbb{Q}^2 \ \ | \ \ a,b \in \bbZ \} \, .
\end{split}
\end{align}
The lattices $\Delta_{\cal Z}$ and $\Delta_{\cal Z}^\vee$ are dual with respect to the Dirac pairing, i.e., ${\cal J}({\bf r}, {\bf q}) \in \bbZ$. This guarantees trivial linking between the Wilson--'t Hooft  operators in $\Delta_{\cal Z}$ and the Gukov--Witten operators labelled by $\Delta_{\cal Z}^\vee$.

Since $\Delta_{\cal Z}$ has determinant $N$, we must add a twisted sector generated by an $N$-fractional vector ${\bf r}_g \in \Delta_{\cal Z}^\vee$, so that the lattice 
\begin{align}
    \Gamma' = \Delta_{\cal Z} + \langle {\bf r}_g \rangle_{\bbZ} \equiv \{ {\bf q} + n \, {\bf r}_g \ | \ {\bf q} \in \Delta_{\cal Z} \, , \ n \in \bbZ \}
\end{align}
of genuine states after gauging is maximally mutually-local. One can check that, by suitably adding vectors in $\Delta_{\mathcal{Z}}$, the twisted sector generator can always be brought to the form ${\bf r}_g = (p,1/N)$, where the choice $p \in \bbZ_N$ fixes the twisted sector.
This choice is also commonly referred to as the ``choice of discrete torsion'' in the gauging of ${\cal Z}$.
The charge lattice $\Gamma'$ after gauging is therefore generated by the two vectors
\begin{align}\label{eq:basis-after-gauging-Z_N}
    {\bf e}'_1 := \begin{pmatrix}
        N \\0
    \end{pmatrix} \, , \quad {\bf e}'_2 := \begin{pmatrix}
        p \\ \tfrac{1}{N}
    \end{pmatrix} \, , \quad p \in \bbZ_N \ \text{ fixed}.
\end{align}
The lattice is indeed unimodular since $\det \left( \begin{smallmatrix}
    N & p \\ 0 & 1/N
\end{smallmatrix} \right) = -{\cal J}({\bf e}'_1, {\bf e}'_2) = 1$, and hence constitutes a maximal mutually-local set of line charges. Accordingly, ${\bf e}_1'$ is the charge vector of a fundamental Wilson line, and ${\bf e}_2'$ that of a fundamental 't Hooft line.

More generally, any finite cyclic subgroup ${\cal Z} \subset U(1)_e \times U(1)_m$ is generated by a topological Gukov--Witten operator labeled by a class ${\bf r}_g \mod \Gamma$. This includes not only the commonly studied case $\mathcal{Z} = \mathbb{Z}_N^{(e)} \times \mathbb{Z}_M^{(m)}$ with co-prime $N$ and $M$, but also general dyonic subgroup, generated by Gukov--Witten operators with both electric and magnetic charges. As in the electric case, the invariant charges form a sublattice $\Delta_{\cal Z} \subset \Gamma$ of index $|{\cal Z}|$, and the choice of discrete torsion amounts to the choice of an explicit representative ${\bf r}_g \in \Delta_{\mathcal{Z}}^\vee \subset \Gamma_{\mathbb{Q}}$ of the generating class. The $\mathbb{Z}$-linear span of this representative together with the invariant sector then yields the charge lattice of the theory after gauging
\begin{align}\label{eq:genuine-after-gauging}
    \Gamma' = \Delta_{\cal Z} + \langle {\bf r}_g\rangle_\bbZ \, .
\end{align}
By construction, $\Gamma'$ is again unimodular, and hence a maximal mutually-local charge lattice; by footnote \ref{footnote:maximal-local-subset}, it can therefore be identified with the genuine line operators of a Maxwell theory at a new value of the complexified gauge coupling. This observation is the starting point of the constructions in the following sections: by combining discrete gauging with a duality transformation that, when possible, returns the coupling $\tau$ to its original value, we obtain the (generally non-invertible) duality defects of Maxwell theory.

\subsection{Topological defects from half-space gauging}\label{ssec:defectsfromgauging}

As the charge lattices of genuine lines before ($\Gamma$) and after ($\Gamma'$) the gauging are unimodular lattices contained in the same vector space $\Gamma_\mathbb{Q}$, there must exist an invertible linear map $R: \Gamma_\mathbb{Q} \rightarrow \Gamma_\mathbb{Q}$ which rotate $\Gamma$ isomorphically into $\Gamma'$, see footnote \ref{footnote:maximal-local-subset}.

Let us, for concreteness, use the gauging of $\mathbb{Z}_N^{(e)}$ discussed above to illustrate how to describe this map.
The idea is that after the gauging, the vectors \eqref{eq:basis-after-gauging-Z_N} --- which are expressed in the canonical basis of $\Gamma$ --- are the new basis of fundamental Wilson and 't Hooft lines in $\Gamma'$.
On $\Gamma_\bbQ$, this therefore acts as a basis change map, satisfying
\begin{align}
     R\begin{pmatrix}
        N \\ 0
    \end{pmatrix} = \begin{pmatrix}
        1 \\ 0
    \end{pmatrix} \, , \quad  R\begin{pmatrix}
        p \\ \tfrac{1}{N}
    \end{pmatrix} = \begin{pmatrix}
        0 \\ 1
    \end{pmatrix} \quad \Leftrightarrow \quad R = \begin{pmatrix}
        N & p \\ 0 & \tfrac{1}{N}
    \end{pmatrix}^{-1} = \begin{pmatrix}
        \tfrac{1}{N} & -p \\ 0 & N
    \end{pmatrix} \, .
\end{align}
That is, $R$ is the inverse of the matrix formed by the basis vectors of $\Gamma'$ expressed in the basis of $\Gamma$.

The effect is the same as if we had performed a field redefinition
\begin{equation}
    \begin{pmatrix}
        \tilde{F} \\ F
    \end{pmatrix} \mapsto\begin{pmatrix}
        \tilde{F}' \\ F'
    \end{pmatrix}= R \begin{pmatrix}
        \tilde{F} \\ F
    \end{pmatrix} \, ,
\end{equation}
provided we also perform a M\"obius transformation on the complexified coupling,
\begin{equation}
    \tau \equiv \frac{\theta}{2\pi} + \frac{i}{g^2} \mapsto\tau' = R \cdot \tau \equiv \frac{a\tau+b}{c\tau+d}, \quad R = \begin{pmatrix}
        a&b\\c&d
    \end{pmatrix} \in SL(2, \mathbb{Q}) \, .
\end{equation}
For example, after gauging $\mathbb{Z}_N^{(e)}$ with discrete torsion $p$ the theory is equivalent to Maxwell theory at the coupling
\begin{align}\label{eq:gaugingtau}
    \tau' = \begin{pmatrix}
        \tfrac{1}{N} & -p \\ 0 & N
    \end{pmatrix} \cdot \tau = \frac{1}{N^2} \tau - \frac{p}{N} \, .
\end{align}

Notice that the value $p$ depends on the chosen lattice bases of $\Gamma$ and $\Gamma'$, i.e., the fundamental Wilson and 't Hooft line charges of the respective theories.\footnote{This is also why there is a priori no canonical choice of discrete torsion.
In 4d gauge theories, it is commonly understood that, by specifying the fundamental Wilson and 't Hooft line operators, or equivalently, the fields $\tilde{F}$ and $F$, one has implicitly made a basis choice for $\Gamma$, and consequently it makes sense to say that one gauges with ``no'', or ``trivial'' discrete torsion.}
This means that, by performing an integral basis change on $\Gamma$, which is nothing but an $SL(2,\bbZ)$ transformation \emph{before} gauging, one can change the value of $p$.
For example, starting from the standard basis ${\bf e}_{1,2}$ and performing the $Np$-th power of the T-duality transformation, given by $M = \left(\begin{smallmatrix}
    1 & Np \\ 0 & 1
\end{smallmatrix}\right)$, we preserve the original lattice $\Gamma$, but have now chosen a new lattice basis ${\bf v}_1 = \left(\begin{smallmatrix}
    1 \\ 0
\end{smallmatrix}\right)$
and ${\bf v}_2 = \left(\begin{smallmatrix}
    Np \\ 1
\end{smallmatrix}\right)$.
The basis change that maps these onto the basis \eqref{eq:basis-after-gauging-Z_N} of $\Gamma'$ is now simply $\tilde{R} = \left(\begin{smallmatrix}
    1/N & 0 \\ 0 & N
\end{smallmatrix}\right)$.
Physically, this means that gauging an electric $\mathbb{Z}_N$ with discrete torsion $p$ is equivalent to first performing the $T^{Np}$ duality transformation, and subsequently gauge with no discrete torsion.

Although the theory at $\tau' = R \cdot \tau$ is in general physically distinct from the original theory at coupling $\tau$, for \emph{special} values of $\tau$ there may exist an \emph{integer} matrix $M \in SL(2, \bbZ)$ mapping the coupling of the gauged theory back to the original one, $M \cdot \tau' = \tau$.
Hence, for those special values of $\tau$, the theory obtained after gauging ${\cal Z}$ at $\tau'$ is equivalent to the original one at $\tau$ up to a duality transformation.
E.g., at $\tau = iN$, gauging ${\cal Z} = \bbZ_N^{(e)}$ with no discrete torsion ($p=0$) produces a theory at $\tau' = i/N$, which is equivalent under S-duality to $\left( \begin{smallmatrix} 0 & -1 \\ 1 & 0 \end{smallmatrix} \right) \cdot \tau' = -1/\tau' = i N = \tau$.

\begin{figure}[ht]
\begin{tikzpicture}
\centering
\node (left) at (0,0)
{\includegraphics[width=0.96\textwidth]{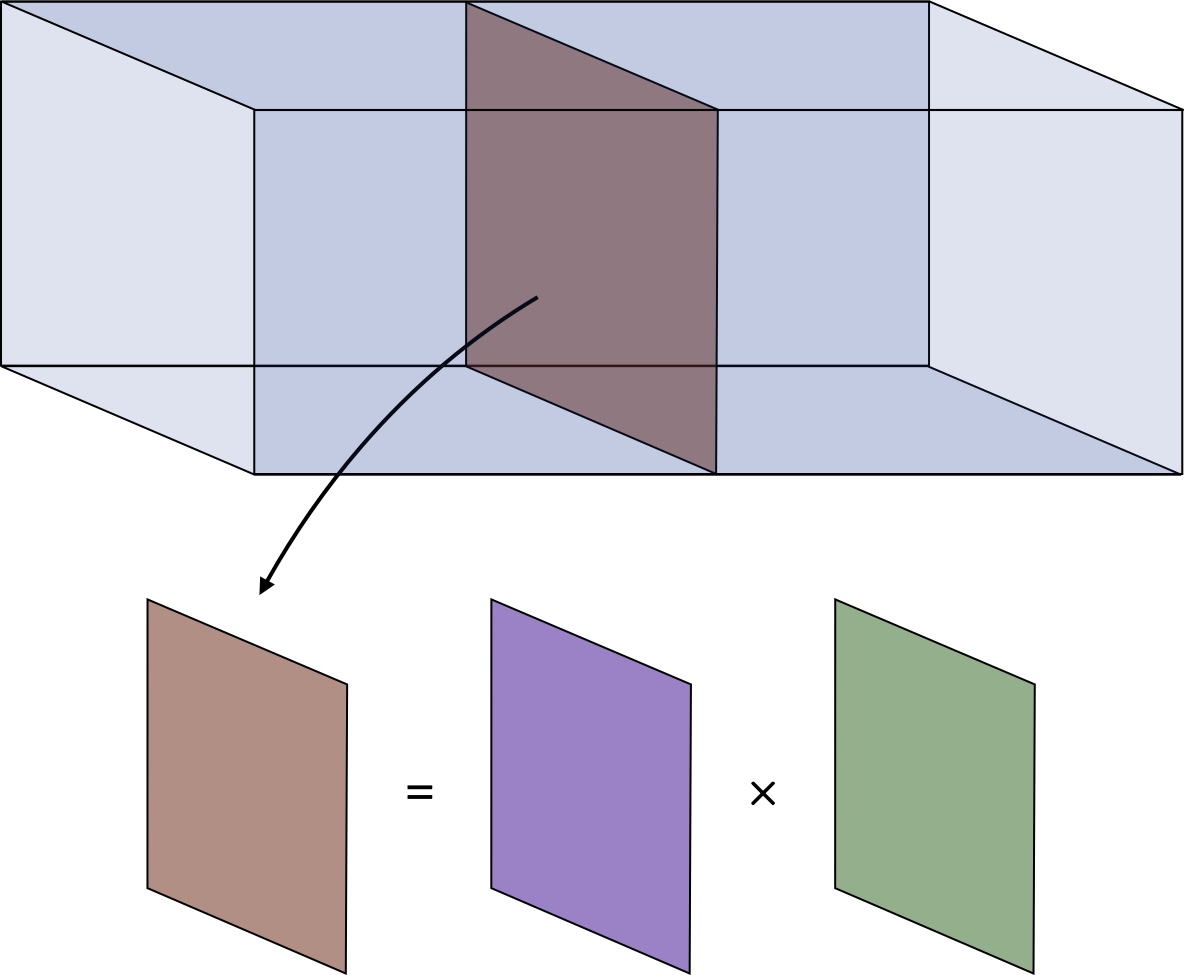}};
\node at (-1.25,5.4) {{$\mathcal D$}};
\node at (-5.15,-1.9) {{$\mathcal D$}};
\node at (-0.95,-1.9) {{$\mathcal W$}};
\node at (3.25,-1.85) {{$\sigma$}};
\node at (-5.75,3) {{$\mathcal{L}(\tau;A)$}};
\node at (5.75,3) {{$\mathcal{L}(\tau;A)$}};
\end{tikzpicture}
\caption{Decomposition of a duality defect as a wall $\sigma$ that performs a half-space gauging combined with an $SL(2,\mathbb{Z})$ duality interface $\mathcal{W}$.}
\label{fig:dual_def_gaug_dual}
\end{figure}

In such cases we can construct a codimension one defect ${\cal D}$ of the theory at $\tau$ by means of a \emph{half-space gauging} as follows.
Starting from gauging ${\cal Z}$ in half of spacetime (say, the right half), we create an interface $\sigma$ to the theory at $\tau' = R_\sigma \cdot \tau$; for generic values of $\tau$ and subgroups ${\cal Z}$ this new theory is physically distinct.
If $\tau = M \cdot \tau'$, the two theories are $SL(2,\bbZ)$-dual, and we can combine $\sigma$ with the duality wall ${\cal W}$ that implements the transformation associated to $M \in SL(2,\bbZ)$, resulting in an interface between the same theory on both sides, i.e., a defect ${\cal D}$ in the theory with coupling $\tau$.
It can be argued that such a defect is topological \cite{Choi:2021kmx, Kaidi:2021xfk} (see fig.~\ref{fig:dual_def_gaug_dual}).

An important remark is that a general gauging interface $\sigma$ can involve the stacking an SPT-phase for ${\cal Z}$ on the half-space in which the gauging takes place.
For a 1-form symmetry in 4d, this SPT-phase is a quadratic counterterm in the background gauge field $B$.\footnote{Explicitly, for ${\cal Z} = \bbZ_N$, the counterterm is proportional to $\frac{2\pi i}{N} \int \frac{N+1}{2} B \cup B$ for odd $N$, and $\frac{2\pi i}{N} \int \frac{1}{2} q(B)$ for $N$ even, with $q(B)$ the Pontryagin-square of $B$.\label{footnote:counterterm}}
For 4d Maxwell theory, it can be shown \cite{Choi:2022zal}\footnote{In the reference it is argued that the gauging/stacking procedure involves a shift of the $\theta$-angle in 4d Maxwell, which results in an equivalent modification of the spectrum of line operators as the discrete torsion parameter; see, e.g., \cite{Kapustin:2014gua, Hsin:2020nts}.} that this gauging/stacking operation is equivalent to gauging with discrete torsion, i.e., a duality transformation following a torsionless gauging.
The situation is analogous to gauging interfaces in 6d (self-dual) tensor gauge theories \cite{Lawrie:2023tdz}, and allows us to capture possible SPT-stackings entirely by $SL(2,\bbZ)$ duality transformations.

\begin{figure}[ht]
\centering
\begin{subfigure}[t]{0.46\textwidth}
    \centering
    \begin{tikzpicture}
        \node[anchor=south west,inner sep=0] (img) at (0,0)
            {\includegraphics[width=\linewidth]{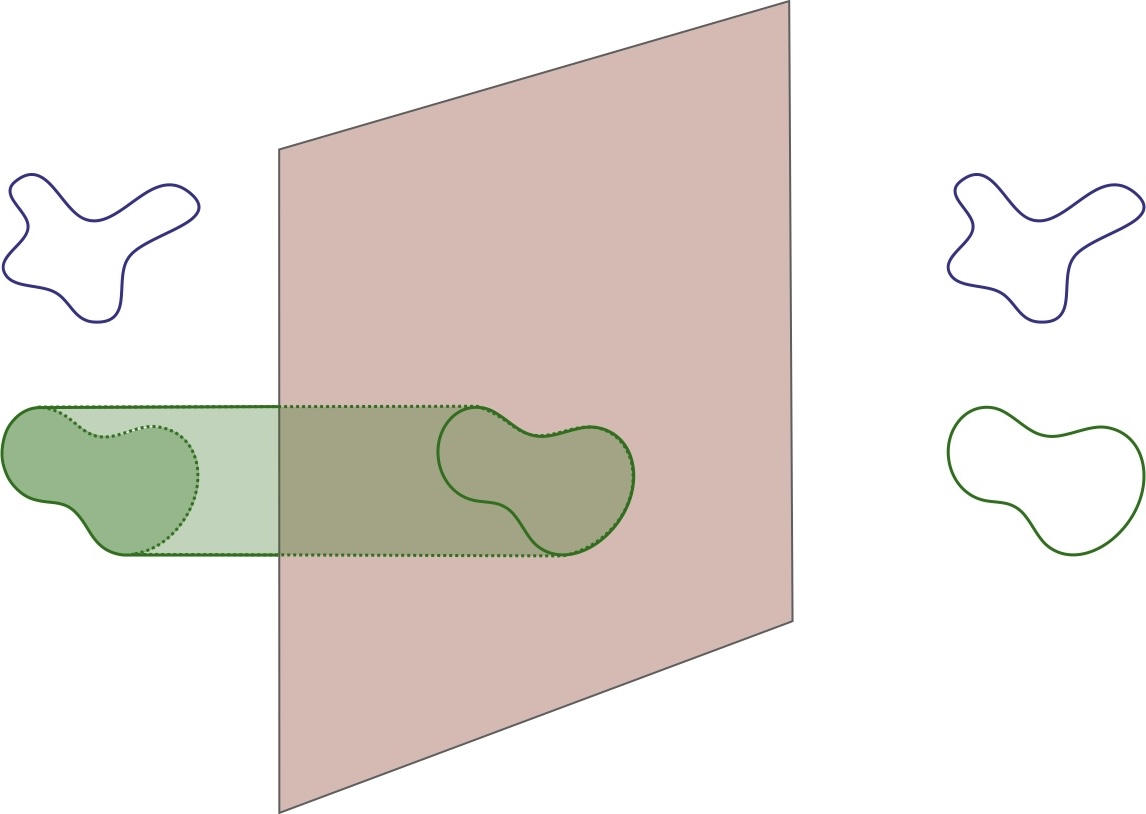}};

        \begin{scope}
            \node at (0.7,4.2) {$D{\bf q} \in \bbZ^2$};
            \node at (0.7,1.3) {$D{\bf p} \notin \bbZ^2$};
            \node at (6.4,4.2) {${\bf q} \in \bbZ^2$};
            \node at (6.4,1.3) {${\bf p} \in \bbZ^2$};
            \node at (4.45,4.5) {{$\mathcal{D}$}};
        \end{scope}
    \end{tikzpicture}
    \caption{Line operators parallel to the defect.}
    \label{fig:ge_non_gen_lines_parallel}
\end{subfigure}
\hfill
\begin{subfigure}[t]{0.46\textwidth}
    \centering
    \begin{tikzpicture}
        \node[anchor=south west,inner sep=0] (img) at (0,0)
            {\includegraphics[width=\linewidth]{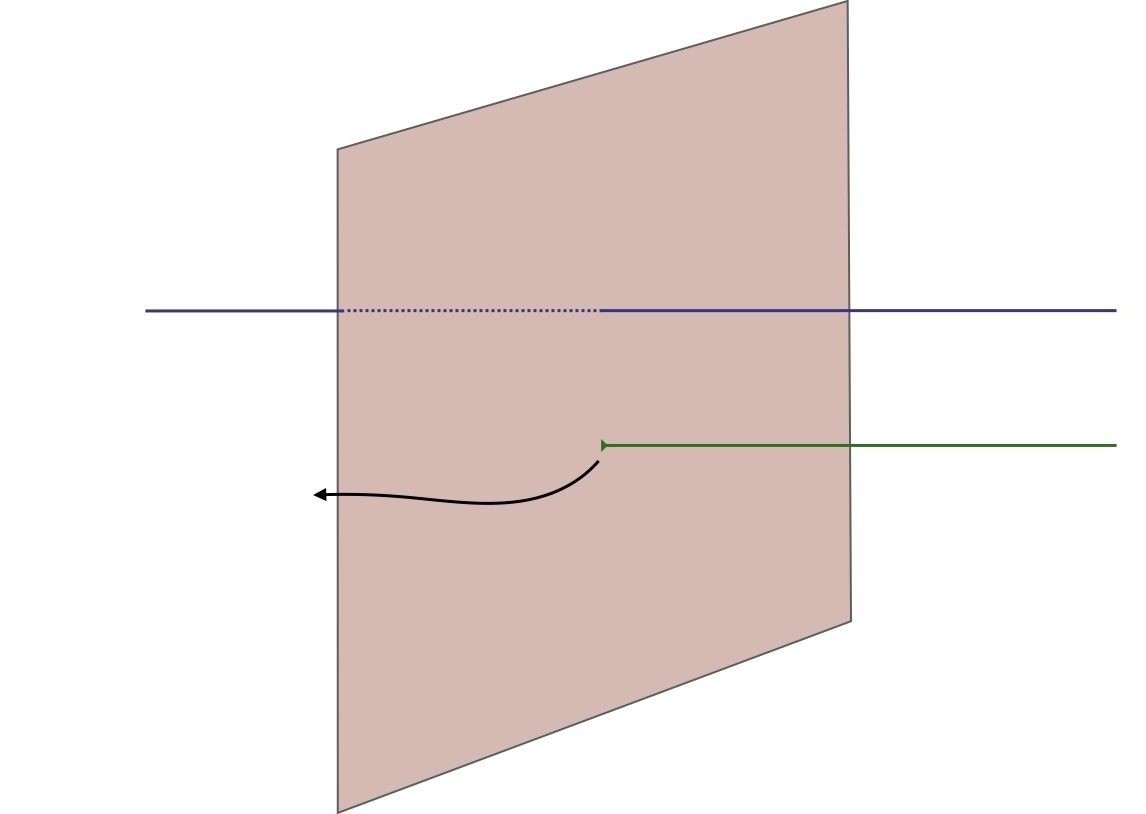}};

        \begin{scope}
            \node at (1.1,3.4) {$D{\bf q} \in \bbZ^2$};
            \node at (1.1,1.95) {$D{\bf p} \notin \bbZ^2$};
            \node at (6.05,3.4) {${\bf q} \in \bbZ^2$};
            \node at (6.05,1.95) {${\bf p} \in \bbZ^2$};
            \node at (4.8,4.5) {{$\mathcal{D}$}};
        \end{scope}
    \end{tikzpicture}
    \caption{Line operators piercing the defect.}
    \label{fig:ge_non_gen_lines_piercing}
\end{subfigure}
\caption{Line operators moved from right to left through the defect $\mathcal{D}$, either remaining genuine or becoming non-genuine. Both parallel (left) and piercing (right) line operators are represented.}
\label{fig:ge_non_gen_lines}
\end{figure}

The overall effect of ${\cal D}$ is captured by its action on the line operators, which is described by the generally rational matrix $D = M R \in SL(2,\mathbb{Q})$. 
In particular, if we pull a genuine line with charge ${\bf q} \in \Gamma$ ``parallel'' through ${\cal D}$, then the resulting line operator $WH_{{\bf q}'}$ is non-genuine iff ${\bf q}' = D {\bf q} \notin \Gamma$ is non-integer, and must be attached to a surface operator $\eta_{{\bf q}'}$.
Conversely, a surface operator $\eta_{\bf r}$ ending on a non-genuine line charge ${\bf r} \in \Gamma_\mathbb{Q}$ becomes a genuine line $WH_{{\bf r}'}$ if ${\bf r}' = D{\bf r} \in \Gamma$.
One can also ``pierce'' ${\cal D}$ with a line charge ${\bf q}$: if ${\bf q}' = D{\bf q} \in \Gamma$ remains integer, then the line continues as a genuine line, albeit with charge ${\bf q}'$; if ${\bf q}' \notin \Gamma$ is fractional, then physically the line ends on the defect. All of these mechanisms have been depicted in fig.~\ref{fig:ge_non_gen_lines} for illustration.

\subsubsection*{Examples: duality and triality defects}

Let us now close by reviewing the ``duality'' and ``triality'' defects of \cite{Choi:2021kmx, Choi:2022zal}. 
We start with the ``duality'' defects at $\tau=i N$.\footnote{We put ``duality'' into quotation marks here because we want to introduce a broader set of non-invertible defects built around the $SL(2,\bbZ)$-duality of the theory.
We shall refer to those as (generalized) duality defects in the following, of which the ``duality'' defect considered here is a special case.
} These arise from combining an interface $\sigma$ for electric gauging $\mathbb{Z}_N^{(e)} \subset U(1)^{(e)}$ (with vanishing discrete torsion: $p=0$) and an S-duality transformation with duality matrix $M$. Explicitly, its action on the charge lattice is given by
\begin{equation}
    D_2 = \underbrace{\begin{pmatrix}
        0 & -1 \\
        1 & 0
    \end{pmatrix}}_{M} \underbrace{\begin{pmatrix}
        \frac{1}{N} & 0 \\
        0 & N
    \end{pmatrix}}_{R} = \begin{pmatrix}
        0 & \frac{1}{N} \\
        - N & 0
    \end{pmatrix}\, .
\end{equation}
Indeed, one straightforwardly verifies that this $SL(2,\mathbb{Q})$ transformation fixes $\tau = i N$.

Next we consider the ``triality'' defects. 
In \cite{Choi:2022zal} this defect was introduced for a broader class of theories with 1-form symmetry $\bbZ_N$, which are self-dual under its gauging (without discrete torsion) after stacking with a quadratic counterterm, see footnote \ref{footnote:counterterm}.
For 4d Maxwell theory, this operation is equivalent to gauging with discrete torsion $p=1$ (for even $N$) or $p=N+1$ (for odd $N$) \cite{Choi:2022zal}.
At $\tau = e^{2 \pi i/3} N$, this can be undone by a suitable modular transformation: for even $N$ it is the S-duality transformation while for odd $N$ it is the combination $ST^{-1}$.
Both cases result in the same defect matrix:
\begin{align}
    D_3 = \underbrace{\begin{pmatrix}
        0 & -1 \\
        1 & 0
        \end{pmatrix}}_{S}
        \underbrace{\begin{pmatrix}
        \frac{1}{N} & 1 \\
        0 & N
    \end{pmatrix}}_{R_{p=1}} = \underbrace{\begin{pmatrix}
        0 & -1 \\
        1 & -1
        \end{pmatrix}}_{ST^{-1}}
        \underbrace{\begin{pmatrix}
        \frac{1}{N} & N+1 \\
        0 & N
    \end{pmatrix}}_{R_{p=N+1}} = \begin{pmatrix}
        0 & - N\\ \tfrac{1}{N} & 1
    \end{pmatrix} \, .
\end{align}

Moreover, as we have explained below \eqref{eq:gaugingtau}, a gauging with discrete torsion can be further decomposed into a modular transformation followed by an untwisted gauging.
In the present case of the triality defects, this manifests itself as
\begin{equation}
\begin{aligned}
    \begin{pmatrix}
        \tfrac{1}{N} & 1 \\ 0 & N
    \end{pmatrix}
    & = \begin{pmatrix}
        \tfrac{1}{N} & 0 \\ 0 & N
    \end{pmatrix}
    \begin{pmatrix}
        1 & N \\ 0 & 1
    \end{pmatrix} \quad &&(\text{relevant for $N$ even}) \, , \\
    \begin{pmatrix}
        \tfrac{1}{N} & N+1 \\ 0 & N
    \end{pmatrix}
    & = \begin{pmatrix}
        \tfrac{1}{N} & 0 \\ 0 & N
    \end{pmatrix}
    \begin{pmatrix}
        1 & N(N+1) \\ 0 & 1
    \end{pmatrix} \quad &&(\text{relevant for $N$ odd}) \, .
\end{aligned}
\end{equation}
We thus have
\begin{equation}\label{eq:smith_decomp_triality}
    D_3 = \underbrace{\begin{pmatrix}
        0 & -1\\
        1 & 0 
    \end{pmatrix}}_{S} \begin{pmatrix}
        \frac{1}{N} & 0 \\
        0 & N
    \end{pmatrix} \underbrace{\begin{pmatrix}
        1 & N \\
        0 & 1
    \end{pmatrix}}_{T^N}
    = \underbrace{\begin{pmatrix}
        0 & -1\\
        1 & -1 
    \end{pmatrix}}_{ST^{-1}} \begin{pmatrix}
        \frac{1}{N} & 0 \\
        0 & N
    \end{pmatrix} \underbrace{\begin{pmatrix}
        1 & N(N+1) \\
        0 & 1
    \end{pmatrix}}_{T^{N(N+1)}}
    \, .
\end{equation}
These equations give another interpretation for the construction of the ``triality'' defect: the outer factors are honest $SL(2,\mathbb{Z})$ duality walls and the middle factor is a torsionless gauging. Clearing the denominator so that $N D_3$ is integral, these factorizations are precisely \textit{Smith decompositions} $N D_3 = S \, \mathrm{diag}(1, N^2) \, T^N = (ST^{-1}) \, \mathrm{diag}(1,N^2) \, T^{N(N+1)}$. As we will see in the next sections, Smith decomposition will play a central role in our discussion.

Two comments are in order.
First, the distinction in \cite{Choi:2022zal} between even and odd $N$ originates from the construction of the ``triality'' defect with counterterms; this was done so because they apply to a broader class of theories.
For 4d Maxwell, we can instead use the $SL(2,\bbZ)$ transformations before and after gauging, which does not depend on the parity of $N$.
Hence, both factorizations in \eqref{eq:smith_decomp_triality}, when interpreted as stackings of duality and gauging interfaces, give rise to a topological defect in the theory, irrespective of the parity of $N$.
This immediately raises a second point, namely, whether two such stackings with different duality interfaces give the same defect.

The ambiguity generically arises in the Smith decomposition $D = U \, \mathrm{diag}(\tfrac{1}{N}, N) \, V$ (with $ND$ an integer matrix), where the diagonal part\footnote{We will argue in section \ref{ssec:action} why the diagonal part for a defect matrix $D$ from half-space gauging a $\bbZ_N$ 1-form symmetry subgroup must take this form.} is unique, but the left/right $SL(2,\bbZ)$-matrices $U,V$ can be different.
Physically they correspond to a different lattice basis, i.e., a different set of fundamental Wilson and 't Hooft lines, in which we do the purely electric gauging.
A priori, we do not have a straightforward argument why this cannot lead to different defects.
However, we will argue in the subsequent sections that this is the case.

In the remainder of the paper, we shall use such decompositions $D = U \, \mathrm{diag}(\tfrac{1}{N}, N) \, V$ with $U,V \in SL(2,\mathbb{Z})$ to \emph{define} a duality defect for every (elliptic) $SL(2,\mathbb{Q})$ transformation $D$.
Namely, it is a torsionless electric gauging defect fused with, from both left and right, an $SL(2, \mathbb{Z})$ duality wall.
By contrast, presenting a duality defect as a single $SL(2,\mathbb{Z})$ duality wall fused to an electric gauging wall with torsion is not always possible; subsection \ref{ssec:tau=3i/2} gives an $SL(2,\mathbb{Q})$ duality defect that fails to admit such a presentation but does admit a Smith decomposition.

\section{Arithmetic characterization of duality defects}\label{sec:CMdefects}

We have seen that any duality defect is associated with an $SL(2,\bbQ)$ matrix.
In this section, we show that there is a natural way to reverse the identification, in such a way to associate to every such (elliptic) matrix a corresponding duality defect, or rather, a generalization of the description of duality defects given above. To do so, we relate their topological nature to the first arithmetic structure we encounter, namely that of complex multiplication symmetries of two-tori. While this article will be self-contained, we refer the reader to \cite{Milne:CM, Cox:Primes, Gukov:2002nw} for reviews on this topic.

\subsection{Topological interfaces from stress tensor continuity}

The starting point is an interface $\mathcal{D}$ supported on a codimension-one wall $W$. For the moment, we characterize this interface purely by its action as a linear transformation $D \in SL(2,\mathbb{Q})$ on the charge lattices of Wilson--'t Hooft lines and Gukov--Witten surfaces. In particular, we keep the complexified couplings $\tau_L$ and $\tau_R$ on its two sides distinct, and do not yet assume that they are related by a modular transformation.
Following \cite{Niro:2022ctq}, we impose that the interface is \textit{topological}, which requires the stress-energy tensor $T_{\mu \nu}$ to be continuous across $W$. Working out what this demands of $D$---both for interfaces ($\tau_L \neq \tau_R$) and defects ($\tau_L = \tau_R)$---is the content of this subsection. 

To this end, let us first recall the stress-energy tensor of Maxwell theory, which can be written succinctly as
\begin{equation}\label{eq:Tmunu}
    T_{\mu \nu } = \frac{1}{2\pi g^2} (F_{\mu \alpha} F_{\nu}^{\ \alpha} - \frac{1}{4} g_{\mu \nu}F_{\alpha \beta}F^{\alpha \beta}) = \frac{1}{4\pi} \text{Im}(\tau) \delta_{IJ} (\mathcal{F})^I_{\mu \alpha} (\mathcal{F}^J)_{\nu}^{\ \alpha}\, ,
\end{equation}
where we have collected the field strengths into the two-component vector $\mathcal{F}^I = (i \ast F, F)$. 
For our purposes, it is more convenient to pass to the quantized field strengths $(\tilde{F}, F)$, which are given by
\begin{equation}\label{eq:Fbold}
    \mathbf{F} = \begin{pmatrix}
        \tilde{F} \\
        F
    \end{pmatrix} = \begin{pmatrix}
        \tau_1 F - \tau_2 \, i \ast F\\
        F
    \end{pmatrix} = \begin{pmatrix}
        -\tau_2 & \tau_1 \\
        0 & 1
    \end{pmatrix} \mathcal{F}\, .
\end{equation}
Rewriting $T_{\mu \nu}$ in terms of $\mathbf{F}$ replaces the $\delta_{IJ}$ by a $\tau$-dependent pairing, we get
\begin{equation}\label{eq:TmunuE}
    T_{\mu \nu}  = \frac{1}{4\pi}\mathcal{E}_{IJ}(\tau)(\mathbf{ F}^I)_{\mu\alpha}
    (\mathbf{F}^J)_{\nu}^{\alpha}\, , \qquad \mathcal{E}_{IJ}(\tau) \equiv \frac{1}{\text{Im}(\tau)}\begin{pmatrix}
        1 & - \text{Re}(\tau) \\
        - \text{Re}(\tau) & |\tau|^2
    \end{pmatrix}\, .
\end{equation}
This pairing is familiar from 2d toroidal CFTs, cf.~\cite{Giveon:1994fu}, where it computes the conformal weights of vertex operators. In 4d $U(1)$ Maxwell theory it plays an analogous role, determining the scaling dimension of Wilson--'t Hooft line operators \cite{Kapustin:2005py},
\begin{equation}\label{eq:scalingdims}
    h: \mathbb{Z}^2 \to \mathbb{R} \, , \qquad \mathbf{q} \to \mathbf{q}^T \mathcal{E}(\tau) \mathbf{q} = \frac{|q_1 - q_2 \tau|^2}{\text{Im}(\tau)}\, .
\end{equation}
Analogously to the discussion in section \ref{ssec:4dMaxwell}, the transformation on the charge lattice induces a transformation on the field strength vector, $\mathbf{F}_L = D {\bf F}_R$.
Continuity of $T_{\mu \nu}$ is therefore not automatic: it holds only if $D$ and the couplings $\tau_L, \tau_R$ satisfy
\begin{equation}
    D^{T} \mathcal{E}(\tau_L) D \overset{!}{=} \mathcal{E}(\tau_R)\, .
\end{equation}
On the other hand, using the explicit form \eqref{eq:Tmunu}, we can verify that for any $D \in SL(2,\mathbb{Q})$ the pairing obeys the general identity $ D^{T} \mathcal{E}(\tau)D =\mathcal{E}(D^{-1} \cdot \tau) $.
Comparing the two sides, we conclude that an interface $\mathcal{D}$ separating half-spaces with couplings $\tau_L$ and $\tau_R$ can be topological only if its duality transformation $D \in SL(2,\mathbb{Q})$ relates them by
\begin{equation}
    \tau_L = D \cdot \tau_R\, .
\end{equation}
In other words, the interface ${\cal D}$ is topological only if it implements a M\"obius transformation on the coupling by the same matrix $D$ that acts on the charge lattice, i.e., it is a duality interface.

To obtain a topological \textit{defect}, we must have $\tau_L =\tau_R = \tau$. 
The topological nature of the defect is then captured by either of two equivalent conditions
\begin{equation}\label{eq:automorphism}
    \tau = D \cdot \tau \,, \qquad \iff \qquad D^T \mathcal{E}(\tau) D = \mathcal{E}(\tau)\, .
\end{equation}
The first condition is the familiar one, stating that $D$ fixes the coupling $\tau$, while the second states that $D$ is a rational automorphism of the Euclidean pairing $\mathcal{E}(\tau)$. As we explain in the following subsection, the latter is precisely the arithmetic structure encoded in the complex multiplication of two-tori. Already at this stage, however, we can draw a useful constraint from the characteristic equation for the eigenvalues $\lambda$ of $D$,
\begin{equation}
    \lambda^2 - \text{tr}(D) \lambda + 1 = 0\, .
\end{equation}
Since $D \, (\tau, 1)^T = (c \tau + d) (D \cdot \tau , 1)^T $ for $D = \left(\begin{smallmatrix}
    a & b \\
    c & d
\end{smallmatrix} \right)$, the condition \eqref{eq:automorphism} is equivalent to $(\tau, 1)^T$ being an eigenvector of $D$ with eigenvalue $\lambda = c\tau+d$. Because $\tau$ lies in the upper-half plane, this eigenvalue cannot be real. And as we have $\det(D) = 1$, its eigenvalues must be complex conjugates with $|\lambda|=1$. This requires a negative discriminant, 
\begin{equation}
    \text{tr}(D)^2-4 < 0\, .
\end{equation}
In the standard classification of $SL(2,\mathbb{R})$ elements as elliptic ($\text{tr}(D)<2$), parabolic ($\text{tr}(D)=2$), and hyperbolic ($\text{tr}(D)>2$), this means that only \textit{elliptic} elements of $SL(2,\mathbb{Q})$ can give rise to duality defects.

Let us briefly pause and compare with the results of \cite{Niro:2022ctq}. There, continuity of $T_{\mu \nu}$ was imposed directly at the level of \eqref{eq:Tmunu} in terms of the fields ${\cal F} = (i *F, F)$, yielding an induced $SO(2)$ rotation that preserves the Kronecker pairing $\delta_{IJ}$. The two formulations describe automorphisms of the same pairing written in different bases, related by the change of frame in \eqref{eq:Fbold}. The advantage of working with $\mathcal{E}(\tau)$ over $\delta_{IJ}$ is that it makes the underlying arithmetic structure manifest: we will find that $\mathcal{E}(\tau)$ takes a special integral form, with $D$ the rational $SL(2)$ matrix acting on its associated charge lattice.

\subsection{Defect matrix from complex multiplication}\label{ssec:construction}

Let us now recast the condition \eqref{eq:automorphism} in terms of arithmetic on a two-torus. 
To this end, note that any complex number $\tau = \text{Re}(\tau)+i \, \text{Im}(\tau) \equiv \tau_1 + i \tau_2$ with $\tau_2>0$ can be used to define a flat positive-definite metric on the complex plane,
\begin{equation}
    \dd s^2 = \frac{1}{\tau_2}|\dd x_1 - \tau \dd x_2|^2 \equiv \mathcal{E}_{IJ}(\tau) \dd x_I \dd x_J\, ,\qquad z = x_1 - \tau x_2 \in \mathbb{C}\, .
\end{equation}
The minus sign is chosen for later consistency with the modular transformations of the charges of Wilson--'t Hooft lines and Gukov--Witten surfaces, cf.~\eqref{eq:SL2WHeta} and the sign conventions in their respective definitions \eqref{eq:etaWG}, \eqref{eq:wilson-thooft-lines-definition}. The point of writing the metric in this form is that the metric components 
$\mathcal{E}_{IJ}$ define precisely the pairing appearing in the stress tensor \eqref{eq:TmunuE}. This trivial reformulation becomes meaningful once we introduce an auxiliary complex two-torus $T^2$ with complex structure $\tau$, defined as the quotient of the complex plane by the lattice $\Lambda = \mathbb{Z}+\tau \mathbb{Z} \subset \mathbb{C}$ spanned by $1$ and $\tau$:
\begin{equation}
    T^2 = \mathbb{C}/(\mathbb{Z}+\tau \mathbb{Z})\, .
\end{equation}
The metric above then descends to the standard flat metric on $T^2$. Considering $\Lambda$ as a subset of $\mathbb{C}$, this lattice naturally inherits the pairing $\mathcal{E}(\tau)$. Thus, the pairing governing the scaling dimensions \eqref{eq:scalingdims} of Wilson--'t Hooft lines is nothing more than the flat metric on this auxiliary torus.

Consider next a linear transformation $D = \left(\begin{smallmatrix}
    a & b \\
    c & d
\end{smallmatrix}\right)\in SL(2,\mathbb{Q})$ on $\mathbb{C}\cong\mathbb{R}^2$ in the $(1,-\tau)$-basis, that is, on the coordinate pair $(x_1, x_2)$:
\begin{equation}\label{eq:SL2onC}
    \mathbb{C} \ni x_1 - \tau x_2 \to \begin{pmatrix}
        a x_1 + b x_2 \\
        c x_1 + d x_2 
    \end{pmatrix} \cong ax_1 + b x_2 - \tau (c x_1 + dx_2)\, .
\end{equation}
We can now match these points and their transformation properties with physical operators. Points inside the unit cell of $T^2$, parametrized by $(x_1, x_2) \in \mathbb{R}^2/\mathbb{Z}^2$, label the Gukov--Witten symmetry surfaces. The lattice points $(x_1, x_2) \in \mathbb{Z}^2$ embedded in $\mathbb{C}$ label the charges of Wilson--'t Hooft lines instead. For both, the transformation rule of $(x_1,x_2)$ also matches their modular transformations \eqref{eq:SL2WHeta}. 

In contrast to the physical setting, where both the physical charges and the complexified gauge coupling transformed actively under $SL(2,\mathbb{Q})$, we have kept the torus complex structure $\tau$ fixed in the transformation \eqref{eq:SL2onC}. Since $SL(2,\mathbb{Q})$ does not preserve integral lattice sites, the transformation of the coordinates $(x_1,x_2)$ takes us to another lattice. To read off its effect on the torus, we act on the lattice generators  $(1,0)\cong 1$ and $(0,1) \cong -\tau$:
\begin{equation}
    1 \mapsto a -c \tau \, , \qquad \tau \mapsto d\tau-b\, .
\end{equation}
This maps the lattice $\Lambda$ to a new lattice
\begin{equation}
    \Lambda' = (a-c\tau) \mathbb{Z}+ (d\tau-b)\mathbb{Z}\, . 
\end{equation}
Equivalently, it maps the torus with complex structure $\tau$ onto a new torus $\mathbb{C}/\Lambda'$, with complex structure
\begin{equation}
    \tau' = \frac{d\tau-b}{a-c \tau}= \begin{pmatrix}
        a & b \\
        c & d
    \end{pmatrix}^{-1}\cdot \tau \qquad \text{(with $ad-bc=1$)}\, .
\end{equation}
Comparing with section \ref{ssec:4dMaxwell}, we see that $\tau$ here transforms passively---with $D^{-1}$---rather than actively---with $D$---as before.\footnote{To see why this difference arises here, it is instructive to compare with the physical process: when a Gukov--Witten operator is moved through a defect, the charges and field strengths transform such that
$$x_1' F' - x_2' \tilde{F}' = -\mathcal{J}(D(x_1, x_2), D(\tilde{F}, F)) =  -\mathcal{J}((x_1, x_2), (\tilde{F}, F)) = x_1 F - x_2 \tilde{F}\, .$$
In contrast, when we took a point $z = x_1 - \tau x_2 \in \mathbb{C}$, we only transformed $x_1,x_2$. By writing this point by using the Dirac pairing, we see that
$$z = - \mathcal{J}((x_1 , x_2 ), (\tau, 1)) \to \mathcal{J}(D(x_1, x_2), (\tau, 1)) = \mathcal{J}((x_1, x_2), D^{-1}(\tau, 1))\, . $$} In the remainder of this section we switch back to $\tau \mapsto D \tau$ to align with our physics-derived conventions; the difference does not matter for our purposes anyway, since we will require $\tau = D \tau$.

Let us now return to the duality defects, and interpret the physical condition \eqref{eq:automorphism} they impose from a geometric point of view. Such a defect induces an elliptic $SL(2,\mathbb{Q})$ transformation that fixes $\tau$, or equivalently, preserves the pairing $\mathcal{E}(\tau)$. From the geometrical perspective of the torus, this means that the duality defect gives rise to a \emph{lattice automorphism} of $\Lambda$
\begin{equation}
    SO(\mathcal{E}, \mathbb{Q}) = \{ D \in SL(2,\mathbb{Q})\, | \, D^T \mathcal{E}D = \mathcal{E}  
    \}  \, .
\end{equation}
In other words, the rational automorphisms of $\Lambda$ are determined by its positive pairing $\mathcal{E}$. As we will see shortly, depending on what lattice $\Lambda$ we have (i.e.~the precise value of $\tau$), the group $SO(\mathcal{E}, \mathbb{Q})$ contains only trivial elements or infinitely many distinct defect matrices. Indeed, for a generic lattice $\Lambda$ this group consists of just $\{\pm \mathbbm{1}_2\}$, and only at special points it enhances to additional transformations.

Now, to give a succinct characterization of these non-trivial automorphisms---and thereby any defect matrix $D$---we use the fact that they have a well-known equivalent description in terms of a \emph{complex multiplication symmetry} of the torus $T^2 = \mathbb{C} /\Lambda$. Let us work over the reals $\mathbb{R}$ first instead of the rationals $\mathbb{Q}$. Taking $D \in SO(\mathcal{E},\bbR)$ as a linear map on $\mathbb{R}^2 \cong \mathbb{C}$ that fixes $\Lambda$, it naturally descends to a linear map on the torus.
Over $\mathbb{C}$, any $SO(2,\bbR)$ action may be represented by multiplication by some complex number $\alpha \in U(1) \subset \mathbb{C}$, thus explaining the name of the symmetry.
In addition, to fix $\alpha$ as a rational map, we have to require that there exist some rational numbers $r_{1,2}, s_{1,2} \in \bbQ$ such that
\begin{align}
    \alpha \cdot 1 = r_1 + s_1 \tau \, , \quad \alpha \cdot \tau = r_2 + s_2 \tau \, .
\end{align}
For generic values of $\tau$ one finds only $\alpha = \pm 1$ as possible solutions. However, by clearing denominators and substituting the first equation into the second, one finds that there are additional automorphisms when 
\begin{equation}\label{eq:CM}
    a \tau^2 + b \tau + c = 0\,   \quad \Leftrightarrow \quad \tau = \frac{-b+i\sqrt{4ac-b^2}}{2a}\, ,
\end{equation}
where $a,b,c \in \mathbb{Z}$ are co-prime, and we write $d=4ac-b^2>0$ for the discriminant. Restricting $\tau$ to lie in the fundamental domain, i.e.~$|\tau_1| \leq 1/2$, $\tau_2\geq0$ and $|\tau|^2 \geq 1$, amounts to the ranges $|b| \leq a \leq c$. At these values \eqref{eq:CM} for $\tau$, the pairing \eqref{eq:TmunuE} takes the form
\begin{equation}
    \mathcal{E} \big|_{\tau = \eqref{eq:CM}}= \frac{1}{\sqrt{4ac-b^2}} \begin{pmatrix}
        2a & b \\
        b & 2c
    \end{pmatrix}\, .\label{eq:binary}
\end{equation}
Up to a squareroot factor $\sqrt{d}$, such a pairing is also called a \textit{binary quadratic form}. Non-invertible defects in 4d U(1) Maxwell theory thus arise \emph{precisely} when the complexified gauge coupling takes a complex multiplication value of the form \eqref{eq:CM}; moreover, the corresponding defect matrices are rational automorphisms of this binary quadratic form \eqref{eq:binary}.

In order to characterize the defects that make up $SO(\mathcal{E},\mathbb{Q})$, it is instructive to briefly introduce some mathematical notions related to complex multiplication. While we have considered complex multiplication  as a rational automorphism of a lattice so far, in mathematics one usually considers it as the (integral) endomorphisms of $T^2$ itself: these are maps specified by $\alpha \in \mathbb{C}$ that preserve the integral lattice $\mathbb{Z}+\tau \mathbb{Z}$ with pairing \eqref{eq:TmunuE}. This gives rise to the endomorphism ring
\begin{equation}\label{eq:endZ}
    \text{End}_{\mathbb{Z}}(T^2) = \begin{cases}
    \mathbb{Z} \oplus a \tau \mathbb{Z} \qquad &\text{$\tau$ satisfies  \eqref{eq:CM}}\, , \\
    \mathbb{Z} \qquad &\text{else}\, ,
    \end{cases}
\end{equation}
in analogy to \eqref{eq:O(T2)}. However, these integral endomorphisms generally do \textit{not} preserve the pairing \eqref{eq:binary}, but rescale it. For the purposes of constructing non-invertible defects, we first have to pass to the rational endomorphism algebra
\begin{equation}
    \text{End}_{\mathbb{Q}}(T^2) = \text{End}_{\mathbb{Z}}(T^2) \otimes \mathbb{Q} = \mathbb{Q}(a\tau)\, ,
\end{equation}
which is the algebraic number field generated by $a \tau$.
Due to \eqref{eq:CM}, this is isomorphic to the quadratic field $\mathbb{Q}(a\tau) \cong \mathbb{Q}(i \sqrt{d})$, but we choose to stick with the former, because its close relation to the integral endomorphisms \eqref{eq:endZ} provides a more natural parametrization for the defect matrices in the following. To determine the rational endomorphisms that give rise to these topological defects, we need to demand that they do not rescale $\mathcal{E}$.
But these are nothing other than linear transformations with unit determinant, i.e., elements of $SL(2,\mathbb{Q})$! In terms of the complex multiplication (CM) element $\alpha$, it amounts to requiring a unit norm 
\begin{equation}\label{eq:alpha}
    \alpha = \frac{1}{k} (\underbrace{m+ n \, a \tau}_{\in \text{End}_\bbZ (T^2) }) \in \mathbb{Q}(a \tau ) = \text{End}_\bbZ(T^2) \otimes \bbQ , \qquad \text{with }|\alpha|^2 = 1\, ,
\end{equation}
where we chose to parametrize $\alpha$ by integers $m,n, k \in \mathbb{Z}$ with $m,n$ co-prime. Upon inserting the CM value \eqref{eq:CM}, the unit norm constraint gives rise to a Diophantine problem
\begin{equation}\label{eq:diophantine}
    m^2 -b mn + ac n^2 = k^2\, ,
\end{equation}
which, in general, has infinitely many primitive solutions. The corresponding elements of $SL(2,\mathbb{Q})$ are given by 
\begin{equation}\label{eq:defectCM}
    D_{m,n,k}= \frac{1}{k}\begin{pmatrix}
        m-b n & -c n\\
        a n& m
    \end{pmatrix} \in SL(2,\mathbb{Q})\, .
\end{equation}
These matrices \eqref{eq:defectCM} are exactly the elements of $SL(2,\mathbb{Q})$ that stabilize the CM value \eqref{eq:CM} of $\tau$. In analogy with \eqref{eq:endZ}, we can thus write the rational automorphism group classifying the defects as
\begin{equation}\label{eq:O(T2)}
\begin{split}
  SO(\mathcal{E}, \mathbb{Q}) = \begin{cases}
    \{ D_{m,n,k \in \mathbb{Z}} | m^2 - b\,mn + ac\,n^2 = k^2,\ \gcd(m,n)=1\big\}
      & \text{if $\tau$ satisfies \eqref{eq:CM}}\, ,\\[4pt]
    \mathbb{Z}_2 \cong \{ \pm \mathbbm{1}_2 \} & \text{else}\, .
  \end{cases}
\end{split}
\raisetag{18pt}
\end{equation}
See also appendix C of \cite{Grimm:2024fip} for an alternative derivation of these automorphisms from a group-theoretic point of view.

\subsection{(Generalized) Duality defects from complex multiplication symmetries}\label{ssec:action}

Having classified the defect matrices from complex multiplication symmetries, we now want to associate a concrete topological defect to each matrix in the parametrization \eqref{eq:defectCM}.
Specifically, we claim that any such matrix has a realization in terms of a single half-space gauging of a cyclic finite 1-form symmetry, which generalizes the construction we reviewed in Section \ref{ssec:defectsfromgauging}.

The key for the construction, as anticipated in our review of the triality defect, is the Smith decomposition of the defect matrix \eqref{eq:defectCM}.
More precisely, by taking out the least common multiple denominator $k$, $D = \frac{1}{k} \tilde{D}$, we perform the Smith decomposition on the integer matrix $\tilde{D}$,
\begin{align}
    \tilde{D} = U \begin{pmatrix}
        \delta_1 & 0 \\ 0 & \delta_2
    \end{pmatrix} V \, .
\end{align}
Here $\delta_{1,2} \in \bbZ$ are integers with $\delta_1 | \delta_2$ fixed by $\tilde{D}$, while $U,V \in SL(2,\mathbb{Z})$ are generally non-unique.
From the divisibility assumptions regarding $a,b,c, m,n \in \bbZ$ that go into the definition of \eqref{eq:defectCM}, the entries in $\tilde{D}$ are coprime, which implies $\delta_1 = 1$.
Furthermore, since $D = \tilde{D}/k \in SL(2,\bbQ)$, we have $\delta_2 = \det(\tilde{D}) = \det(k D) = k^2 \det(D) = k^2$.
Thus, we have an associated decomposition for the defect matrix $D$,
\begin{equation}\label{eq:smith}
    D = \frac{1}{k} \tilde{D} = U \begin{pmatrix}
        \frac{1}{k} & 0 \\
        0 & k
    \end{pmatrix} V\, , 
\end{equation}
for some (non-unique) $U,V \in SL(2,\mathbb{Z})$. 

\begin{figure}[ht]
\begin{tikzpicture}
\centering
\node (left) at (0,0)
{\includegraphics[width=0.99\textwidth]{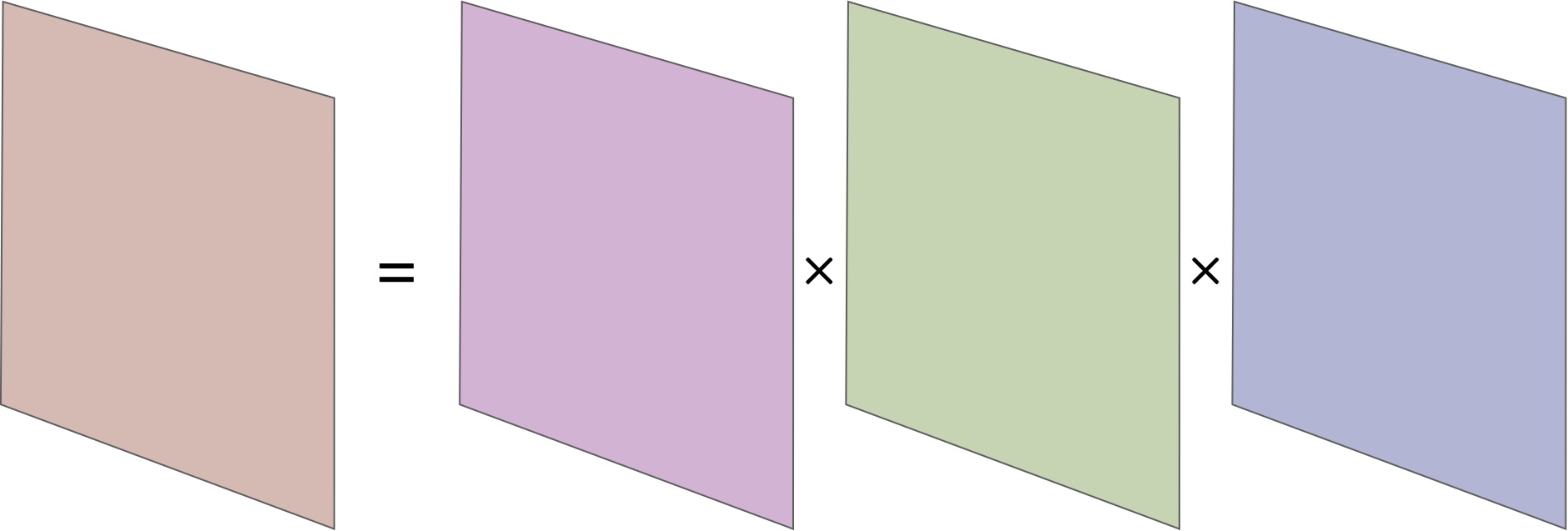}};
\node at (-7.1,1.9) {\LARGE{$\mathcal D$}};
\node at (-2.65,1.9) {\LARGE{$\mathcal U$}};
\node at (1.07,1.9) {\LARGE{$\mathcal G_\text{el}$}};
\node at (4.7,1.9) {\LARGE{$\mathcal V$}};
\end{tikzpicture}
\caption{Representation of the Smith decomposition of a duality defect $\mathcal{D}$ in terms of two invertible duality walls $\mathcal{U}$ and $\mathcal{V}$ surrounding from the left and the right, respectively, a wall $\mathcal{G}_\text{el}$ gauging an electric discrete subgroup $\bbZ_k\subset U(1)_e^{(1)}$.}
\label{fig:Smith_deco_defects}
\end{figure}

This decomposition has a simple physical interpretation, as depicted in fig.~\ref{fig:Smith_deco_defects}: the defect matrix $D$ can be realized by two invertible duality walls $\mathcal{U}, \mathcal{V}$ that implement the transformations $U,V$ located on the left and the right, while in the middle we have a wall $\mathcal G_{el}$ that gauges an electric discrete subgroup $\mathbb{Z}_k \subset U(1)^{(1)}_e$ to its left side.
When pushed together, they form a defect ${\cal D}$ that implements the half-space gauging of a (generally dyonic) 1-form symmetry subgroup of order $k$.
In fact, we can perform a duality transformation $V$ on the theory, thereby transforming the defects through conjugation as
\begin{equation}\label{eq:VDV}
    D \to V D V^{-1} = V U \begin{pmatrix}
    \frac{1}{k} & 0 \\
    0 & k
    \end{pmatrix}\, ,
\end{equation}
which can be understood as stacking an invertible duality wall $\mathcal{V}$ from the left and $\mathcal{\bar V}$ from the right.

The conjugation relation \eqref{eq:VDV} also explains why we have found infinitely many non-invertible duality defects at a given CM value of $\tau$. Let us start from duality defects realized as a half-space gauging with an (invertible) duality transformation on only one side; for instance, defects of the form \eqref{eq:smith} but with $V=1$. Indeed, in this case only finitely many of these defects can arise at a given point. However, we also inherit such defects located at other points $\tau'$ on the upper-half plane, related by some duality transformation $V \in SL(2,\mathbb{Z})$ as $\tau = V \cdot \tau'$. Then following \eqref{eq:VDV}, a defect $\mathcal{D}'$ at $\tau'$ yields a defect $\mathcal{D} $ at $\tau$ that acts by $D = V D' V^{-1}$. In this way, for a given point $\tau$, we inherit an infinite set of defects through its infinitely many dual frames under SL$(2,\mathbb{Z})$.

For the moment, we will take definition of the defects ${\cal D} = {\cal U} \times {\cal G} \times {\cal V}$ as dependent on $U$ and $V$, and discuss their action on the line and surface operators of the theory.
We will address the ambiguity coming from the non-uniqueness of the Smith decomposition subsequently.

\paragraph{Action on line and surface operators.} 
The action of our defects on line operators follows straightforwardly from the action described in fig.~\ref{fig:ge_non_gen_lines} for defects of the form ${\cal U} \times \sigma$, where $\sigma$ is a generic gauging defect and ${\cal S}$ an integer duality interface.
By letting $\sigma = {\cal G}$ be the torsionless electric gauging, we see that a duality defect of the form ${\cal U} \times {\cal G} \times {\cal V}$ simply acts additionally with an invertible duality transformation $V$ which rotates the charges but keeps (non-)genuine lines (non-)genuine.

As a result, the action of a defect defined by a Smith decomposition $D = U \, \text{diag}(\tfrac{1}{k}, k) \, V$ on line operators is identical to the situation described in fig.~\ref{fig:ge_non_gen_lines}:
As we pull genuine lines with charge ${\bf q}$ through ${\cal D}$, those with integral transformed charges ${\bf q}' = D{\bf q}$ remain genuine, while those with fractional charges become non-genuine.
When genuine lines ``pierce'' ${\cal D}$, they are trivialized if $D{\bf q}$ is fractional.

The sublattice of genuine line operators
\begin{equation}
    \Gamma(D) = \mathbb{Z}^2 \cap D^{-1} \mathbb{Z}^2
\end{equation}
classifies those on the right-hand side of the defect that remain genuine as they are moved to its left-hand side. 
Viewed from the opposite perspective, the lattice of line operators on the left that remain genuine as they cross to the right are given by $D\Gamma(D) = \Gamma(D^{-1})$; this can equivalently be understood as considering the orientation-reversed defect $\mathcal{\bar D}$ which acts by $D^{-1}$. 
In turn, Wilson--'t Hooft loops on the right that become non-genuine on the left as they are pulled through the defect,
\begin{equation}\label{eq:genuine-to-nongenuine}
    \mathcal{D}: \quad WH_{\mathbf{q}}(\gamma) \to \eta_{D{\bf q}}(S)\, , \quad \text{with} \quad \partial S = \gamma \, ,
\end{equation}
are classified by the quotient as
\begin{equation}
    0 \neq [\mathbf{q}] \in \mathbb{Z}^2 /\Gamma(D) \, .
\end{equation}
For the orientation reversed defect $\overline{\cal D}$ (or if we pull the loops from left to right) we replace $D$ by $D^{-1}$.

This also characterizes the action of ${\cal D}$ on surface operators.
By reading \eqref{eq:genuine-to-nongenuine} in the other direction, we see that surface operators $\eta_{\bf r} (\Sigma)$ \emph{on the left} can end on ${\cal D}$, i.e., become trivial on the right, if ${\bf r} \in D\bbZ^2$.
Likewise, surface operators terminate on ${\cal D}$ on the right if ${\bf r} \in D^{-1}\bbZ^2$.

These sublattices and quotients can be determined straightforwardly through the Smith normal form of $D$ given in \eqref{eq:smith}. Since $U,V$ are unimodular matrices and therefore leave $\mathbb{Z}^2$ invariant, it follows directly that the lattice quotient is cyclic
\begin{equation}
    \mathbb{Z}^2 /\Gamma(D) \cong \mathbb{Z}_k\, .
\end{equation}
The basis that spans $\Gamma(D)$ can be read off from the column vectors of $V$; explicitly, we may write $V^{-1} = (\mathbf{v}_1, \mathbf{v}_2)$ so that 
\begin{equation}
    \Gamma(D) = \langle k \mathbf{v}_1,  \mathbf{v}_2 \rangle\, .
\end{equation}
Note that the column vectors $\mathbf{v}_1,\mathbf{v}_2$ define a complete basis for $\mathbb{Z}^2$ since $\det V = 1$, making manifest that $\Gamma(D)$ forms an index-$k$ sublattice.

Now recall that the defect $\mathcal{D}$ is realized by gauging some discrete global symmetry subgroup $\mathcal{Z} \subset U(1)^{(1)}_e \times U(1)_m^{(1)}$ in the left half of spacetime. 
We can detect this subgroup precisely from the surface operators $\eta_{\alpha, \beta}$ that can end on the defect from the right: these surface operators would cease to exist on the left side, since they generate a global symmetry subgroup that has been gauged there. Thus, the global symmetry subgroup that is gauged on the left-hand side of the defect is given by
\begin{equation}\label{eq:defZ}
    \mathcal{Z}(D) = \frac{\mathbb{Z}^2 + D^{-1}\mathbb{Z}^2}{\mathbb{Z}^2} \subset U(1)_e^{(1)} \times U(1)_m^{(1)}\, ,
\end{equation}
which classify the surface operators that can end on $\mathcal{D}$ from the right. 
Indeed, since the defect sends the charges of surface operators $\eta_{\alpha,\beta}$ to $D \left(\begin{smallmatrix} \alpha \\ \beta \end{smallmatrix} \right)$, and integral charges correspond to trivial surfaces operators, the lattice $\mathbb{Z}^2+ D^{-1}\mathbb{Z}^2$ classifies those surface operators that become trivial as they are pulled to the left side of $\mathcal{D}$.
From the point of view of complex multiplication, the group of Gukov--Witten operators $\mathcal{Z}(D)$ corresponds precisely to the kernel of the CM map given by $D$: it takes the points on the torus $x \in \mathbb{R}^2/\mathbb{Z}^2$ which are mapped to lattice sites $D x \in \mathbb{Z}^2$.
Thus, the invertibility of ${\cal D}$ as a symmetry defect goes hand in hand with the invertibility of $D$ as an endomorphism on $T^2$. From the opposite perspective, surface operators that can end from the left on $\mathcal{D}$ are given by $D\mathcal{Z}(D) = \mathcal{Z}(D^{-1})$, which hence defines the global symmetry subgroup gauged on the right-hand side of $\mathcal{D}$; or equivalently, on the left-hand side of the orientation-reversed defect $\mathcal{\bar D}$. 

Using again the Smith normal form \eqref{eq:smith} and writing $V^{-1} = (\mathbf{v}_1,\mathbf{v}_2)$, the global symmetry subgroup that is gauged by $\mathcal{D}$ reads
\begin{equation}
    \mathcal{Z} = \frac{\langle \mathbf{v}_1, \frac{1}{k}\mathbf{v}_2 \rangle}{\langle \mathbf{v}_1, \mathbf{v}_2 \rangle} \cong \mathbb{Z}_k \subset U(1)_e^{(1)} \times U(1)_m^{(1)}\, ,
\end{equation}
so in other words, $\mathcal{Z}(D)$ is generated by $\frac{1}{k}\mathbf{v}_2$. The fact that $\mathbf{v}_2$ need not be a purely electric or magnetic basis vector highlights the fact that we are gauging more general discrete 1-form symmetries than factorwise subgroups $\mathbb{Z}_{N_e}  \subset U(1)_e^{(1)} $ and $ \mathbb{Z}_{N_m} \subset U(1)_m^{(1)}$ with $\gcd(N_e, N_m)=1$.

An interesting alternative interpretation arises when inspecting the relation between the non-genuine charges on the left and right.
To do so, we take a surface operator $\eta_{\mathbf{x}}$ with $0 \neq [\mathbf{x}]\in \mathcal{Z}(D)$ that links with the line operator of charge ${\bf q}$. 
Recalling the phase picked up through this linking from \eqref{eq:linking-maxwell_lines-surfaces}, we find indeed that
\begin{equation}
    e^{2\pi i \mathcal{J}(\mathbf{x}, \mathbf{q})} = 1 \quad \text{unless} \quad 0 \neq [\mathbf{q}] \in \mathbb{Z}^2 /\Gamma(D)\, ,
\end{equation}
which follows from the fact that $\Gamma(D)$ and $\mathbb{Z}^2 + D^{-1}\mathbb{Z}^2$ are dual lattices with respect to the Dirac pairing.\footnote{Equivalently, we can write out the representative as $\mathbf{x} = \frac{1}{k}\mathbf{v}_2$ in terms of the generator of $\mathcal{Z}(D)\cong \mathbb{Z}_k$, and similarly take $\mathbf{q} = \mathbf{v}_1$ as generator for $\mathbb{Z}^2/\Gamma(D) \cong \mathbb{Z}_k$; unless we take either of them to be the trivial element of $\mathbb{Z}_k$, we get a non-trivial phase.}
Thus, line operators that become non-genuine are precisely those that are charged under the global symmetry subgroup $\mathcal{Z}(D)$ gauged on the left-hand side of $\mathcal{D}$. Now let us take one of these, say with charge vector $\mathbf{v}_1$, and pull it through the defect. Writing out $U= (\mathbf{u}_1,\mathbf{u}_2)$ into column vectors similar to $V^{-1}$ before, we find that the line operator gets the rational charges
\begin{equation}
    D \mathbf{v}_1 = \frac{1}{k}\mathbf{u}_1 \in \mathbb{Z}^2 + D \mathbb{Z}^2\, .
\end{equation}
The identification as a particular lattice element here was on purpose: from the perspective of the left-hand side of the defect, this charge vector $\frac{1}{k}\mathbf{u}_1$ generates the discrete global symmetry subgroup $\mathcal{Z}(D^{-1})$ that is gauged on the right half-space of the defect. In fact, this is analogous to how $\frac{1}{k}\mathbf{v}_2$ generates the subgroup $\mathcal{Z}(D)$ gauged in the left half-space, as described below \eqref{eq:defZ}.\footnote{The fact that now $\frac{1}{k}\mathbf{u}_1$ instead of $\frac{1}{k}\mathbf{u}_2$ (like $\frac{1}{k}\mathbf{v}_2$) generates the discrete symmetry comes from the fact that inverting \eqref{eq:smith} to get $D^{-1}$ also swaps the diagonal entries.} 
Thus, as we move a line operator charged under $\mathcal{Z}(D)$ through the defect, it becomes attached to a surface operator of the dual symmetry $\widehat{\mathcal{Z}(D)} = \mathcal{Z}(D^{-1})$.

\subsubsection*{Uniqueness of defect for given defect matrix}

As foreshadowed in our discussion of the triality defect at the end of section \ref{ssec:defectsfromgauging}, the above definition inherits the ambiguity from Smith decomposition.
Namely, if $D = U \, \text{diag} (\tfrac{1}{k},k) \, V = U' \, \text{diag} (\tfrac{1}{k},k) \, V'$, then the fusion products ${\cal D} = {\cal U} \times {\cal G}_\text{el} \times {\cal V}$ and ${\cal D}' = {\cal U}' \times {\cal G}_\text{el} \times {\cal V}'$ could a priori be different topological defects with the same matrix $D$.
For the efficacy of our construction, we must therefore demonstrate that the two defects agree.\footnote{The same issue arises in principle also for the defect construction in \cite[Appendix A]{Niro:2022ctq}, where a different factorization for a given matrix $D \in SL(2,\bbQ)$ is employed.}

At this point, we can easily establish that ${\cal D}$ and ${\cal D}'$ act identically on line and surface operators, since for defects obtained by stacking a gauging interface between two duality interfaces this action only depends on the $SL(2,\mathbb{Q})$ matrix $D$.
However, the defects could differ by an invertible TQFT or decoupled sector which we cannot determine from the action on the charge lattice alone.
To properly address this ambiguity and show that different Smith decompositions of the same matrix $D \in SL(2,\mathbb{Q})$ give rise to the same topological defect, we now turn to the explicit description of the defect worldvolume in terms of 3d Chern--Simons theories.

\section{Worldvolume Chern--Simons theory of duality walls}\label{sec:worldvolumeaction}

In this section we describe the 3d worldvolume action for both invertible and non-invertible duality walls. We consider both invertible $SL(2,\mathbb{Z})$ walls that are interfaces between 4d U(1) Maxwell theories at different couplings, and non-invertible $SL(2,\mathbb{Q})$ defects that are topological defects between 4d U(1) Maxwell theories at the same coupling. In both cases, the wall hosts a 3d Chern--Simons theory which can have $U(1)$ gauge fields living on it. Examples of such worldvolume theories have appeared in the literature both for invertible $SL(2,\mathbb{Z})$ duality walls, e.g.,~\cite{Gaiotto:2008ak, Kapustin:2009av}, and for non-invertible $SL(2,\mathbb{Q})$ defects in e.g.,~\cite{Cordova:2023ent, Kan:2024fuu}.
We will use the explicit worldvolume action to validate our construction in the previous section, i.e., that the defect construction in fig.~\ref{fig:Smith_deco_defects} using Smith decomposition \eqref{eq:smith} does not depend on the integer matrices $U$ and $V$.

\subsection{General worldvolume action}
Let us begin by describing the setting in its totality. We have two copies of 4d $U(1)$ Maxwell theory at possibly different values of the coupling $\tau_L$ and $\tau_R$ living in the left and right half-space, with a 3d wall $\mathcal{W}$ which we take to be located at $x=0$. On this duality wall, we have $N$ 3d $U(1)$ gauge fields $a^I$, where $I=1,\ldots, N$, and we take $N$ to be arbitrary. The action of this system has the form
\begin{equation}\label{eq:full_4d_action_defect}
    S = \int_{x<0} \mathcal{L}_{\rm Maxwell}(A_L, \tau_L) + \int_{x=0} \mathcal{L}_{\mathcal{W}}(a^I, A_L, A_R) +\int_{x>0}\mathcal{L}_{\rm Maxwell}(A_R, \tau_R)\,.
\end{equation} 
Before we describe the Chern--Simons couplings associated to the 3d gauge fields $a^I$, let us first assume they have been integrated out; we will reverse-engineer their Chern--Simons theory at a later stage, depending on the type of defect under consideration. After solving the equations of motion of the $a^I$, we are left with effective Chern--Simons couplings between $A_L, A_R$ that may be rational. In general, this effective action takes the form
\begin{equation}\label{eq:Leff}
    S_{\mathcal{W}, {\rm eff}} = \frac{i}{4\pi}\int_{x=0} \alpha \, A_L \wedge \dd A_L + \beta \, A_R \wedge \dd A_R + 2 \gamma \, A_L \wedge \dd A_R\, ,
\end{equation}
where $\alpha, \beta, \gamma \in \mathbb{Q}$. The fact that these coefficients can be rational implies that $S_{\mathcal{W}, {\rm eff}}$ is not well-defined under large gauge transformations of $A_L, A_R$; this is exactly what the 3d gauge fields $a^I$ will remedy. Nevertheless, the effective action suffices to derive the response matrix between the field strengths (and their duals) on the left and right-hand side. It is obtained by deriving the equations of motion for $A_L$ and $A_R$, including boundary terms on the defect coming from the left and right bulk Maxwell actions, yielding the relations
\begin{equation}\label{eq:FLRrelations}
\begin{aligned}
    \tilde{F}_L &= -\alpha F_L - \gamma F_R \\  \tilde{F}_R &= \gamma F_L + \beta F_R 
    \end{aligned}
    \qquad \iff \qquad \begin{pmatrix}
        \tilde{F}_L \\
        F_L
    \end{pmatrix} = \frac{1}{\gamma}\begin{pmatrix}
    -\alpha & \alpha \beta -\gamma^2\\
    1 & - \beta 
\end{pmatrix} \begin{pmatrix}
    \tilde{F}_R \\
    F_R
\end{pmatrix}\, ,
\end{equation}
where we identified the induced $SL(2,\mathbb{Q})$ transformation. Equivalently, we can express the effective Chern--Simons coupling $\alpha,\beta,\gamma$ in terms of the matrix coefficients of some $\left(\begin{smallmatrix} a & b \\
c & d
\end{smallmatrix}\right) \in SL(2,\mathbb{Q})$ transformation as
\begin{equation}\label{eq:CSeffectiveSL2Q}
    \alpha = -\frac{a}{c}\, , \qquad \beta = -\frac{d}{c} \, , \qquad \gamma = \frac{1}{c}\, .
\end{equation}
Relations \eqref{eq:FLRrelations} and \eqref{eq:CSeffectiveSL2Q} may be used both for $SL(2,\mathbb{Z})$ interfaces between theories at different couplings $\tau_L \neq \tau_R$ as well as $SL(2,\mathbb{Q})$ duality defects with $\tau_L = \tau_R$. In both of these cases, it may happen that the effective Chern--Simons coefficients are rational $\alpha, \beta, \gamma \not\in \mathbb{Z}$; for instance, an $SL(2,\mathbb{Z})$ duality interface with $ c \neq 1$. In this case, the effective Chern--Simons theory is ill-defined under large gauge transformations of $A_L, A_R$. When $c=0$, as will happen for duality walls implementing $T$-transformations \eqref{eq:CSTn} or gauging walls \eqref{eq:K_matrix_gauging}, worldvolume fields cannot be left out in the description of the wall.

To obtain a well-defined action for the duality walls under large gauge transformations, let us now write down the full Chern--Simons theory including 3d worldvolume $U(1)$ gauge fields. In general, it takes the form
\begin{equation}\label{eq:CStheory}
    S_{\mathcal{W}} = \frac{i}{4\pi} \int_{x=0} \Big( K_{IJ} a^I \wedge \dd a^J + 2Q_{I}^{\; r} a^I \wedge \dd A_r + P^{rl} A_r \wedge \dd A_l \Big)\, ,
\end{equation}
where $K_{IJ}, Q_{I}^{\; r}, P^{rl}$ are integral with $K_{IJ}, P^{rl}$ being symmetric and $r, l=\{L,R\}$. For brevity, we sometimes collect all fields into one collection $\mathcal{A}^{\mathcal{I}} = (A_L, a^I, A_R)$, such that the defect action can be specified by a single $\mathcal{K}$-matrix
\begin{equation}\label{eq:Kbig}
    S_{\mathcal{W}} = \frac{i}{4\pi }\int_{W} \mathcal{K}_{\cI\cJ} \mathcal{A}^{\cI} \wedge \dd \mathcal{A}^{\cJ}\, , \qquad \mathcal{K}_{\mathcal{I}\mathcal{J}} = \begin{pmatrix}
        P^{LL} & Q_J^{\; L} & P^{LR} \\
        Q_I^{\; L} & K_{IJ} & Q^{\; R}_{I}\\
        P^{RL} & Q^{\; R}_{J} & P^{RR}
    \end{pmatrix}\, .
\end{equation}
Integrating out the 3d gauge fields $a^i$, we are left with the effective action
\begin{equation}
    S_{\mathcal{W}, \rm eff} = \frac{i}{4\pi}\int_{x=0} \left(P^{rl} - K^{IJ} Q_I^{\;r} Q_{J}^{\; l}\right) A_r \wedge \dd A_l\, ,
\end{equation}
where $K^{IJ}$ denotes the inverse of $K_{IJ}$. From a linear algebraic perspective, the effective Chern-Simons couplings arise by taking the Schur complement of the matrix $\mathcal{K}$ with respect to the outer entries; for a more detailed discussion relating the Schur complement to integrating out fields, we refer to \eqref{Schur_complement_integrating_internal_fields} in the appendix. Matching with the effective  action \eqref{eq:Leff}, we find that the worldvolume theory should have Chern--Simons couplings such that
\begin{equation}\label{eq:effconstr}
    P - Q^T K^{-1} Q = \begin{pmatrix}
        \alpha & \gamma \\
        \gamma & \beta
    \end{pmatrix}\, .
\end{equation}
Clearly, there are different choices of couplings $K, Q , P$ that yield the same effective Chern--Simons action through \eqref{eq:effconstr}. For instance, we could add decoupled sectors that do not couple to $A_L, A_R$ or the other 3d $U(1)$ gauge fields. While this particular example is somewhat trivial, it illustrates that for a given choice of $K, Q, P$, it is difficult to assess whether only the ``necessary'' Chern--Simons sectors have been added. This is especially relevant in the context of the fusion rules discussed in section \ref{sec:fusion}, since unnecessary sectors might erroneously indicate decoupled sectors. For this reason, our choice of Chern--Simons theories for the defects will be set up so that it is free of such ``unnecessary'' sectors by construction, that is to say, minimal.

\subsection[Invertible SL(2,Z) duality walls]{Invertible $SL(2,\mathbb{Z})$ duality walls}\label{ssec:SL2walls}
Our strategy for developing the worldvolume action of the $SL(2,\mathbb{Q})$ duality defects leverages the Smith decomposition \eqref{eq:smith}. 
In this subsection, we first construct the Chern--Simons theory of the invertible duality interfaces associated with $U,V \in SL(2,\mathbb{Z})$, for which we explicitly verify from their fusion rules that they do not involve unnecessary degrees of freedom. 
This relies on expanding any $M \in SL(2,\bbZ)$ in a ``word'' $M = T^{p_1} S \cdots S T^{p_r}$ of $S$ and $T$ matrices, and computing the worldvolume action of ${\cal M}$ as the corresponding stacking of ${\cal T}$ and ${\cal S}$ duality interfaces.
Interestingly, such word expansions are not unique, similarly to the non-uniqueness of Smith decompositions. 
By using fixed representations of the worldvolume action of ${\cal T}$ and ${\cal S}$ interfaces, we will explicitly show that different word expansions lead to truly equivalent 3d Chern--Simons theories for ${\cal M}$, including the SPT-phases (i.e., invertible TQFT factors) for the bulk gauge fields.

Thereafter, in the next subsection, we combine these invertible duality walls with the usual gauging wall of a discrete subgroup of the electric 1-form symmetries. 
By construction, this approach then yields minimal defects.
Generalizing the proof for the physical equivalence of different word expansions, we then also show that stackings ${\cal D}_i={\cal U}_i \times {\cal G}_\text{el} \times {\cal V}_i$ (with $i=1,2$) based on different Smith decompositions $D = U_i \, \text{diag}(\tfrac{1}{k}, k) \, V_i$ are physically identical defects, thereby justifying our construction of duality defects in section \ref{ssec:action}.

\paragraph{Invertible duality walls for $T,S \in SL(2,\mathbb{Z})$.} Let us first describe how to construct the Chern--Simons theory for invertible duality walls implementing $SL(2,\mathbb{Z})$-transformations. We start from the generators $S,T \in SL(2,\mathbb{Z})$, which were set up in \cite{Gaiotto:2008ak, Kapustin:2009av}. The duality defect $\mathcal{S}$ can be represented by the Chern--Simons coupling
\begin{equation}\label{eq:CSS}
    S_{\mathcal{S}}[A_L, A_R] = \frac{i}{2\pi}\int_{W} A_L \wedge \dd A_R\, ,
\end{equation}
where $W$ represents the worldvolume of the defect. From the general effective action \eqref{eq:Leff} and its equations of motion \eqref{eq:FLRrelations}, one may verify that this indeed reproduces the continuity equations corresponding to $S$. For the $\mathcal{T}^p$-transformation, a single 3d $U(1)$ gauge field $a$ is added as a Langrange multiplier that implements $A_L=A_R$. The resulting Chern--Simons action for $\mathcal{T}^p$ reads
\begin{equation}\label{eq:CSTn}
    S_{\mathcal{T}^p}[A_L, A_R, a] = \frac{i}{4\pi}  \int_{W} \Big(-p \, A_R \wedge \dd A_R +2\, a \wedge \dd (A_R - A_L) \Big)\, .
\end{equation}
Indeed, by varying $a$ we recover the continuity equation $F_L = F_R$. In contrast, varying $A_L$ and $A_R$ and combining their constraints, we find that
\begin{equation}
    f = \tilde{F}_L  \, , \quad f = \tilde{F}_R + p\, F_R\quad \implies \quad \tilde{F}_L = \tilde{F}_R + p\, F_R\, ,
\end{equation}
where we wrote $f = \dd a$. Taken together with the identification $F_L = F_R$, we indeed recover the transformation $T^p = \left( \begin{smallmatrix}
    1 & p \\
    0 & 1
\end{smallmatrix} \right)$ on the field strengths in accordance with \eqref{eq:defSL2}.

One can straightforwardly check that both of these invertible interfaces indeed yield a trivial identity defect when fused with their orientation reversal. For $\mathcal{S}$ and $\mathcal{\bar{S}}$ the Lagrangian of the fusion product reads
\begin{equation}
    \mathcal{L}_{\mathcal{S}}(A_L, A_I) + \mathcal{L}_{\bar{\mathcal{S}}}(A_I, A_R) = \frac{i}{2\pi} A_I \wedge \dd (A_L - A_R)\, ,
\end{equation}
where $\mathcal{L}_{\bar{\mathcal{S}}}$ has the same Lagrangian as $\mathcal{L}_{\mathcal{S}}$ but with opposite sign.
The intermediate 4d gauge field $A_I$ between the defects reduces to a 3d field living on $W$ because of the fusion. Its equation of motion implements the identity wall action $A_L = A_R$ without any further degrees of freedom. As defects, we can therefore write the fusion rule $\mathcal{S} \times \mathcal{\bar S} = 1$. 

We can similarly analyze the fusion of the $\mathcal{T}^p$ and $\mathcal{\bar{T}}^{p}$-walls. To obtain the latter, we take $\mathcal{L}_{\mathcal{T}^p}$ and reverse the orientation by exchanging $A_L$ and $A_R$ and applying an overall minus sign. In total, the Lagrangian for this fusion product reads
\begin{align}
    \mathcal{L}_{\mathcal{T}^p}(A_L, A_I, a_1) + \mathcal{L}_{\bar{\mathcal{T}}^{p}}(A_I, A_R, a_2) = \frac{i}{2\pi} \big( & A_I\wedge \dd(a_1-a_2) -a_1 \wedge \dd A_L + a_2 \wedge \dd A_R\big) \, . \nonumber
\end{align}
We can again freely integrate out $A_I$, which now sets $a_2=a_1$. In turn, the action reduces to $\frac{i}{2\pi }a_1 \wedge \dd (A_R - A_L)$, which just describes the identity defect that sets $A_L = A_R$. So also for these defects, we can write $\mathcal{T}^p \times \mathcal{\bar T}^p = 1$.

\paragraph{Continued fractions and $SL(2,\mathbb{Z})$.} Starting from the Chern--Simons description of the elementary duality walls $\mathcal{S}$ and $\mathcal{T}$, we can construct the defect action of an arbitrary $M \in SL(2,\mathbb{Z})$ transformation by decomposing it into the generators $S$ and $T$ as
\begin{equation}\label{eq:SL2Zexpand}
    M = \begin{pmatrix}
        a & b \\
        c & d
    \end{pmatrix} = T^{p_1} S T^{p_2}  \cdots  T^{p_{r-1}} S T^{p_r} \in SL(2,\mathbb{Z})\, .
\end{equation}
The exponents $p_i$ follow from a classical algorithm governed by the arithmetic of the first column $(a,c)$; see e.g.~\cite{ConradSL2Z}. Concretely, the leading exponents $p_1,\ldots,p_{r'}$ (with $r' \leq r-1$) are the partial quotients of the negative continued fraction
\begin{equation}\label{eq:acratio}
    \frac{a}{c} = p_1 - \cfrac{1}{p_2 - \cfrac{1}{p_3 - \ddots}}\,.
\end{equation}
Since distinct $SL(2,\mathbb{Z})$ elements can share the same ratio $a/c$, this determines $M$ only up to a residual factor, obtained by stripping off the corresponding prefix
\begin{equation}\label{eq:stripoff}
    \bigl(T^{p_1}\, S \cdots T^{p_{r'}}\, S\bigr)^{-1} M
    = \varepsilon\, T^{p_{r'+1}}\,, \qquad \varepsilon = \pm 1\,,
\end{equation}
for some exponent $p_{r'+1} \in \mathbb{Z}$. For $\varepsilon = +1$ the expansion \eqref{eq:SL2Zexpand} terminates with $r = r'+1$; for $\varepsilon = -1$ the sign is absorbed via $S^2 = -1$ by appending two trivial factors $p_{r'+2} = p_{r'+3} = 0$, such that $S^2=T^{p_{r'+2}}ST^{p_{r'+3}}S$ and $r = r'+3$.

\paragraph{General $SL(2,\mathbb{Z})$-duality walls.} Given the expansion \eqref{eq:SL2Zexpand}, the 3d Chern--Simons theory of a general $SL(2,\mathbb{Z})$ defect $\mathcal{M}$ is assembled from the building blocks \eqref{eq:CSS} and \eqref{eq:CSTn}. Intuitively, we fuse the sequence of defects $\mathcal{T}^{p_1}, \mathcal{S}, \ldots, \mathcal{S}, \mathcal{T}^{p_r}$: each $\mathcal{S}$ introduces a 3d gauge field with an off-diagonal BF coupling to its neighbors, while each $\mathcal{T}^{p_i}$ contributes a diagonal Chern--Simons term for the field of the region it acts on. Since the outermost factors $\mathcal{T}^{p_1}$ and $\mathcal{T}^{p_r}$ act on the bulk gauge fields directly, only $p_2, \ldots, p_{r-1}$ are accompanied by internal 3d gauge fields $a_1, \ldots, a_{r-2}$, and the defect action takes the compact form
\begin{equation}\label{eq:LDSL2Z}
    \mathcal{L}_{\mathcal{W}}
    = \frac{i}{4\pi}\, \cK_{\cI\cJ}\, \mathcal{A}^\cI \wedge \dd \mathcal{A}^\cJ\,,
    \qquad
    \mathcal{A}^\cI = \bigl(A_L,\, a_1,\, \ldots,\, a_{r-2},\, A_R\bigr)\,,
\end{equation}
where $A_{L,R}$ denote the restrictions of the 4d gauge fields to the wall, and all couplings are encoded in the tridiagonal $\cK$-matrix
\begin{equation}\label{eq:KmatSL2Z}
    \cK = \begin{pmatrix}
        -p_1 & 1 &  &  \\
        1 & -p_2 & 1 &  \\
         & 1 & -p_3 & \ddots  \\
        & & \ddots & \ddots
    \end{pmatrix}\,.
\end{equation}
Abelian Chern--Simons theories of this type are commonplace in the literature on the fractional quantum Hall effect, cf.\ the review \cite{Tong:2016kpv}; similar constructions have appeared in the high-energy literature in \cite{Ganor:2014pha, Heckman:2017uxe}. In the quantum Hall context there is a single external electromagnetic field, and integrating out the internal gauge fields produces a filling fraction given by a continued fraction of precisely the form \eqref{eq:acratio}. The same mechanism operates here for the two external fields $A_L, A_R$: integrating out all 3d fields $a_1, \ldots, a_{r-2}$ amounts to taking the so-called \textit{Schur complement} \eqref{eq:effconstr} of $\cK$ onto its corner entries. In the end, this yields a Chern--Simons theory with rational couplings
\begin{equation}\label{eq:effcouplings}
    \mathcal{L}_{\mathcal{W}}^{\rm eff}
    = \frac{i}{4\pi\, c} \begin{pmatrix} A_L \\ A_R \end{pmatrix}^{\!T}
    \begin{pmatrix} -a & 1 \\ 1 & -d \end{pmatrix}
    \wedge \dd \begin{pmatrix} A_L \\ A_R \end{pmatrix}\,,
\end{equation}
where we used that the internal block of $\mathcal{K}$ has determinant $\det K = c$, the continuant underlying the continued fraction expansion \eqref{eq:acratio}. In particular, the diagonal couplings $a/c$ and $d/c$ are themselves the negative continued fractions in $p_1, \ldots, p_{r-1}$ and $p_r, \ldots, p_2$ respectively---the direct analogues of the quantum Hall filling fraction. Varying the action with respect to the 4d gauge fields, including the boundary contribution of the bulk Maxwell action, one verifies that this effective action indeed implements the transformation $T^{p_1} S T^{p_2} \cdots S T^{p_r}$ on the field strengths $(\tilde{F}, F)$, in accordance with \eqref{eq:FLRrelations}.

\paragraph{Illustrative examples.} To illustrate this general discussion, let us briefly study some simple examples of duality walls. Consider first the duality transformation $M = T^p \in SL(2,\mathbb{Z})$. We see that the tri-diagonal $\cK$-matrix in \eqref{eq:KmatSL2Z} reduces to a number $\mathcal{K} = p$: this gives a somewhat simplified representation of the worldvolume Lagrangian, writing just $ -\tfrac{i p}{4\pi}A \wedge \dd A$ instead of the system \eqref{eq:CSTn}. The difference is that \eqref{eq:CSTn} includes a worldvolume field $a$ whose EOM sets $A_L =A_R$, while \eqref{eq:KmatSL2Z} implicitly assumes that we have already imposed this. 

Next we take the duality transformation transformation $M= T^{p_1} S T^{p_2}$ as a more representative example of the general decomposition of $SL(2,\mathbb{Z})$ elements in \eqref{eq:SL2Zexpand}. In this case $r=2$, so the Lagrangian \eqref{eq:LDSL2Z} still does not involve any worldvolume fields, and is given in terms of the bulk fields $A_L, A_R$ as
\begin{equation}
    \mathcal{L}_{\mathcal{T}^{p_1}\mathcal{S}\mathcal{T}^{p_2}} = \frac{i}{4\pi} \big( -p_1 A_L \wedge \dd A_L - p_2 A_R \wedge \dd A_R + 2 A_L \wedge \dd A_R\big)\, , \quad \mathcal{K} = \begin{pmatrix}
        -p_1 & 1 \\
        1 & -p_2
    \end{pmatrix}\, .
\end{equation}
One may check that these integral Chern--Simons couplings reproduce the correct continuity equations between $(\tilde{F}_L, F_L)$ and $(\tilde{F}_R, F_R)$: specializing  \eqref{eq:FLRrelations} to $\alpha = -p_1, \beta=-p_2, \gamma = 1$, we indeed recover the transformation $T^{p_1}ST^{p_2} = \left( \begin{smallmatrix}
p_1 & p_1 p_2 - 1 \\
1 & p_2 \end{smallmatrix}\right)$ acting on the field strengths.

\paragraph{Invertibility of $SL(2,\mathbb{Z})$ duality walls.}
Let us also comment on the invertibility of $SL(2,\mathbb{Z})$ duality walls in the realization \eqref{eq:LDSL2Z}. In principle, the $SL(2,\mathbb{Z})$ action on the field strengths $\tilde F_L, F_L$ and $\tilde F_R, F_R$ can be realized with fewer 3d gauge fields; see for instance \cite{Paznokas:2025auw, Arbalestrier:2025jsg}, using one or two 
worldvolume gauge fields. Such more economical realizations, however, are not always invertible as defects: fusing the wall with its orientation reversal, $\mathcal{D} \times \bar{\mathcal{D}}$, need not yield the identity defect but can leave behind a decoupled topological sector. By contrast, our sequential construction manifestly preserves invertibility: each elementary wall in \eqref{eq:CSS} and \eqref{eq:CSTn} is invertible, with $\mathcal{S} \times \bar{\mathcal{S}} = \mathcal{T}^{p} \times \mathcal{\bar{T}}^{p} = 1$, and the orientation reversal of the composite wall is simply the reversed sequence $\bar{\mathcal{D}} = \mathcal{\bar{T}}^{p_r} \times \bar{\mathcal{S}} \times \cdots \times \bar{\mathcal{S}} \times \mathcal{\bar{T}}^{p_1}$, mirroring the inverse of the group element \eqref{eq:SL2Zexpand}, so that $\mathcal{D} \times \bar{\mathcal{D}} = 1$. Thus, despite our description of duality walls involves many fields, it is in fact the economical way to keep invertible fusion rules free of unwanted decoupled sectors. Another way to see this is by looking at the anyon groups of the Chern--Simons theories, whose orders are computed by $\det \mathcal{K}$. We will leverage this perspective in the next subsection in our study of non-invertible defects.

\paragraph{Equivalence of Chern--Simons descriptions.} As stressed above, a given duality transformation $M \in SL(2,\mathbb{Z})$ admits many word expansions, $M = T^{p_1} S \cdots S T^{p_r}$, and each of them yields a different presentation \eqref{eq:KmatSL2Z} of the Chern--Simons theory on the corresponding duality wall. For the wall itself to be well-defined, these presentations must all describe the same theory. We can check this straightforwardly by leveraging the structure of $SL(2,\mathbb{Z})$ as an abstract group, cf.~for related discussions in the physics literature \cite{Seiberg:2018ntt, Hsieh:2020jpj, Debray:2021vob, Debray:2023yrs}. To be precise, it is generated by the order-four element $S$, and $T$, subject only to the relations
\begin{equation}
    SL(2,\mathbb{Z}) = \langle S, T \ | \ S^2 = (ST)^3 = -1\rangle = \mathbb{Z}_4 \ast_{\mathbb{Z}_2} \mathbb{Z}_6\, .
\end{equation}
Sometimes $T$ is replaced by the order-six element $U = S T$ as generator, related to the second rewriting as a so-called amalgamated free product. For our purposes, the main take-away is that any two words in $S$ and $T$ representing the same element of $SL(2,\mathbb{Z})$ are related by applications of the identities $S^2 = (ST)^3 = -1$. At the level of the duality walls, it therefore suffices to establish the corresponding identities
\begin{equation}\label{eq:wallidentities}
    \mathcal{S}^2=\mathcal{C}\, \qquad (\mathcal{S}\mathcal{T})^3 = \mathcal{C}\, ,
\end{equation}
where $\mathcal{C}$ is the charge conjugation wall that implements $-\mathbbm{1}$. 

Consider first $\mathcal{S}^2$. In the word expansion \eqref{eq:SL2Zexpand} it has exponents $p_1=p_2=p_3=0$, so the $\mathcal{K}$-matrix and corresponding worldvolume Lagrangian read
\begin{equation}
    \mathcal{K} = \begin{pmatrix}
        0 & 1 & 0 \\
        1 & 0 & 1 \\
        0 & 1 & 0
    \end{pmatrix}\, , \qquad \mathcal{L}_{\mathcal{S}^2} = \frac{i}{2\pi} \, a \wedge \dd(A_L+A_R)\, .
\end{equation}
The worldvolume field $a$ acts as a Lagrange multiplier, imposing $F_L = - F_R$. In turn, varying $A_L$ and $A_R$ relates the dual field strengths as $\tilde{F}_L =  - f = - \tilde{F}_R$, where we wrote $f = \dd a$ for the field strength of the 3d Chern--Simons gauge field. Thus, $\mathcal{S}^2$ acts as $- \mathbbm{1}$ on $(\tilde{F}, F)$, and hence is equivalent to the charge conjugation wall $\mathcal{C}$.

Next we consider $(\mathcal{S}\mathcal{T})^3 $. Its word expansion \eqref{eq:SL2Zexpand} has exponents $p_1=0$ and $p_2=p_3=p_4=1$, which gives
\begin{equation}
    \mathcal{K} = \left(\begin{smallmatrix}
        0 & 1 & 0 & 0 \\
        1 & -1 & 1 & 0 \\
        0 & 1 & -1 & 1 \\
        0 & 0 & 1 & -1
    \end{smallmatrix}\right) , \, \mathcal{L}_{(\mathcal{S}\mathcal{T})^3} = \frac{i}{4\pi}\Big(2 \, A_L \wedge \dd a - a \wedge \dd a + 2 a \wedge \dd b - b \wedge \dd b + 2 b \wedge \dd A_R - A_R \wedge \dd A_R\Big) . \nonumber
\end{equation}
Redefining $b \to b+ a + A_R$, we find that the worldvolume Lagrangian simplifies to 
\begin{equation}
    \mathcal{L}_{(\mathcal{S}\mathcal{T})^3} = \frac{i}{2\pi} a \wedge \dd(A_L+A_R) - \frac{i}{4\pi} b\wedge \dd b\, .
\end{equation}
The Chern--Simons field $b$ decouples and it has trivial level $k=-1$, i.e.~it describes an invertible TQFT that is not coupled to the bulk gauge fields $A_{L/R}$. This leaves us with the coupling of the worldvolume field $a$ to the bulk fields $A_L, A_R$, which we recognize as that of the charge conjugation wall $\mathcal{C}$ from before. Both identifications $\mathcal{S}^2 = \mathcal{C}$ and $(\mathcal{S}\mathcal{T})^3 = \mathcal{C}$ therefore hold at the level of the worldvolume theories. Thereby, any two words representing the same element of  $SL(2,\mathbb{Z})$ define equivalent Chern--Simons descriptions of the same duality wall.

\subsection{Chern--Simons theory of non-invertible defects}\label{ssec:CSnoninvertible}

Using the Chern--Simons description \eqref{eq:LDSL2Z} of invertible $SL(2,\mathbb{Z})$ duality walls, we proceed to construct the non-invertible walls implementing the $SL(2,\mathbb{Q})$ transformation \eqref{eq:defectCM}. However, in order to construct the full non-invertible duality wall, we first review the Chern--Simons theory gauging an electric discrete subgroup $\bbZ_k$.

\paragraph{Electric $\bbZ_k$ subgroup gauging wall.} As we saw around~\eqref{eq:smith}, a generic non-invertible defect can always be described in terms of the gauging of an electric subgroup $\bbZ_k$ of the form ${\rm diag}(1/k,\,k)$; here we derive the Chern--Simons theory of such a gauging wall. Similarly to the $\mathcal{T}^p$-transformation~\ref{eq:CSTn}, the world-volume description of such a gauging wall requires the introduction of an auxiliary field $c$ implementing the identification $F_L=k\,F_R$ on the two sides of the defect. This leads to a Chern--Simons action and $\cK$-matrix of the form:
\begin{equation}\label{eq:K_matrix_gauging}
    \mathcal{L}_{\bbZ_k}=\frac{i}{2\pi}\, c \wedge \dd(A_L-k\, A_R),\quad \cK=\begin{pmatrix}
        0&1&0\\1&0&-k\\0&-k&0
    \end{pmatrix};
\end{equation}
indeed, it is possible to check that by varying $c$ we get the continuity condition $A_L=kA_R$. At the same time, by considering the full action introduced in~\eqref{eq:full_4d_action_defect}, and varying $A_L$ and $A_R$, we get the additional constraints:
\begin{equation}
    dc=\tilde{F}_L\,,\; dc=\frac{1}{k}\tilde{F}_R\implies \tilde{F}_L=\frac{1}{k}\tilde{F}_R\,.
\end{equation}
In this way, we recover the overall transformation between the fields on the left and right-sides of the wall
\begin{equation}
    \begin{pmatrix}
        \tilde{F}_L\\F_L
    \end{pmatrix}=\begin{pmatrix}
        \frac{1}{k}&0\\0&k
    \end{pmatrix}\begin{pmatrix}
        \tilde{F}_R\\F_R
    \end{pmatrix}\,,
\end{equation}
which precisely describes the gauging of a discrete electric subgroup $\bbZ_k$.

\paragraph{Non-invertible defect walls.} Now that we have obtained the Chern--Simons theory associated with the gauging of an electric subgroup, we can finally construct the world-volume theory of a generic non-invertible defect. To do so, recall the Smith decomposition \eqref{eq:smith} into $U, V \in SL(2,\mathbb{Z})$ and $\mathrm{diag}(1/k, k) \in SL(2,\mathbb{Q})$, and expand
\begin{equation}
    U = T^{p_1}\, S\, T^{p_2} \cdots S\, T^{p_r}\,, \qquad
    V = T^{q_1}\, S\, T^{q_2} \cdots S\, T^{q_s}\,.
\end{equation}
Fusing the sequence of walls $\mathcal{U}$, the $\mathbb{Z}_k$-gauging wall, and $\mathcal{V}$, the outermost exponents $p_1$ and $q_s$ again act directly on the restrictions $A_R$ and $A_L$ of the 4d gauge fields, while the 3d gauge fields comprise the internal fields $a_1, \ldots, a_{r-1}$ and $b_1, \ldots, b_{s-1}$ of the two $SL(2,\mathbb{Z})$ chains---where $a_{r-1}$ and $b_1$ arise from the strips separating $\mathcal{U}$ and $\mathcal{V}$ from the gauging wall---together with the field $c$ implementing the $\mathbb{Z}_k$ gauging. The defect action again takes the form
\begin{equation}\label{eq:LDSL2Q}
    \mathcal{L}_{\mathcal{D}}
    = \frac{i}{4\pi}\, \cK_{\cI\cJ}\, \mathcal{A}^\cI \wedge \dd \mathcal{A}^\cJ\,,
    \qquad
    \mathcal{A}^\cI = \bigl(A_L,\, a_1, \ldots, a_{r-1},\, c,\,
    b_1, \ldots, b_{s-1},\, A_R\bigr)\,,
\end{equation}
with the tridiagonal $\cK$-matrix
\begin{equation}\label{eq:Kmatfull}
    \cK =
        \begin{pmatrix}
        -p_1 & 1 &  &  & & & \\
        1 & \ddots & \ddots &  & & & \\
         & \ddots & -p_r & 1 & & & \\
         & & 1 & 0 & -k & & \\
         & & & -k & -q_1 & \ddots & \\
         & & & & \ddots & \ddots & 1 \\
         & & & & & 1 & -q_s
    \end{pmatrix}\,.
\end{equation}
The outer sub-blocks are the $\cK$-matrices \eqref{eq:KmatSL2Z} of the $SL(2,\mathbb{Z})$ walls $\mathcal{U}, \mathcal{V}$, while the middle sub-block encodes the gauging of the electric $\mathbb{Z}_k$ 1-form symmetry. Integrating out the internal gauge fields $a_i, c, b_j$---i.e.\ taking the Schur complement of $K$ onto its corner entries---yields the effective rational Chern--Simons couplings
\begin{equation}\label{eq:QeffSL2Q}
    \mathcal{L}_{\mathcal{D}}^{\rm eff}
    = \frac{i}{4\pi\, an}
    \Bigl( 2 k\, A_L \wedge \dd A_R - (m-b n)\, A_L \wedge \dd A_L - m\, A_R \wedge \dd A_R \Bigr)\,,
\end{equation}
where $m/k$ and $(m-b n)/k$ are the diagonal entries and $an/k$ one of the off-diagonal entries of the $SL(2,\mathbb{Q})$ transformation $D = U\, \mathrm{diag}(1/k,k)\, V $ as given explicitly in \eqref{eq:defectCM}. Using \eqref{eq:FLRrelations}, one can verify that this effective action indeed reproduces the correct continuity equations on the field strengths $(\tilde{F}, F)$. In deriving \eqref{eq:QeffSL2Q}, we used that the internal block $K$ has integer determinant $|\det K |=| a n|$, generalizing the continuant identity of the invertible $SL(2,\mathbb{Z})$ case. In fact, such determinants double as orders of the anyon groups associated with the abelian Chern--Simons theory. Considering the case of the 3d gauge fields in isolation and the combined system including the 4d gauge fields, we find respectively
\begin{equation}
    |\mathbb{Z}^{N+2}/\cK \mathbb{Z}^{N+2}| = |\det \cK_{\cI\cJ}| = |c \, n|\, , \quad  |\mathbb{Z}^{N}/K \mathbb{Z}^{N}| = |\det K_{IJ}| = |a\,  n|\, ,
\end{equation}
where $N=r+s-1$ denotes the total number of 3d worldvolume gauge fields. The anyon count $| an |$ is the minimal order required to generate the denominator $an$ in \eqref{eq:QeffSL2Q}; no Chern--Simons sector of smaller order can reproduce this effective coupling, independently of how the anyons are distributed among the 3d gauge fields. This minimal order should be contrasted with the number of gauge fields, which is not itself an invariant: the construction \eqref{eq:Kmatfull} may introduce many fields, but the physical anyon content remains fixed at $|an|$. 

\subsection{Uniqueness of duality defects from Smith decomposition}
The Smith decomposition \eqref{eq:smith} of defect matrices $D \in SL(2,\mathbb{Q})$ has played a crucial role in our physical understanding and construction of duality defects so far. However, as stressed before, the Smith decomposition of such a matrix is not unique. It therefore gives rise to multiple representations of the defect, e.g., $\mathcal{D}_i = \mathcal{U}_i \times \mathcal{G} \times \mathcal{V}_i$ with $i=1,2$, which a priori could be distinct physical defects. By arguing that their Chern--Simons theories are equivalent, we will show here that different Smith decompositions lead to the same defect. In this subsection we discuss a simple example and summarize the general argument, while we refer to appendix \ref{app:equiv_chern_simo_smith_deco} for the details.

Let us begin by considering an example of two Smith decompositions of the same defect. Concretely, recall the decomposition \eqref{eq:smith_decomp_triality} of the triality defect, given by
\begin{equation}
    D_3 = S \,  \text{diag}(\tfrac{1}{N}, N) \, T^N = S T^{-1} \, \text{diag}(\tfrac{1}{N}, N) \, T^{N(N+1)}\, .
\end{equation}
Originally, these two decompositions were reserved for separate cases, the first for $N$ even and the second for $N$ odd, due to the formulation using counterterm stackings. 
Here we argue for any $N$ that the two decompositions give rise to the same defect in 4d Maxwell. To show the equivalence, we first disregard the factors of $S$ and $T^N$ at the ends, and focus on the middle part
\begin{equation}
    T^{-1} \, \text{diag}(\tfrac{1}{N}, N) \, T^{N^2} = \text{diag}(\tfrac{1}{N}, N) \, .
\end{equation}
If we can indeed show that the wall $\mathcal{T}^{-1} \times \mathcal{G}_{\rm el} \times \mathcal{T}^{N^2}$ is equivalent to the electric gauging wall $\mathcal{G}_{\rm el}$, the equivalence of the defects associated to the two Smith decompositions in \eqref{eq:smith_decomp_triality} follows automatically, just by applying the same duality transformations before and after the gauging wall.

The Chern--Simons theory of $\mathcal{T}^{-1} \times \mathcal{G}_{\rm el} \times \mathcal{T}^{N^2}$ takes the simple form
\begin{equation}
    \mathcal{K} = \begin{pmatrix}
        1 & 1 & 0 \\
        1 & 0 & - N \\
        0 & -N & -N^2
    \end{pmatrix}\, , \qquad \mathcal{L}= \frac{i}{4\pi} \Big( A_L \wedge \dd A_L -N^2 \, A_R \wedge \dd A_R + 2 c \wedge (A_L - N A_R)\Big)\, .
\end{equation}
The only difference with the worldvolume theory \eqref{eq:K_matrix_gauging} of the gauging wall $\mathcal{G}_{\rm el}$ is the presence of the Chern--Simons couplings $A_L \wedge A_L$ and $-N^2  A_R \wedge \dd A_R$. However, upon imposing the constraint $A_L = N A_R$ imposed by $c$ as Lagrange multiplier, these two terms cancel. We are thus left to conclude that the two walls are equivalent
\begin{equation}
    \mathcal{T}^{-1} \times \mathcal{G}_{\rm el}\times \mathcal{T}^{N^2} = \mathcal{G}_{\rm el}\, .
\end{equation}
Hence, the two Smith decompositions of the triality defect describe the same theory (at fixed $N$). 

Let us now turn to the general case and sketch the argument of appendix \ref{app:equiv_chern_simo_smith_deco}. Instead of $\mathcal{T}$-transformations, we get the more general situation in which we need to show the equivalence of walls
\begin{equation}
    \mathcal{U}_{21} \times \mathcal{G}_{\rm el} \times \mathcal{V}_{12} = \mathcal{G}_{\rm el}\, ,
\end{equation}
where $\mathcal{G}_{\rm el}$ denotes a $\mathbb{Z}_k$ gauging wall. The matrices $U_{21}, V_{12} \in SL(2,\mathbb{Z})$ associated to the two duality walls relate the two different Smith decompositions. They are related by $V_{12} = \text{diag}(k, 1/k) U_{21}^{-1} \text{diag}(1/k, k)$;\footnote{Moreover, the integrality of $V_{12}$ constrains the other to the Hecke congruence subgroup $U_{21} \in \Gamma_0(k^2) \subset SL(2,\mathbb{Z})$. Vice versa, integrality of $U_{21}$ constrains $V_{12} \in \Gamma^0(k^2)$.} notice that $V_{12}=T^{N^2}$ and $U_{12}=T^{-1}$ indeed satisfy this relation for $k=N$. By using the word expansions of $U_{21}, V_{12}$ into generators $S,T$, we can describe the wall $\mathcal{U}_{21} \times \mathcal{G}_{\rm el} \times \mathcal{V}_{12}$ by some Chern--Simons theory
\begin{equation}
    \mathcal{L}_{\mathcal{U}_{21} \times \mathcal{G}_{\rm el} \times \mathcal{V}_{12}} = \frac{i}{4\pi}\mathcal{K}_{\mathcal{I}\mathcal{J}} A^{\mathcal{I}} \wedge \dd A^{\mathcal{J}}
\end{equation}
whose tri-diagonal $\mathcal{K}$-matrix has been given explicitly in \eqref{eq:chern_simons_initial_proof_smith}. In general, this worldvolume theory involves many fields $\mathcal{A}_I = (A_L; a_1, \ldots, a_n; \phi, b_2, \psi; c_1, \ldots, c_m; A_R)$, in contrast to the single worldvolume field of $\mathcal{G}_{\rm el}$ whose equation of motion $A_L = k A_R$ implements the $\mathbb{Z}_k$ gauging. To show that the two walls are equivalent as Chern--Simons theories, we thus have to manipulate the worldvolume fields of $\mathcal{L}_{\mathcal{U}_{21} \times \mathcal{G}_{\rm el} \times \mathcal{V}_{12}}$ accordingly. Concretely, in appendix \ref{app:equiv_chern_simo_smith_deco} we first integrate out worldvolume fields that are trivial Lagrange multipliers, and subsequently redefine the remaining worldvolume fields to decouple all but one from the bulk fields $A_L, A_R$. We omit these manipulations here, and skip straight to the final form 
\begin{equation}
    \mathcal{L}_{\mathcal{U}_{21} \times \mathcal{G}_{\rm el} \times \mathcal{V}_{12}} = \frac{i}{2\pi} c \wedge \dd (A_L - k A_R) + \frac{i}{4\pi} \hat{\mathcal{K}}^{(\hat{1})}_{ij} x^i \wedge \dd x^j\, ,
\end{equation}
where $x^i$ denote the redefined worldvolume fields. The hats on $\hat{\mathcal{K}}^{(\hat{1})}$ indicate that we have removed the rows and columns of $\mathcal{K}$ corresponding to the bulk fields $A_L,A_R$ and the Lagrange multiplier $c$. Thus the only remaining difference between the $\mathcal{U}_{21} \times \mathcal{G}_{\rm el} \times \mathcal{V}_{12}$ wall and the gauging wall $\mathcal{G}_{\rm el}$ is this decoupled TQFT. The computation \eqref{eq:detK=1} of the associated determinant shows that $| \det \hat{\mathcal{K}}^{(\hat{1})}| =1$. Thus this sector has no anyon group and is therefore an invertible TQFT not coupled to the bulk fields $A_{L/R}$, which means the two walls are indeed equivalent at the level of their Chern--Simons theories.

\section{Fusion rules and condensation defects}\label{sec:fusion}
In this section we study the fusion of non-invertible duality defects arising from complex multiplication symmetry. We describe these fusion products in two complementary ways. First, in subsection \ref{ssec:fusionlattice}, through lattices of symmetry surfaces and line operators stretched between the duality defects and ending upon them. Second, in subsection \ref{ssec:fusionworldvolume}, through the 3d Chern--Simons theory living on the fusion of the two defects. We illustrate these observations on various examples both here and in the next section.

\subsection{Condensation defects and decoupled TQFTs from fusion}
Let us first describe our general expectation for the fusion product of two duality defects. From related studies of gauge theories in the literature, cf.~\cite{Shao:2023gho}, we expect that there are three sorts of defects that can appear: condensation defects $\mathcal{C}$ \cite{Roumpedakis:2022aik, Bhardwaj:2022lsg, Lin:2022xod, Bartsch:2022mpm}, (non-trivial) decoupled TQFTs $\mathcal{T}$, and non-invertible duality defects themselves. In general, we expect all these objects to appear together when fusing two non-invertible duality defects. Moreover, the ordering of the condensation defect with respect to the new duality defect matters: the duality defect rotates the discrete subgroup that the condensation subgroup projects along. In other words, we obtain the fusion of a duality defect and a condensation defect, with coefficients given by decoupled TQFTs. Altogether, we can summarize this fusion rule as
\begin{equation}\label{eq:fusionrule}
	\mathcal{D}_1 \times \mathcal{D}_2 = \mathcal{T}_{\rm dec} \ \mathcal{C}_{\mathcal{Z}_{12}^L} \times \mathcal{D}_3 = \mathcal{T}_{\rm dec} \  \mathcal{D}_3 \times C_{\mathcal{Z}_{12}^R}\, .
\end{equation}
The two different condensator subgroups $\mathcal{Z}_{12}^L$ and $\mathcal{Z}_{12}^R$ are related by the defect matrix $D_3$, which will be made precise in subsection \ref{ssec:fusionlattice}. The candidate duality defect $\mathcal{D}_3$ is identified by multiplying the defect matrices
\begin{equation}
    D_3 = D_1 D_2 \in SL(2,\mathbb{Q})\, .
\end{equation}
The combined degrees of freedom associated with the decoupled TQFT $\mathcal{T}_{\rm dec}$ and condensation defect $\mathcal{C}_{\mathcal{Z}_{12}^{L/R}}$ can be estimated from this matrix multiplication. Recall that these defect matrices have overall denominators $k_i$, such that the rescaled defect matrix has integral entries with no common divisors. However, considering the right-hand side above, the rescaled product $k_1 k_2 D_1 D_2$ may have a common divisor among its entries. So in general, we find that a measure for these additional degrees of freedom is given by
\begin{equation}\label{eq:Nextra}
    N_{\rm extra} = \frac{k_1 k_2}{k_3} = \gcd(k_1 k_2 D_1 D_2)\, ,
\end{equation}
where the right-hand side denotes the greatest common divisor among all matrix entries. From the perspective of these duality defects as half-space gaugings, the subgroups $\mathbb{Z}_{k_1}$ and $\mathbb{Z}_{k_2}$ gauged by $\mathcal{D}_1$ and $\mathcal{D}_2$ combine into a smaller group $\mathbb{Z}_{k_3}$ gauged by $\mathcal{D}_3$, due to alignment between these dyonic 1-form symmetries. The remaining degrees of freedom counted by $k_1 k_2/k_3$ result in a condensation defect and/or a decoupled TQFT. In the following subsection, we make precise how to identify the gauging and condensation subgroups as well as decoupled sectors.

\subsection{Fusion rules from lattice arithmetics}\label{ssec:fusionlattice}
Here we make precise the fusion rule \eqref{eq:fusionrule} of the duality defects from their action on symmetry surfaces. Based on whether a symmetry surface can end on both of the defects, or be stretched between them, we argue that it forms the condensation defect or decoupled TQFT resulting from the fusion. We identify the lattice intersections and quotients that characterize these symmetry subgroups, reproducing the expected dimension \eqref{eq:Nextra} of these operators.

\begin{figure}
\begin{tikzpicture}
\centering
\node (left) at (0,0)
{\includegraphics[width=0.96\textwidth]{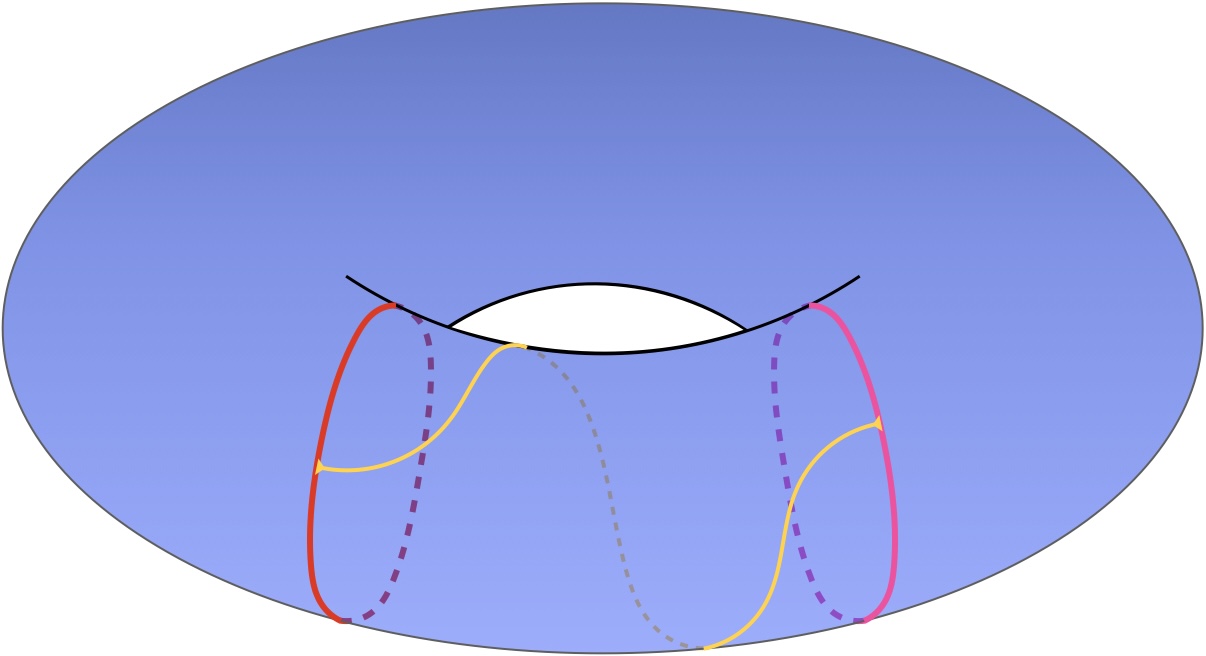}};
\node at (-3.4,-3.8) {\LARGE{$\mathcal D_1$}};
\node at (3.6,-3.8) {\LARGE{$\mathcal D_2$}};
\node at (1.45,-2.5) {\LARGE{$\mathcal \eta_{\bf r}(\Sigma)$}};
\end{tikzpicture}
\caption{Representation of a Gukov--Witten operator $\eta$ of charge $\bf r$ supported over a 2-chain $\Sigma$ ending on the defects $\mathcal D_1$ and $\mathcal D_2$ from the right and the left, respectively. For clarity, the 2-chain is drawn as a curve while the four-dimensional spacetime is depicted as a two-torus. }
\label{fig:Line_winding_between_def}
\end{figure}

\begin{figure}[ht]
\centering

\begin{subfigure}[t]{0.55 \textwidth}
    \centering
    \caption{Gukov--Witten operators between defects.}
    \label{fig:decomoposit_cond_TQFT_defe_top}
    \begin{tikzpicture}[remember picture]
        \node[anchor=south west,inner sep=0] (topimg) at (0,0)
            {\includegraphics[width=\linewidth]{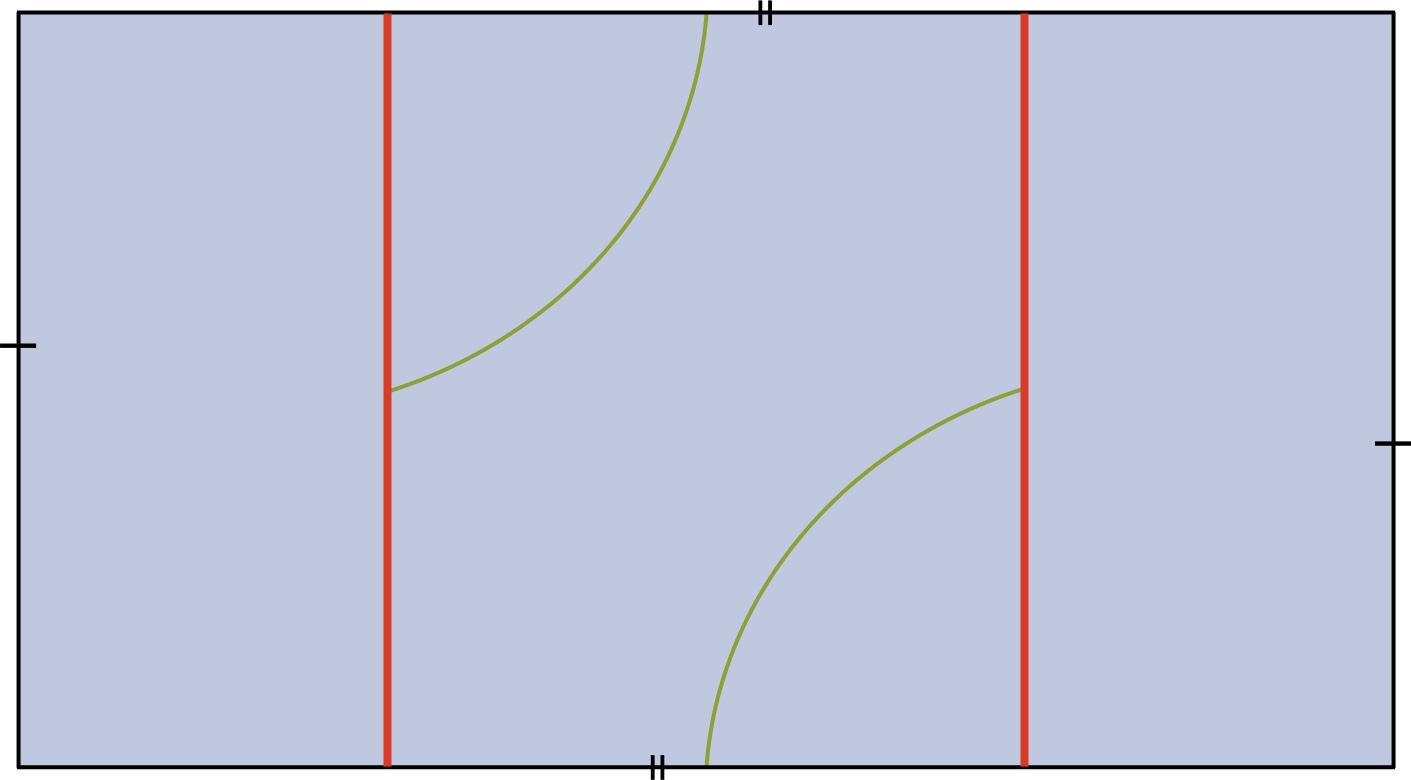}};
        \begin{scope}
            \node at (6.35,0.4) {$\mathcal D_2$};
            \node at (2,0.4) {$\mathcal D_1$};
            \node at (6.95,2.25) {$\eta_{D_2\mathbf{q}_R+\mathbf{r}_R}$};
            \node at (1.3,2.25) {$\eta_{D_1^{-1}\mathbf{q}_L+\mathbf{r}_L}$};
        \end{scope}
    \end{tikzpicture}
\end{subfigure}

\vspace{4em} 

\begin{subfigure}[t]{0.48\textwidth}
    \centering
    \begin{tikzpicture}[remember picture]
        \node[anchor=south west,inner sep=0] (blimg) at (0,0)
            {\includegraphics[width=\linewidth]{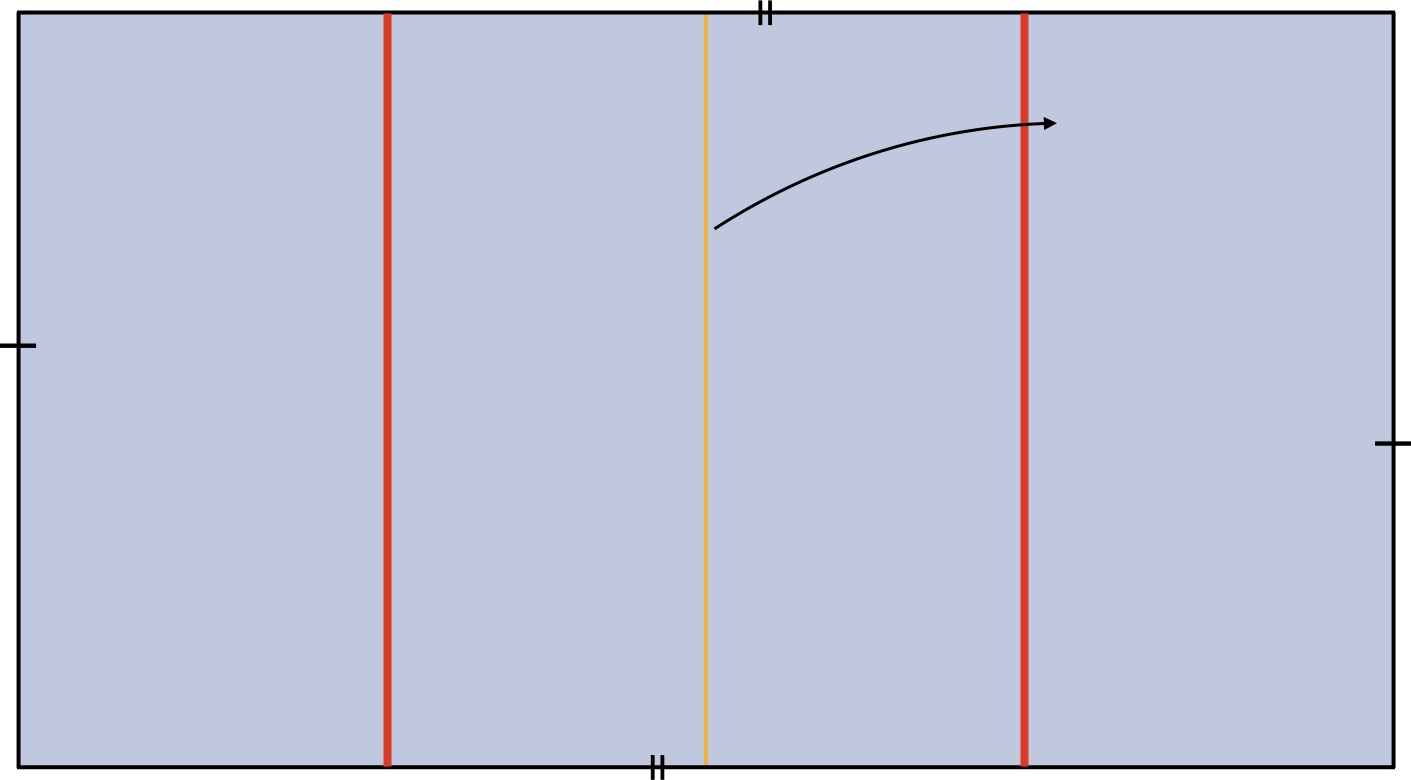}};
        \begin{scope}
            \node at (5.6,0.4) {$\mathcal D_2$};
            \node at (1.7,0.4) {$\mathcal D_1$};
            \node at (6.5,3.25) {\small{$\eta_{D_2\mathbf{q}_R=D_1^{-1}\mathbf{q}_L}$}};
        \end{scope}
    \end{tikzpicture}
    \caption{Condensation sector.}
    \label{fig:decomoposit_cond_TQFT_defe_bottom_left}
\end{subfigure}
\hfill
\begin{subfigure}[t]{0.48\textwidth}
    \centering
    \begin{tikzpicture}[remember picture]
        \node[anchor=south west,inner sep=0] (brimg) at (0,0)
            {\includegraphics[width=\linewidth]{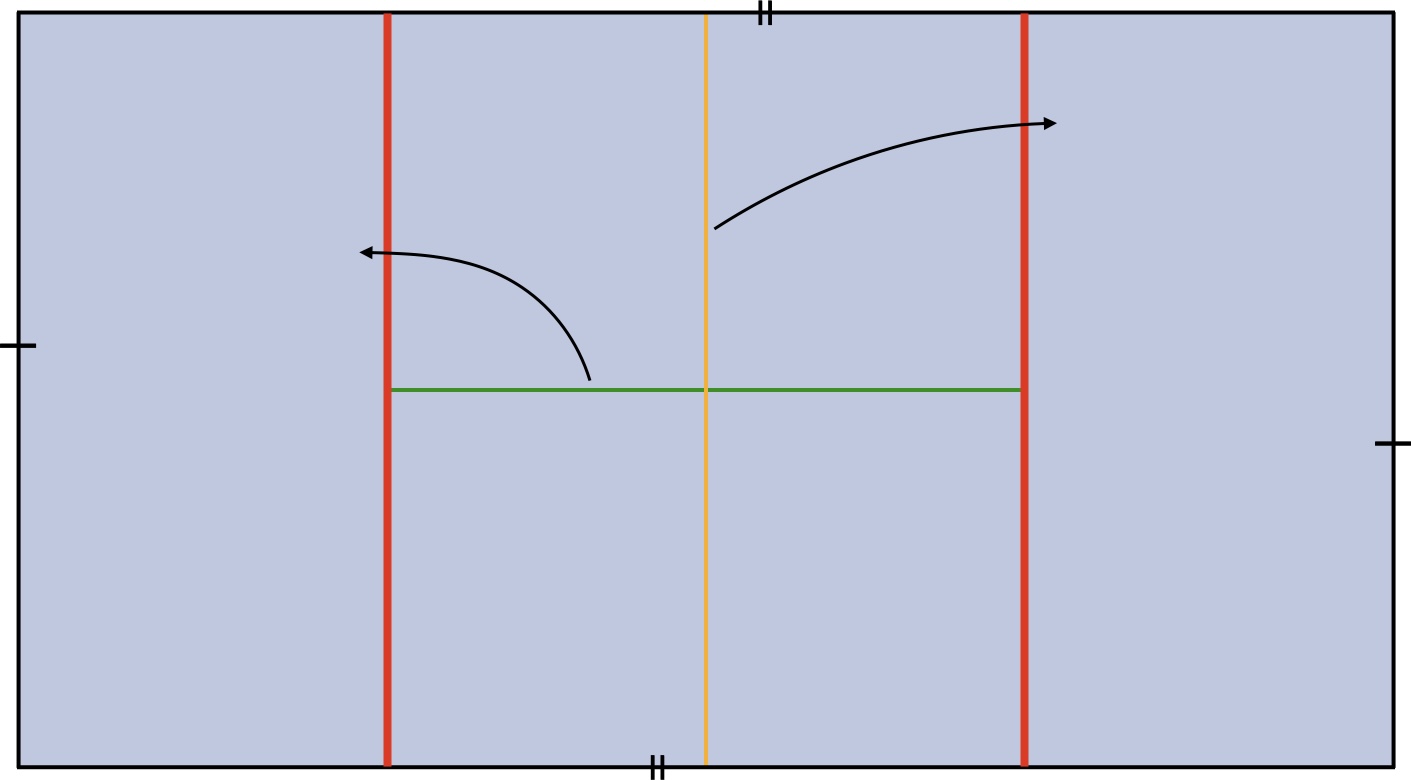}};
        \begin{scope}
            \node at (5.6,0.4) {$\mathcal D_2$};
            \node at (1.7,0.4) {$\mathcal D_1$};
            \node at (6.5,3.25) {\small{$\eta_{D_2\mathbf{q}_R=D_1^{-1}\mathbf{q}_L}$}};
            \node at (1.3,2.65) {\small{$\eta_{\mathbf{r}_L-\mathbf{r}_R}$}};
        \end{scope}
    \end{tikzpicture}
    \caption{Condensation and decoupled TQFT sectors.}
    \label{fig:decomoposit_cond_TQFT_defe_bottom_right}
\end{subfigure}
\begin{tikzpicture}[remember picture,overlay]
\draw[thick,->] ($(topimg.south) - (1,0.1)$) -- ($(blimg.north) - (-0.3,-0.1)$);
\node at ($(topimg.south) - (3.4,0.65)$) {$\mathbf{r}_L=\mathbf{r}_R$};
\draw[thick,->] ($(topimg.south) - (-1,0.1)$) -- ($(brimg.north) - (0.3,-0.1)$);
\node at ($(topimg.south) + (3.3,-0.65)$) {$\mathbf{r}_L\neq\mathbf{r}_R$};
\end{tikzpicture}

\caption{Gukov--Witten operator in the presence of two parallel defects. An operator supported on a 2-cycle parallel to the defects gives rise to the condensation defect, while a 2-chain bounded by the two defects yields the decoupled TQFT. For clarity, the 2-chains are drawn as curves and the four-dimensional spacetime as a two-torus. }
\label{fig:decomoposit_cond_TQFT_defe}
\end{figure}

Let us start by explaining how the symmetry surfaces living between the defects can give rise to such condensation defects and decoupled TQFTs. To this end, recall that we have a 4d gauge field $A_I$ living between $\mathcal{D}_1$ and $\mathcal{D}_2$. Fusing these defects amounts to integrating out $A_I$, meaning we have to sum over all possible field configurations. We can view these field configurations as symmetry surfaces that are traced out by Wilson--'t Hooft lines, both non-genuine and genuine, that move from one defect to the other. The boundary conditions imposed by $\mathcal{D}_1$ and $\mathcal{D}_2$---i.e.~the equations of motion arising from integrating over their worldvolume degrees of freedom---then require that these symmetry surfaces can end on both of these defects (see fig.~\ref{fig:Line_winding_between_def}). Using \eqref{eq:defZ}, surfaces ending from the right on $\mathcal{D}_1$ generate the quotient $\mathcal{Z}(D_1) = (\mathbb{Z}^2 + D_1^{-1} \mathbb{Z}^2)/\mathbb{Z}^2$, while those ending from the left on $\mathcal{D}_2$ generate $\mathcal{Z}(D_2^{-1}) = (\mathbb{Z}^2 + D_2 \mathbb{Z}^2)/\mathbb{Z}^2$. Requiring a surface to end on both defects therefore amounts to the intersection
\begin{equation}\label{eq:Z12}
    \mathcal{Z}_{12} = \frac{(\mathbb{Z}^2 + D_1^{-1} \mathbb{Z}^2)\cap(\mathbb{Z}^2 + D_2\mathbb{Z}^2)}{\mathbb{Z}^2} \, .
\end{equation}
In order to understand which of these symmetry surfaces correspond to the condensation defect and which to the decoupled TQFT, we have to consider what surfaces they actually wrap. Let us write our fusion setup as $W \times I$, where $I = [0,\epsilon]$ is the interval between the defects and $W$ the worldvolume of the defects introduced in~\eqref{eq:CSS}. We can consider two kinds of 2-chains as surfaces in this setting: 2-cycles $\sigma_2 \subset W$ of the wall itself, or 2-chains $\sigma_1 \times I \subset W \times I$ where $\sigma_1 \subset W$ bounds the 2-chain by line operators on the two defects (see fig.~\ref{fig:decomoposit_cond_TQFT_defe_top}). In fact, using K\"unneth theorem,\footnote{At the level of homology, this can be made precise as the exact sequence
\begin{equation}
    0 \to H_2(W) \to H_2^{\text{rel}}(W \times I, \gamma_0 \sqcup \gamma_\epsilon) \to H_1(\gamma_0 \sqcup \gamma_\epsilon) \to  0
\end{equation}
where $\gamma_i = \sigma_1 \times \{i\}$ with $i=0,\epsilon$. } any relative 2-chain $\Sigma_2 \subset W \times I$ (with the same boundary on $\mathcal{D}_1$ and $\mathcal{D}_2$ but opposite orientation) can be decomposed as
\begin{equation}
    \Sigma_2 = (\sigma_2 \times \{pt\}) + (\sigma_1 \times I )\, ,
\end{equation}
where $\{pt\}$ indicates that the first component is pointlike on $I$. 

The next question is what symmetry charges can be carried by the surface on these components. Starting with $\sigma_2 \times \{pt\}$, we have a symmetry surface that is parallel to the defects. Since a surface swept wholly through $\mathcal{D}_1$ and $\mathcal{D}_2$ cannot be absorbed by a line operator living on these defects, these are generated by the subgroup
\begin{equation}
    \mathcal{Z}_{12, \rm cond} = \frac{\mathbb{Z}^2 + D_1^{-1} \mathbb{Z}^2 \cap D_2 \mathbb{Z}^2}{\mathbb{Z}^2} \subseteq \mathcal{Z}_{12} \, .
\end{equation}
To understand how surfaces with these charges can be moved through the defects, it is instructive to write out the representative classes explicitly
\begin{equation}\label{eq:reprZ12cond}
    [\mathbf{q}] = [D_1^{-1} \mathbf{q}_L+\mathbf{r}] = [D_2 \mathbf{q}_R+\mathbf{r}] \in \mathcal{Z}_{12, \rm cond}\, ,
\end{equation}
where $\mathbf{q}_{L/R},\mathbf{r} \in \mathbb{Z}^2$ and $D_1^{-1} \mathbf{q}_L= D_2 \mathbf{q}_R \notin \mathbb{Z}^2$. Because we quotient by $\mathbb{Z}^2$, we are free to set $\mathbf{r}=0$ (fig.~\ref{fig:decomoposit_cond_TQFT_defe_bottom_left}). This makes it clear that these surfaces become trivial as they are moved through $\mathcal{D}_1$ from the right and $\mathcal{D}_2$ from the left, since only integral charges $\mathbf{q}_R$ or $\mathbf{q}_L$ will be left then. We associate this subgroup to the condensation defect because it is generated by symmetry surfaces wrapping all possible 2-cycles of the 3d worldvolume $W$; this follows closely the general construction of condensation defects through higher-gauging defined in \cite{Roumpedakis:2022aik}. 

The remaining charge of the symmetry surfaces on $\Sigma_2$ is then left on the tube $\sigma_1 \times I$, which is classified by the quotient
\begin{equation}
    \mathcal{Z}_{12, \rm dec} = \frac{(\mathbb{Z}^2 + D_1^{-1} \mathbb{Z}^2)\cap(\mathbb{Z}^2 + D_2\mathbb{Z}^2)}{\mathbb{Z}^2 + D_1^{-1} \mathbb{Z}^2 \cap D_2 \mathbb{Z}^2} \subseteq \mathcal{Z}_{12}\, .
\end{equation}
The fact that this quotient group is not trivial follows from the failure of the intersection to be distributive, i.e., $A \cap (B+C) \neq A\cap B + A\cap C$. To understand this subtlety and the nature of these tubular symmetry surfaces, it is instructive to look at their representative classes. First, let us take a general class among all surfaces in \eqref{eq:Z12}
\begin{equation}
    [\mathbf{q}] = [D_1^{-1}\mathbf{q}_L+\mathbf{r}_L] = [D_2\mathbf{q}_R +\mathbf{r}_R] \in \mathcal{Z}_{12}\, ,
\end{equation}
where again $\mathbf{q}_{L/R} \in \mathbb{Z}^2$ and $\mathbf{r}_{L/R} \in \mathbb{Z}^2$. While we can set $\mathbf{r}_L = 0$ or $\mathbf{r}_R = 0$ because we quotient by $\mathbb{Z}^2$, their difference always remains present. Comparing with \eqref{eq:reprZ12cond}, we find that we can generate the tubular symmetry surfaces $\mathcal{Z}_{12, \rm dec}$ by representatives
\begin{equation}
    [\mathbf{r}_L - \mathbf{r}_R] \in \mathcal{Z}_{12 , \rm dec}\, .
\end{equation}
This mismatch is precisely the integral charge carried by the line operators induced on $\mathcal{D}_1$ and $\mathcal{D}_2$ by the symmetry surface (fig.~\ref{fig:decomoposit_cond_TQFT_defe_bottom_right}). We view these line operators as generating the decoupled TQFT arising from the fusion product. 

Given the above characterization of the condensation sector and decoupled sector, let us compute the orders of these quotient groups. In particular, we want to relate these orders to the common factor $N_{extra}$ that canceled out in the multiplication of the defect matrices in \eqref{eq:Nextra}. While we refer the reader to appendix \ref{app:orders} for a detailed calculation of the group orders, we just recall here the fact that
\begin{equation}
    N_{\rm extra} = \frac{k_1 k_2}{k_3} = |\mathcal{Z}_{12}| |\mathcal{Z}_{12, \rm cond}| = |\mathcal{Z}_{12, \rm cond}|^2 |\mathcal{Z}_{12, \rm dec}|\, .
\end{equation}
Thus, the order of the condensation subgroup appears quadratically, while the decoupled sector appears linearly. 

In order to gain further intuition for these symmetry subgroups, let us consider the case of fusing a defect $\mathcal{D}$ to its orientation reversal $\mathcal{\bar D}$. In this case we have $D_1 = D$ and $D_2 = D^{-1}$, so the lattice quotients reduce to
\begin{equation}\label{eq:DDbarlattice}
    \mathcal{Z}_{12} = \mathcal{Z}_{\rm 12, cond} = \frac{\mathbb{Z}^2+D^{-1}\mathbb{Z}^2}{\mathbb{Z}^2} = \mathcal{Z}(D)\, , \qquad \mathcal{Z}_{\rm 12, dec} = \{1\}\, .
\end{equation}
Indeed, this only results in a condensation defect for the symmetry subgroup that $\mathcal{D}$ gauges, and no further decoupled sector. For an instructive example of the opposite case with only a decoupled TQFT and no condensation defect, we refer to section \ref{ssec:dual_trial_defe_chern_sim_lattice}, where we discuss the triality defects of \cite{Choi:2022zal} from the lattice perspective.

While we have so far defined the symmetry subgroups from the interior region, the condensation subgroup should conventionally be defined from the left or the right of the defect, in alignment with the fusion rule \eqref{eq:fusionrule}. There, we can understand the condensation defect as a projection operator acting on charged states. To work out what this symmetry subgroup is, it is useful to first recast $\mathcal{Z}_{\rm 12, cond}$ in terms of the charged lines as
\begin{equation}
    (\mathcal{Z}_{\rm 12, cond})^\vee = \frac{D_1^{-1} \mathbb{Z}^2 \cap D_2\mathbb{Z}^2}{\mathbb{Z}^2 \cap D_1^{-1} \mathbb{Z}^2 \cap D_2\mathbb{Z}^2}\, ,
\end{equation}
using the so-called second isomorphism theorem. Rotating by $D_1$ or $D_2^{-1}$ to interpret the condensation subgroup from the left or the right of the fusion, we get
\begin{equation}
     (\mathcal{Z}^L_{\rm 12, cond})^\vee = \frac{\mathbb{Z}^2 \cap D_3 \mathbb{Z}^2}{\mathbb{Z}^2 \cap D_3 \mathbb{Z}^2 \cap D_1 \mathbb{Z}^2}\, , \qquad (\mathcal{Z}^R_{\rm 12, cond})^\vee = \frac{\mathbb{Z}^2 \cap D_3^{-1}\mathbb{Z}^2}{\mathbb{Z}^2 \cap D_3^{-1} \mathbb{Z}^2 \cap D_2^{-1}\mathbb{Z}^2} \, .
\end{equation}
These lattices describe Wilson--'t Hooft line operators whose charges remain integral when swept through both defects, but that pick up surface attachments in the intermediate region between $\mathcal{D}_1$ and $\mathcal{D}_2$. To be precise, $(\mathcal{Z}^L_{\rm 12, cond})^\vee$ describes line operators that remain genuine as they are pulled through $\mathcal{D}_3$, but picked up surface attachments after being pulled through $\mathcal{D}_2$ in the intermediate region; vice versa, $(\mathcal{Z}^R_{\rm 12, cond})^\vee$ classifies line operators that remain genuine as they are pulled through $\mathcal{D}_3$ from the left to the right, but became non-genuine after being pulled through $\mathcal{D}_1$. In order to obtain the dual symmetry surfaces, we can take the Pontryagin duals of these lattice quotients
\begin{equation}
 \mathcal{Z}^L_{\rm 12, cond} = \frac{\mathbb{Z}^2 + D_3 \mathbb{Z}^2 + D_1 \mathbb{Z}^2}{\mathbb{Z}^2 + D_3 \mathbb{Z}^2}\, , \qquad \mathcal{Z}^R_{\rm 12, cond}= \frac{\mathbb{Z}^2 + D_3^{-1} \mathbb{Z}^2 + D_2^{-1}\mathbb{Z}^2}{\mathbb{Z}^2 + D_3^{-1}\mathbb{Z}^2} \, .
\end{equation}
Here $\mathcal{Z}^L_{\rm 12, cond}$ classifies symmetry surfaces ending on $\mathcal{D}_1$ from the left which cannot end on $\mathcal{D}_3$; equivalently, the condensation subgroup arises from the one-form symmetry subgroup that is gauged by $\mathcal{D}_1$ but not $\mathcal{D}_3$. By viewing it from the right, we arrive at a similar conclusion: $\mathcal{Z}^R_{\rm 12, cond}$ contains the symmetry surfaces ending on $\mathcal{D}_2$ from the right which cannot end on $\mathcal{D}_3$, or equivalently, the one-form symmetry subgroup that is gauged by $\mathcal{D}_2$ but not $\mathcal{D}_3$. The two lattice quotients are also directly related by multiplication by $D_3$ when we move them from the right region to the left region of the fusion product. 

Lastly, it is also remarkable that many of the quotient groups discussed here are cyclic, i.e., they are generated by a single element. Indeed, we already know that $\mathcal{Z}(D) = (\mathbb{Z}^2+D^{-1} \mathbb{Z}^2)/\mathbb{Z}^2$ is cyclic, which follows straightforwardly from bringing $D$ to its Smith normal form $D = U \text{diag}(1/k, k) V$. In turn, it follows that the condensation and decoupled TQFT quotient groups are cyclic themselves, since they are subgroups of $\mathcal{Z}(D)$
\begin{equation}
    \mathcal{Z}_{\rm 12, cond} \subseteq \mathcal{Z}_{12} \subseteq \mathcal{Z}_{\rm }(D_2)\, ,\quad \mathcal{Z}_{\rm dec} = \mathcal{Z}_{12}/\mathcal{Z}_{\rm 12, cond} \subseteq \mathcal{Z}_{\rm 12}\subseteq \mathcal{Z}_{\rm }(D_2)\,,
\end{equation}
and subgroups of cyclic groups must be cyclic as well.

\subsection{Fusion rules from worldvolume theory}\label{ssec:fusionworldvolume}
In this section we describe the Chern--Simons theory of the fusion of non-invertible duality defects. Since these individual defects are described by Chern--Simons theories with tridiagonal $K$-matrices, the worldvolume theory of their fusion admits a similar description. Here we derive some general features of this fusion Chern--Simons theory, such as the anyon group order. We refer to section \ref{sec:examples} for explicit examples, where properties such as the decoupled TQFT sector are extracted. 

Let us begin by describing the Lagrangian of the worldvolume theory. We take the Chern--Simons theories defined in subsection \ref{ssec:CSnoninvertible} for two duality defects $\mathcal{D}_1, \mathcal{D}_2$ associated to $D_1, D_2 \in SL(2,\mathbb{Q})$. When fusing as $\mathcal{D}_1 \times \mathcal{D}_2$, we identify the 4d gauge field $A_R$ on the right side of $\mathcal{D}_1$ with the gauge field $A_L$ on the left side of $\mathcal{D}_2$; moreover, it reduces to a 3d gauge field living on the worldvolume of the fused defects. Altogether, this gives
\begin{equation}
    \mathcal{L}_{\mathcal{D}_1 \times \mathcal{D}_2}(A_L, \alpha_I, \gamma, \beta_J, A_R) = \mathcal{L}_{\mathcal{D}_1}(A_L, \alpha_I, \gamma) + \mathcal{L}_{\mathcal{D}_2}(\gamma, \beta_I, A_R)\, , 
\end{equation}
where $\gamma$ denotes the 3d gauge field arising from the 4d gauge field living between the defects, $\alpha_I,\beta_J$ the collections of worldvolume gauge fields living on the respective defects, and $A_L,A_R$ the 4d gauge fields to the left- and right-hand side of the fusion. 

We can manipulate the worldvolume gauge fields to simplify the worldvolume action slightly. Specifically, we can consider the 3d gauge fields living between the two $\mathbb{Z}_{k_1}$ and $\mathbb{Z}_{k_2}$ gauging walls. They describe the fusion of the $SL(2,\mathbb{Z})$ duality walls $\mathcal{V}_1$ and $\mathcal{U}_2$ arising from the Smith decomposition \eqref{eq:smith} of the respective defects. Their fusion also gives an $SL(2,\mathbb{Z})$ duality wall, described by the multiplication of their matrices,
\begin{equation}
    V_1 U_2 = T^{r_1} S T^{r_2} S \cdots S T^{r_u} \in SL(2,\mathbb{Z})\, .
\end{equation}
This means that, instead of representing both $\mathcal{V}_1$ and $\mathcal{U}_2$ by their individual Chern--Simons blocks, we can merge them into one block. This shortens the tri-diagonal $\cK$-matrix of the defect fusion to
\begin{equation}
    \cK_{12} = \scalebox{0.9}{$\begin{pmatrix}
         \ddots & 1 & & \\
         1 & -p_s & 1 & \\
         & 1 & 0 & -k_2 & \\
         & & -k_2 & -r_1 & 1 &  \\
         & & & 1 &  \ddots & 1 \\
         & & & & 1 & -r_{u}& 1 & \\
         & & & & & 1 & 0 & -k_1 \\
         & & & & & & -k_1 & -q_1 & 1 \\
         & & & & & & & 1 & \ddots 
    \end{pmatrix}$}\, .
\end{equation}
This general form of the Chern--Simons theory looks rather daunting, but we can perform some consistency checks and extract useful invariants. Let us begin with the effective Chern--Simons action that it induces for $A_L, A_R$ when integrating out the 3d worldvolume fields. Knowing that the worldvolume theory of $\mathcal{D}_1$ yields the rational Chern--Simons action implementing $D_1 \in SL(2,\mathbb{Q})$, and similarly that of $\mathcal{D}_2$ yields $D_2 \in SL(2,\mathbb{Q})$, one concludes for the fusion that
\begin{equation}
    \mathcal{L}_{\mathcal{D}_1 \times \mathcal{D}_2 , {\rm eff}} = \frac{i}{4\pi \, a n_3} \Big( 2k_3 \, A_L \wedge \dd A_R  -(m_3-bn_3) A_L \wedge \dd A_L -m_3\, A_R \wedge \dd A_R  \Big)\, , 
\end{equation}
which is the rational Chern--Simons action inducing $D_3 \in SL(2,\mathbb{Q})$. This may in principle also be verified by working out the continued fraction expansions arising when integrating out the worldvolume fields. 

In a similar vein, the lattice quotient $\mathbb{Z}^N / \cK\mathbb{Z}^N$ describes the anyon group associated with the worldvolume theory. While computing this through a Smith decomposition in general is rather tedious, it is straightforward to compute its order
\begin{equation}
    |\mathbb{Z}^{N_{12}} / \cK_{12}\mathbb{Z}^{N_{12}}| = \det \cK_{12} = k_1 k_2 \frac{a n_3}{k_3}
\end{equation}
For comparison, recall that the anyon group of the individual duality wall $D_3 \in SL(2,\mathbb{Q})$ is of order $\det \mathcal{K}_3 = c n_3$. This suggests that there is an additional amount of degrees of freedom given by
\begin{equation}
    N_{\rm extra} = \frac{|\mathbb{Z}^{N_{12}} / \cK_{12}\mathbb{Z}^{N_{12}}|}{|\mathbb{Z}^{N_3} / \cK_3\mathbb{Z}^{N_3}|} = \frac{k_1 k_2}{k_3}\, .
\end{equation}
This coincides precisely with the common factor that canceled out in the matrix multiplication in \eqref{eq:Nextra}, which was also reproduced from the symmetry surfaces stretched between the defects in \eqref{eq:Nextra2}, corresponding to the condensation subgroup and the decoupled
sector. 

Let us now specialize to the fusion of a defect $\mathcal{D}$ to its orientation reversal $\mathcal{\bar D}$: $\mathcal{D}\times\mathcal{\bar{D}}$. In this case we want to reproduce the condensation defect resulting from their fusion. While this readily follows from the lattice arithmetic in \eqref{eq:DDbarlattice}, we now want to show that the Chern--Simons description matches this result. To this end, we are free to choose the duality frame in which we work. In particular, we may choose $U=1$ for the Smith decomposition \eqref{eq:smith} by conjugating the defect matrix as $D \to U^{-1} D U$; at the level of defects, this amounts to fusing by $\mathcal{\bar U}$ from the left and $\mathcal{U}$ from the right. In this frame, the duality defect reduces to an $SL(2,\mathbb{Z})$-duality transformation $\mathcal{V}' = \mathcal{V}\mathcal{U}$, followed by gauging an \textit{electric} $\mathbb{Z}_k$ one-form symmetry subgroup. The $\cK$-matrix for the fusion product then takes the simple form
\begin{equation}
    \cK_{|\mathcal{D} |^2}^{\rm el} = \scalebox{0.9}{$\begin{pmatrix}
        0 & -k &  &  & \\
        -k & 0 & 1 &  & \\
         & 1 & 0 & 1 &  \\
         &  & 1 & 0 & -k \\
         &  &  & -k & 0
    \end{pmatrix}$} , \quad \mathcal{L}^{\rm el}_{|\mathcal{D} |^2} = \frac{i}{2\pi} \big(  b \wedge \dd (a+c)- k 
    \, a \wedge \dd A_L - k \, c\wedge \dd A_R \big)\, ,
\end{equation}
where the first and last block are trivial because we conjugated by $U$, while the middle block between the $\mathbb{Z}_k$ gaugings becomes trivial by using that $\mathcal{V}' \mathcal{\bar V}' = \mathcal{V}\,\mathcal{U}\,\mathcal{\bar U}\, \mathcal{\bar V} = 1$. In writing out the Lagrangian, we labeled the gauge fields as $\mathcal{A}^I = (A_L, a, b, c,A_R)$. We can view the field $b$ simply as a Lagrange multiplier whose equation of motion sets $c = -a$. Upon this identification, the Lagrangian reduces to
\begin{equation}\label{eq:LDDbar}
    \mathcal{L}_{|\mathcal{D}|^2}^{\rm el} = -\frac{k}{2\pi} a \wedge \dd (A_L - A_R)\, .
\end{equation}
This indeed describes the condensation defect of higher-gauging a $\mathbb{Z}_k$ \textit{electric} 1-form symmetry. It is electric rather than dyonic,  because we have chosen to rotate to a different duality frame by applying $U \in SL(2, \mathbb{Z})$. We may obtain the general dyonic result by conjugating by $\mathcal{U}$ and $\mathcal{\bar U}$, which yields a condensation defect described by the $K$-matrix 
\begin{equation}
    \cK_{|\mathcal{D}|^2} = \scalebox{0.9}{$\begin{pmatrix}
         \ddots & 1 & & \\
         1 & -p_s & -k & \\
         & -k & 0 & k &  \\
         & & k & -p_1 & 1 \\
         & & & 1 & \ddots 
    \end{pmatrix}$}\, .
\end{equation}
The part between the $\mathbb{Z}_k$ gaugings stays trivial, because we can still use $\mathcal{V} \mathcal{\bar V} =1$. In fact, we have used the same rewriting as before into \eqref{eq:LDDbar}, which simplified the electric condensation defect. It is also instructive to point out that the anyon group in the general case still has order
\begin{equation}
    |\mathbb{Z}^{2s+3}/\cK\mathbb{Z}^{2s+3} | = k^2\, .
\end{equation}
This is the same as for the electric condensation defect, so the additional worldvolume gauge fields of the $SL(2,\mathbb{Z})$ duality walls do not introduce new degrees of freedom.

\section{Examples}\label{sec:examples}
In this section we work through explicit examples of duality defects. These examples serve two purposes. First, they show how the arithmetic methods we developed classify duality defects in practice. We reproduce the duality and triality defects of \cite{Choi:2021kmx, Choi:2022zal}, but also illustrate new examples of infinite-order that go beyond these $N$-ality constructions. Second, they allow us to study the fusion rules for these duality defects, describing both from lattice arithmetic and Chern--Simons theories how the condensation defects and decoupled TQFT sectors arise.

\subsection{Duality and triality defects}\label{ssec:dual_trial_defe_chern_sim_lattice}
To warm up, we consider the duality and triality defects of \cite{Choi:2021kmx, Choi:2022zal}. In this setting, we illustrate how our lattice arithmetics reproduce their fusion rules, complementary to the Chern--Simons description. It also makes apparent how we find many more defects for the same gauge couplings, although we defer a more detailed analysis to the subsequent subsection \ref{ssec:examplestau=i}. 

\paragraph{Duality defects.} Let us start with the ``duality'' defects of \cite{Choi:2021kmx}. To avoid confusion, let us stress that ``duality'' here indicates that the defect squares to a condensation defect; in the rest of this paper, duality refers to defects associated to any $SL(2,\mathbb{Q})$ transformation. They arise at $\tau = i N$, which corresponds to setting $a=1$ and $c=N^2$, while $b=0$. The family \eqref{eq:defectCM} of defect matrices are then given by
\begin{equation}\label{eq:tauiN}
    D = \frac{1}{k}\begin{pmatrix}
        m & - N^2 n  \\
        n & m
    \end{pmatrix}\, , \qquad m^2 + N^2 n^2 = k^2\, .
\end{equation}
In general, there are infinitely many (primitive) solutions to this Diophantine problem; for further discussion, we refer to subsection \ref{ssec:examplestau=i}, where we study the Pythagorean family arising at $\tau=i$. The special solution $(m,n,k) = (0,1,N)$ corresponds to the non-invertible ``duality'' defects constructed in \cite{Choi:2021kmx} that combine S-duality and the gauging of a $\mathbb{Z}_N$ electric one-form symmetry. In this case, we can indeed write out the defect matrix as
\begin{equation}
    D = \frac{1}{N}\begin{pmatrix}
        0 & -N^2\\
        1 & 0
    \end{pmatrix} = 
    \begin{pmatrix}
       0 & -1 \\
       1 & 0
    \end{pmatrix}\begin{pmatrix}
        \frac{1}{N} & 0 \\
        0 & N
    \end{pmatrix}\, ,
\end{equation}
reflecting its sequential construction. Following section \ref{sec:worldvolumeaction}, the Chern--Simons theory of this defect has $K$-matrix
\begin{equation}
    \mathcal{K}_{\rm duality} = \begin{pmatrix}
    0 & 1 & \\
    1 & 0& 1 &  \\
     & 1 & 0 & -N  \\
    &  & -N & 0 
\end{pmatrix}\, , \qquad \mathcal{L}_{\mathcal{D}} = \frac{i}{2\pi}\big(A_L \wedge \dd a + a \wedge \dd b - N \, b \wedge \dd A_R\big)\, ,
\end{equation}
where we used that $V = 1$ (so $q_1 = 0$) and $U = S$ (so $p_1=p_2=0$), following the notation around~\eqref{eq:Kmatfull}. The field $a$ simply acts as a Langrange multiplier for the constraint $b = - A_L$, so we straightforwardly reduce the worldvolume action to
\begin{equation}
    \mathcal{L}_{\mathcal{D}} = \frac{ i N}{2\pi} A_L \wedge \dd A_R\, .
\end{equation}
This result indeed agrees with the worldvolume action given in \cite{Choi:2022zal}. 

Let us next turn to the fusion of this defect to itself (or equivalently, its orientation reversal, since $D^2 = 1$). We find for the lattices of symmetry surfaces stretched between the defects that
\begin{equation}
    \mathcal{Z}_{12} \cong \mathcal{Z}_{\rm 12, cond} \cong \mathbb{Z}_N \, , 
\end{equation}
while their relative quotient is therefore trivial, implying $\mathcal{Z}_{\rm 12, dec} = 1$. The condensation defect is generated by symmetry surfaces $\eta_{0,1/N}$ stuck between the two duality defects, which become trivial as they are moved to the left or right exterior. This condensation subgroup indeed agrees with the fusion observed in \cite{Choi:2022zal} from the defect Lagrangian, which yields
\begin{equation}
    \mathcal{L}_{|\mathcal{D}|^2} = \frac{ i N}{2\pi} a \wedge \dd(A_R-A_L)\, ,
\end{equation}
where $a$ denotes the 3d gauge field that descends from the 4d bulk field living between the defects. This also matches with our general result \eqref{eq:LDDbar} for the electric condensation defect arising from fusing a defect to its orientation reversal. The fusion rule can thus be summarized as
\begin{equation}
    \mathcal{D} \times \mathcal{D} = \mathcal{C}_{N, \rm el}\, ,
\end{equation}
where the subscript indicates that it is the $\mathbb{Z}_N$ electric condensation defect.

\paragraph{Triality defects.} Let us next turn to the ``triality'' defects of \cite{Choi:2022zal}, where ``triality'' here indicates that the cube of the defect matrix in $SL(2,\mathbb{Q})$ is proportional to the identity. These defects arise at $\tau = e^{2\pi i/3} N$, corresponding to $(a,b,c) = (1,N,N^2)$. The Diophantine family of duality defects is given by 
\begin{equation}
    D = \frac{1}{k}\begin{pmatrix}
        m- N n & -n N^2 \\
        n & m
    \end{pmatrix}\, , \qquad m^2 - N m n + N^2 n^2 = N^2
\end{equation}
For $N=1$ these reduce to the Eisenstein triples. For any $N$, there are infinitely many primitive solution. A special case is given by $(m,n, k) = (N,1,N)$, which reproduces the triality defect of \cite{Choi:2022zal}. The corresponding defect matrix $\left( \begin{smallmatrix}
    0 & -N \\
    1/N & 1
\end{smallmatrix}\right)$ can be decomposed into Smith normal form as described in \eqref{eq:smith_decomp_triality}, with either $U= S$ and $V=T^N$ for $N$ even, and $U=ST^{-1}$ and $V=T^{N(N+1)}$ for $N$ odd. The $\mathcal{K}$-matrix of the Chern--Simons theory follows straightforwardly from these decompositions
\begin{equation}
\begin{aligned}
    N \text{ even:}\quad \mathcal{K}_{\rm triality} = \scalebox{0.8}{$\begin{pmatrix}
        0 & 1 \\
        1 & 0 & 1 \\
         & 1 & 0 & -N \\
         &   & -N & -N 
    \end{pmatrix}$}\, , \quad N\text{ odd:}\quad \mathcal{K}_{\rm triality} = \scalebox{0.8}{$\begin{pmatrix}
        0 & 1 \\
        1 & 1 & 1 \\
         & 1 & 0 & -N \\
         &   & -N & -N(N+1) 
    \end{pmatrix}$}\, ,   
\end{aligned}
\end{equation}
where for $N$ even we have exponents $q_1=q_2=0$ and $p_1=N$, while for $N$ odd we have exponents $q_1=0, q_1=-1$ and $p_1=N(N+1)$. The Chern--Simons theories involve two worldvolume fields, but in both cases we can integrate out one to eliminate the other. Writing $\mathcal{A} = (A_L, a, b, A_R)$, we set $b=-A_L$ in the first case, and $a=N A_R$ in the second. The resulting couplings for the bulk fields are the same for both $N$ even and odd,
\begin{equation}
    \mathcal{L}_{\rm triality} =  \frac{i N}{2\pi} A_L \wedge \dd A_R - \frac{iN}{4\pi} A_R \wedge \dd A_R \, ,
\end{equation}
in agreement with the form obtained in \cite{Choi:2022zal}.

Let us next turn to the fusion rule of this defect with itself. To be precise, we consider $\mathcal{D} \times \mathcal{D}$, which is different from the fusion $\mathcal{D} \times \mathcal{\bar D}$ that always results in a condensation defect. From the point of view of the lattice intersection \eqref{eq:Z12}, we find that
\begin{equation}
    \mathcal{Z}_{12} \cong \mathcal{Z}_{\rm 12, dec} \cong \mathbb{Z}_{N}\, ,
\end{equation}
while $\mathcal{Z}_{\rm 12, cond} = (\mathbb{Z}^2 + D\mathbb{Z}^2 \cap D^{-1} \mathbb{Z}^2)/ \mathbb{Z}^2 = 1$. The decoupled TQFT is generated by the symmetry surfaces $\eta_{0,1/N}$ stretched between the defects. To see more explicitly why they generate a decoupled TQFT rather than a condensation defect, we can check what happens as we pull them through one of the two defects. Pulling the symmetry surfaces $\eta_{0,1/N}$ through the $\mathcal{D}$ on the right, they map under $D^{-1}$ to $\eta_{1,0}$ (see fig.~\ref{fig:decomoposit_cond_TQFT_defe} for a depiction of this process; in this example, both $\mathcal D_1$ and $\mathcal D_2$ correspond to $\mathcal{D}$). However, to end them on the other $\mathcal{D}$ on the left, we have to use the equivalence $\eta_{0,1/N}\simeq \eta_{-1, 1/N}$: $D$ maps the latter to $\eta_{-1,0}$, while the former charges would be mapped to $(-1,1/N)$. We view this integral charge $(-1,0)$ as the line operator that the surface emits on the wall, generating the decoupled $\mathbb{Z}_N$ TQFT.

This effect is corroborated by the worldvolume Lagrangians of the duality defects, as was originally worked out in \cite{Choi:2022zal}. We find for the fusion that
\begin{equation}
    \mathcal{L}_{\mathcal{D}\times \mathcal{D}} = -\frac{i N}{4\pi} A_I \wedge \dd A_I +\frac{i N}{2\pi}A_L \wedge \dd A_I -\frac{i N}{4\pi} A_R \wedge \dd A_R + \frac{i N}{2\pi}A_I \wedge \dd A_R\, ,
\end{equation}
where $A_I$ denotes the intermediate gauge field between the two defects. Since $A_I$ reduces to a 3d gauge field in the fusion, we are free to shift it by the external fields as $A_I \to A_I + A_L+A_R$, which makes the decoupled sector manifest
\begin{equation}
    \mathcal{L}_{\mathcal{D}\times \mathcal{D}} = \frac{i N}{4\pi} A_L \wedge \dd A_L  + \frac{i N}{2\pi}A_L \wedge \dd A_R-\frac{i N}{4\pi} A_I \wedge \dd A_I\, .
\end{equation}
The first two terms correspond to the Lagrangian of the ``triality'' defect associated with $D^2 = - D^{-1}\in SL(2,\mathbb{Q})$, while the last describes a decoupled TQFT $\mathcal{T}_{\mathbb{Z}_N}$. The fusion rule can hence be summarized as
\begin{equation}
    \mathcal{D} \times \mathcal{D} = \mathcal{T}_{\mathbb{Z}_N}\,  \mathcal{C} \times  \mathcal{\bar D} \, ,
\end{equation}
where $\mathcal{C}$ denotes the charge conjugation defect, i.e.~it acts by $-1 \in SL(2,\mathbb{Z})$.

\subsection{Pythagorean defects at \texorpdfstring{$\tau = i$}{tau = i}}\label{ssec:examplestau=i}
Having discussed the ``duality'' and ``triality'' defects, we next turn to the infinite Diophantine families that arise at any complex multiplication point for the gauge coupling. In particular, this gives rise to defects of infinite order, instead of the $N$-ality defects we have seen so far in this section. We will also encounter fusion rules of two duality defects that produce a condensation defect and a new non-invertible duality defect at the same time.

To find defects exhibiting these features, it suffices to look at the $S$-dual point $\tau=i$. For the binary quadratic form discussed in section \ref{ssec:construction}, we set $a=c=1$ and $b=0$. The defect matrices are then given by
\begin{equation}\label{eq:Dtaui}
    D = \frac{1}{k} \begin{pmatrix}
        m & -n  \\
        n & m
    \end{pmatrix}\, , \qquad m^2+n^2 = k^2\, .
\end{equation}
Hence, the topological defects are classified by the primitive Pythagorean triples $(m,n,k)$. This includes the special case $(m,n,k)=(0,1,1)$, which is just the usual S-duality transformation. However, we also find new non-invertible defects at this point $\tau=i$ by considering all other non-trivial Pythagorean triples. In fact, this leads to infinitely many non-invertible defects, since there are known to be infinitely many primitive Pythagorean triples. 

\begin{figure}
    \centering
    \includegraphics[width=0.5\linewidth]{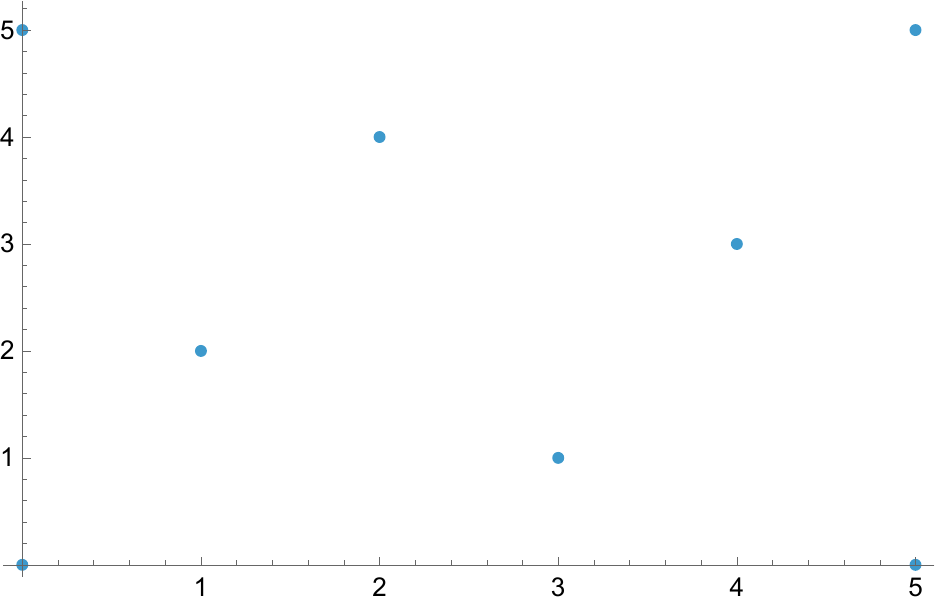}
    \caption{Lattice of genuine operators for the CM defect $D= \frac{1}{5}\left(
\begin{smallmatrix}
 3 & -4 \\
 4 & 3\\
\end{smallmatrix}
\right)$ at $\tau=i$.}
    \label{fig:taui}
\end{figure}

As an example of such a non-invertible defect, we will consider the case $(m,n,k) = (3,4,5)$, which we will refer to as the 345-defect:
\begin{equation}
    D_{3,4,5} = \frac{1}{5}\begin{pmatrix}
        3 & -4 \\
        4 & 3
    \end{pmatrix}\, .
\end{equation}
From the action of $D_{3,4,5}$ on the lattice we infer that the lattice of genuine line operators $\Gamma(D_{3,4,5}) = \mathbb{Z} \cap D_{3,4,5}^{-1} \mathbb{Z}$ is a rank-five sublattice, or equivalently, it gauges a $\mathbb{Z}_5$ dyonic subgroup of the 1-form global symmetry; this sublattice has been depicted in figure \ref{fig:taui}. The defect $\mathcal{D}_{3,4,5}$ is not an $N$-ality defect like we saw before. This can be seen clearly from the eigenvalue $\alpha = (3+4i)/5$ of its defect matrix: it is not a discrete root of unity, i.e., it does not satisfy $\alpha^N = 1$ for any $N \in \mathbb{N}$. Instead, this defect is of infinite order: by multiplying $D_{3,4,5}$ with itself as many times as we like, we can generate arbitrarily many distinct duality defects.

We next construct the Chern--Simons theory underlying the defect explicitly following subsection \ref{sec:worldvolumeaction}. First, we perform the Smith decomposition of the defect matrix, which may be written as $U = T S T^4 S$ and $V = T^7$. In turn, the $K$-matrix of the 3d gauge fields takes the form
\begin{equation}
    \mathcal{K}_{\text{3-4-5}} = \left(
\begin{array}{ccccc}
 -1 & 1 & 0 & 0 & 0 \\
 1 & -4 & 1 & 0 & 0 \\
 0 & 1 & 0 & 1 & 0 \\
 0 & 0 & 1 & 0 & -5 \\
 0 & 0 & 0 & -5 & -7 \\
\end{array}
\right)\, .
\end{equation}
One may verify explicitly that integrating out the 3d gauge fields produces the rational Chern--Simons couplings for $A_L,A_R$ as
\begin{equation}
\begin{aligned}
    \mathcal{L}_{\rm eff} &= \frac{i}{4\pi}( -\frac{3}{4}A_L \wedge \dd A_L - \frac{3}{4} A_R \wedge \dd A_R + \frac{5}{2} A_L \wedge \dd A_R )\, ,
\end{aligned}
\end{equation}
which agrees with the effective action described in \eqref{eq:Leff}.

Next, we study fusion within this Pythagorean family of defects. We know generally from section \ref{ssec:fusionworldvolume} that the fusion of a defect with its orientation reversal produces a condensation defect. Instead, here we focus on the fusion with other defects in the family. Products of Pythagorean matrices \eqref{eq:Dtaui} again produce a matrix of the same type, so it is clear that we stay within the family. However, the fusion need not be limited to duality defects, and can also produce condensation defects on top of this. We illustrate both a case with and a case without a condensation defect. To be precise, we choose the defect matrices
\begin{equation}
    D_{5, 12, 13} = \frac{1}{13}\begin{pmatrix}
        5 & -12 \\
        12 & 5
    \end{pmatrix}, \qquad D_{7, 24, 25} = \frac{1}{25}\begin{pmatrix}
        7 & -24 \\
        24 & 7
    \end{pmatrix}.
\end{equation}
From the denominators, we see that these defects gauge $\mathbb{Z}_{13}$ and $\mathbb{Z}_{25}$ dyonic 1-form symmetries. For both defects we want to determine the fusion with the $\mathcal{D}_{3,4,5}$. 

For the first case, we find from multiplying their defect matrices that
\begin{equation}
    D_{3,4,5} D_{5,12,13} = \frac{1}{65} \begin{pmatrix}
        -33 & -56 \\
        56 & -33
    \end{pmatrix} = D_{-33,56,65}\, ,
\end{equation}
which is indeed another Pythagorean triple. No common factors canceled out, as one may have expected from the fact that the gauging orders have no common divisor: $\gcd(5, 13) = 1$. From the 1-form symmetry subgroups one may verify that there is indeed no condensation defect or decoupled TQFT $\mathcal{Z}_{12, \rm cond} = \mathcal{Z}_{12, \rm dec} = 1$. Thus, the fusion rule reads simply
\begin{equation}
    \mathcal{D}_{3,4,5} \times \mathcal{D}_{5,12,13} = \mathcal{D}_{-33, 56, 65}\, .
\end{equation}

However, for the defect $\mathcal{D}_{7,24,25}$ there is potential overlap with the 1-form symmetry subgroup of $\mathcal{D}_{3,4,5}$, since their gauging orders have a common divisor: $\gcd(5, 25) = 5$. This is also reflected by the multiplication of their defect matrices
\begin{equation}
    D_{3,4,5} D_{7,24,25} = \frac{1}{5}\begin{pmatrix}
        -3 & -4 \\
        4 & -3
    \end{pmatrix} = D_{-3, 4, 5}\, ,
\end{equation}
where a factor of $N_{\rm extra} = 25$ canceled out. This can be traced back to a condensation defect, since for the symmetry surfaces between the defects we have that
\begin{equation}
    \mathcal{Z}_{12} \cong \mathcal{Z}_{12, \rm cond} \cong \mathcal{Z}_5\, , 
\end{equation}
while $\mathcal{Z}_{12, \rm dec} = 1$. The generator for the symmetry surfaces between the defects is $\eta_{1/5, 3/5}$. From the left or right-side perspective, it is spanned by
\begin{equation}
    \mathcal{Z}_{12, \rm cond}^L = \frac{\langle (\frac{1}{25}, \frac{7}{25})\rangle}{\langle (\frac{1}{5}, \frac{2}{5}) \rangle}\, , \qquad \mathcal{Z}_{12, \rm cond}^R = \langle (0,\tfrac{1}{5})\rangle \, ,
\end{equation}
where all symmetry operators are understood to modulo $\mathbb{Z}^2$. The denominator indicates the symmetry surfaces that can end on $\mathcal{D}_3$ from the left. This gives for the fusion rule
\begin{equation}
    \mathcal{D}_{3,4,5} \times \mathcal{D}_{7,24,25} = \mathcal{C}_{\mathcal{Z}_{12, \rm cond}^L} \times \mathcal{D}_{-3, 4, 5} = \mathcal{D}_{-3, 4, 5} \times \mathcal{C}_{\mathcal{Z}_{12, \rm cond}^R}\, .
\end{equation}
One may wonder whether we can also find fusion rules for defects at $\tau=i$ with decoupled TQFTs. We expect this to be impossible due to the underlying arithmetic structure, so we look at another value of the gauge coupling in the next subsection.

\subsection{Fusion with condensation and decoupled TQFT at \texorpdfstring{$\tau = i \sqrt{11}$}{tau = i sqrt(11)}}
Here we illustrate an example of a fusion product that results in a non-trivial condensation defect and a decoupled TQFT sector at the same time. We show how to see these sectors both from the lattices of Gukov--Witten surfaces and from the worldvolume Chern--Simons theory. Concretely, we take the point $\tau = i \sqrt{11}$, corresponding to $a=1, b=0, c=11$. The duality defects we consider are given by
\begin{equation}
    D_1 = \frac{1}{6}\begin{pmatrix}
        -5 & -11 \\
        1 & -5
    \end{pmatrix}\, , \qquad D_2 = \frac{1}{18}\begin{pmatrix}
        -7 & 55 \\
        -5 & -7
    \end{pmatrix}\, ,
\end{equation}
corresponding to the parameter choices $(m,n,k)=(-5,1, 6)$ and $(m,n,k)=(-7,-5, 18)$. Their product yields the defect
\begin{equation}
    D_3 = D_1 D_2 = \frac{1}{6} \begin{pmatrix}
        5 & -11 \\
        1 & 5
    \end{pmatrix} = - D_1^{-1}\, ,
\end{equation}
which has $(m,n,k)=(5,1,6)$. In the matrix multiplication a factor of  $N_{\rm cond+dec} = 18$ canceled out. We will find that this factor decomposes as 
\begin{equation}
    N_{\rm cond+dec} = N_{\rm cond}^2 N_{\rm dec} = 3^2 \cdot 2\, ,
\end{equation}
with $N_{\rm cond} = 3$ and $N_{\rm dec} = 2$. We can recover this $\mathbb{Z}_3$ condensation defect and decoupled $\mathbb{Z}_2$ sector explicitly. Indeed, considering the quotient lattices of symmetry surfaces between the defects, we find the cyclic groups
\begin{equation}
    \mathcal{Z}_{\rm 12} \cong \mathbb{Z}_6\, , \qquad \mathcal{Z}_{\rm 12, cond} \cong \mathbb{Z}_3 \, , \qquad \mathcal{Z}_{\rm 12, dec} \cong  \mathcal{Z}_{\rm 12}/\mathcal{Z}_{\rm12, cond} \cong \mathbb{Z}_2\, .
\end{equation}
The quotient lattice $\mathcal{Z}_{\rm 12}$ of all Gukov--Witten surfaces that can end on both defects is generated by $\eta_{\frac{1}{6}, -\frac{1}{6}}$. Meanwhile, the subset of Gukov--Witten surfaces generating the condensation sector may be given by $\eta_{\frac{1}{3}, -\frac{1}{3}}$. The quotient between these two generates the $\mathbb{Z}_2$ decoupled TQFT. 

To describe the condensation defect from the outside of the fusion, we rotate either by pulling $\mathcal{Z}_{\rm cond, 12}$ to the left through $\mathcal{D}_1$ or to the right through $\mathcal{D}_2$. The resulting condensation subgroups are given by
\begin{equation}
    \mathcal{Z}_{12, \rm cond}^L = \frac{\langle (\frac{1}{6}, \frac{1}{6}) \rangle}{\langle (\frac{1}{2}, \frac{1}{2})\rangle } \,, \qquad     \mathcal{Z}_{12, \rm cond}^R = \frac{\langle (\frac{1}{18}, \frac{7}{18})\rangle}{\langle ( \frac{1}{6}, \frac{1}{6} ) \rangle}\, ,
\end{equation}
where all lattices are understood modulo $\mathbb{Z}^2$. We had to quotient by surfaces ending from the left or the right on $\mathcal{D}_3$ respectively, as these are not counted for the condensation defect. One may verify straightforwardly that both these lattice quotients give a $\mathbb{Z}_3$ condensation defect. Altogether, we can summarize the fusion rule including the decoupled TQFT sector as
\begin{equation}
    \mathcal{D}_{1} \times \mathcal{D}_2 = \mathcal{T}_{\mathbb{Z}_2}\ \mathcal{C}_{\mathcal{Z}_{12, \rm cond}^L} \times \mathcal{D}_3 = \mathcal{T}_{\mathbb{Z}_2}\ \mathcal{D}_3 \times \mathcal{C}_{\mathcal{Z}_{12, \rm cond}^R}\, .
\end{equation}

Let us now study this system through the Chern--Simons theory of the fusion product. In particular, we would like to extract the decoupled TQFT sector explicitly. The Smith decomposition of the defect matrices yields
\begin{equation}
    U_1 = T^{-5} S T^{-1}, \quad V_1 = T^{31}\, , \qquad U_2 = ST^{-47}S^3\,, \quad V_2 = T^{-7} S T^{-8}
\end{equation}
The resulting Chern--Simons theory then has the $\mathcal{K}$-matrix
\begin{equation}
    \mathcal{K}= \scalebox{0.8}{$\left(
\begin{array}{ccccccccc}
 5 & 1 & 0 & 0 & 0 & 0 & 0 & 0 & 0 \\
 1 & 1 & 1 & 0 & 0 & 0 & 0 & 0 & 0 \\
 0 & 1 & 0 & -6 & 0 & 0 & 0 & 0 & 0 \\
 0 & 0 & -6 & -31 & 1 & 0 & 0 & 0 & 0 \\
 0 & 0 & 0 & 1 & 47 & 1 & 0 & 0 & 0 \\
 0 & 0 & 0 & 0 & 1 & 0 & -1 & 0 & 0 \\
 0 & 0 & 0 & 0 & 0 & -1 & 0 & -18 & 0 \\
 0 & 0 & 0 & 0 & 0 & 0 & -18 & 7 & 1 \\
 0 & 0 & 0 & 0 & 0 & 0 & 0 & 1 & 8 \\
\end{array}
\right)$}\, ,
\end{equation}
where we took the freedom to eliminate the worldvolume fields with vanishing level related to the $S^3$ factor in $U_2$. The gauge fields are sorted as $\mathcal{A} = (A_L, a_1, \cdots, a_7, A_R)$. 

There is an algorithmic way to make the decoupled $\mathbb{Z}_2$ TQFT manifest. The general set of basis transformations that are allowed take the form
\begin{equation}
    B = \begin{pmatrix}
        1 & 0 & \ldots & 0 \\
        b_{1,0} & b_{1,1} & \ldots & b_{1,8} \\
        \vdots & \vdots & \ddots & \ldots \\
        b_{7,0}& b_{7,1}& \ldots & b_{7,8} & \\
        0 & \ldots & 0 & 1
    \end{pmatrix}\, , \qquad \mathcal{K} \to B^T \mathcal{K} B\, .
\end{equation}
Note that this basis transformations only changes the worldvolume gauge fields $a_1, \ldots,a_{7}$, and leaves the gauge fields $a_0=A_L$ and $a_8 = A_R$ unchanged. We now want to identify a 3d gauge field that decouples from all other gauge fields. Let us take it to be the second-to-last column of $B$. Then a suitable basis transformation can typically be found if it satisfies the following conditions
\begin{equation}
    \mathbf{v} = (0, b_{1,7}, \ldots, b_{7,7}, 0)\, : \qquad \mathbf{v}^T \mathcal{K} \mathbf{v} = \pm 2\, , \quad \mathcal{K} \mathbf{v} \in 2 \mathbb{Z}^n\, .
\end{equation}
In our case, such a vector is given by
\begin{equation}
    \mathbf{v} = (0,-2,-2,-1,-1,2,1,0,0)\, .
\end{equation}
We can then complete the basis by considering the orthogonal integral subspace $\langle \mathcal{K} \mathbf{v}/2 \rangle^\perp$ to span the other $8$ columns of $B$. In the end, we take for the basis transformation
\begin{equation}
    B = \scalebox{0.8}{$\left(
\begin{array}{ccccccccc}
 1 & 0 & 0 & 0 & 0 & 0 & 0 & 0 & 0 \\
 2 & 1 & 0 & 0 & 0 & 0 & 0 & -2 & 0 \\
 2 & 0 & 1 & 0 & 0 & 0 & 0 & -2 & 0 \\
 1 & 0 & 0 & 1 & 0 & 0 & 0 & -1 & 0 \\
 1 & 0 & 0 & 0 & 1 & 0 & 0 & -1 & 0 \\
 -2 & -2 & 2 & 21 & -23 & -1 & -9 & 2 & 0 \\
 -1 & 0 & 0 & 0 & 0 & 1 & 0 & 1 & 0 \\
 0 & 0 & 0 & 0 & 0 & 0 & 1 & 0 & 0 \\
 0 & 0 & 0 & 0 & 0 & 0 & 0 & 0 & 1 \\
\end{array}
\right)$}\, , \quad B^T \mathcal{K} B = \scalebox{0.8}{$\left(
\begin{array}{ccccccccc}
 7 & 1 & 0 & 0 & 0 & 0 & 0 & 0 & 0 \\
 1 & 1 & 1 & 0 & -2 & 2 & 0 & 0 & 0 \\
 0 & 1 & 0 & -6 & 2 & -2 & 0 & 0 & 0 \\
 0 & 0 & -6 & -31 & 22 & -21 & 0 & 0 & 0 \\
 0 & -2 & 2 & 22 & 1 & 22 & -9 & 0 & 0 \\
 0 & 2 & -2 & -21 & 22 & 2 & -9 & 0 & 0 \\
 0 & 0 & 0 & 0 & -9 & -9 & 7 & 0 & 1 \\
 0 & 0 & 0 & 0 & 0 & 0 & 0 & -2 & 0 \\
 0 & 0 & 0 & 0 & 0 & 0 & 1 & 0 & 8 \\
\end{array}
\right)$}\, .
\end{equation}
In this basis we can indeed see the decoupled $\mathbb{Z}_2$ sector explicitly: it is the entry $-2$ at position (7,7) of the $\mathcal{K}$-matrix, where all other entries in its row and column vanish. This indicates that it is a $\mathbb{Z}_2$ sector that does not couple to the bulk fields $A_L, A_R$ or the other worldvolume fields living on the wall. In principle, one should be able to further reduce the worldvolume theory to recover the defect $\mathcal{D}_3$ more explicitly, as well as the $\mathbb{Z}_3$ condensation defect. We refrain from doing this analysis here, as it is even more tedious than singling out the decoupled sector.

\subsection{Example beyond electric gauging with torsion at \texorpdfstring{$\tau = 3i/2$}{tau=3i/2}}\label{ssec:tau=3i/2}
Finally, we give a simple example demonstrating that not every duality defect can be realized as an electric gauging (possibly with torsion) followed by a single $SL(2,\mathbb{Z})$ duality transformation. For this defect, duality transformations are needed both before and after the gauging, and neither can be absorbed into the torsion of the electric gauging (though it could be understood as gauging with a suitable counterterm). One could of course pass to a different $SL(2,\mathbb{Z})$ frame and remove one of the two duality walls using \eqref{eq:VDV}. Here, however, we keep the frame fixed so both duality walls are present.

To appreciate this distinction, let us first write down the defect matrix that describes an electric gauging with torsion followed by a duality transformation. The torsion corresponds to adding an upper-triangular component to the electric gauging matrix, such that we have
\begin{equation}\label{eq:Dtorsion}
	D_{\text{torsion}} = U \begin{pmatrix}
		\frac{1}{N} & - p \\
		0 & N
	\end{pmatrix}\, .
\end{equation}
Adding this discrete torsion $p$ to the gauging matrix can be viewed as a duality transformation $V = T^{p N}$, i.e., such that $D_{\text{torsion}} = U \text{diag}(1/N, N) V$. However, this means we cannot consider any duality transformation $V \in SL(2,\mathbb{Z})$ on the right side, but only powers of $T$-transformations. The duality defect discussed in this subsection gives an example where the duality wall $V$ can never be brought to this form.

However, finding a particular Smith decomposition where $V$ has a different form does not suffice; the Smith decomposition is not unique, so there may be another one where it does take the form of a $T$-transformation. To this end, the lattice of genuine operators serves as a smoking gun for when a defect matrix cannot be brought to the form \eqref{eq:Dtorsion}. Namely, the factor $\left(\begin{smallmatrix}
1/N & -p \\
0 & N
\end{smallmatrix}\right)$ always keeps the 't Hooft-line operator, i.e.~with charges $(0,1)$, genuine. In fact, the lattice left integral by this map may be generated by $(N, 0)$ and $(0,1)$. Subsequently, the duality matrix $U \in SL(2,\mathbb{Z})$ merely takes linear combinations of these charges, always keeping integral charges integral, such that
\begin{equation}
    \Gamma(D_{\rm torsion}) = \langle (N, 0), (0,1)\rangle\, .
\end{equation}
In other words, whenever the lattice of genuine line operators does not contain the 't Hooft line operator, the defect does not come from an electric gauging with torsion.

\begin{figure}
    \centering
    \includegraphics[width=0.5\linewidth]{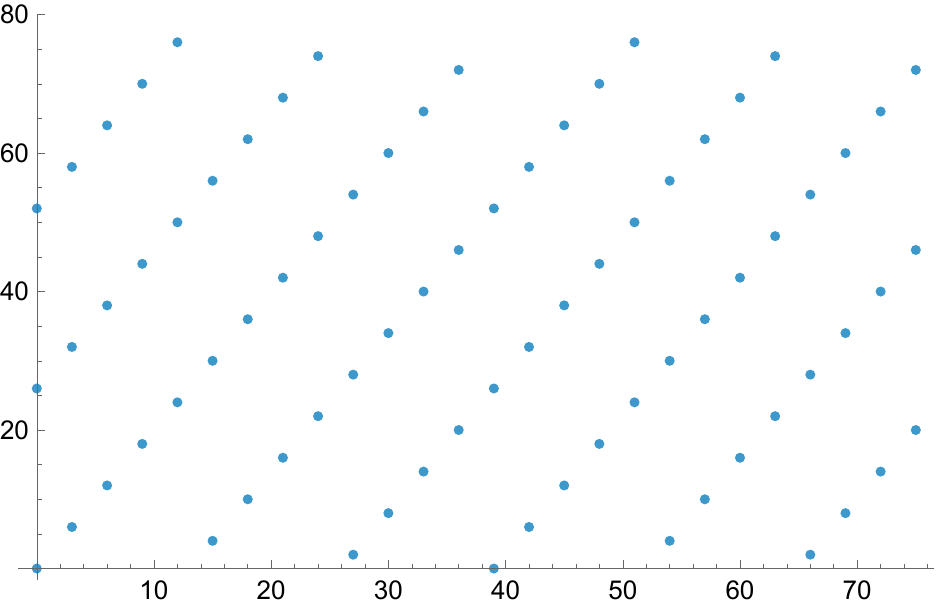}
    \caption{Lattice of genuine operators for the CM defect $D= \frac{1}{78}\left(
\begin{smallmatrix}
 72 & -45 \\
 20 & 72\\
\end{smallmatrix}
\right)$ at $\tau=3i/2$.}
    \label{fig:tau3i/2}
\end{figure}

Concretely, we consider the point $\tau=3i/2$, corresponding to a binary quadratic form with $a=4$, $b=0$, $c=9$. In general, the defect matrices at these points are classified by the Diophantine problem
\begin{equation}
    D = \frac{1}{k}\begin{pmatrix}
        m & -9 n \\
        4 n & m
    \end{pmatrix}\, , \qquad m^2 + 36 n^2 = k^2\, .
\end{equation}
As particular example we choose $m=72$, $n=5$ and $k=78$, thus gauging a $\mathbb{Z}_{78}$-subgroup. Its defect matrix admits the decomposition
\begin{equation}
    D_{72, 5, 78} = \frac{1}{78}\begin{pmatrix}
        72 & -45 \\
        20 & 72
    \end{pmatrix} = \left(
\begin{array}{cc}
 -50175 & -11 \\
 -13684 & -3 \\
\end{array}
\right) \begin{pmatrix}
    1/78 & 0 \\
    0 & 78
\end{pmatrix}\left(
\begin{array}{cc}
 4 & 927 \\
 -3 & -695 \\
\end{array}
\right)\, .
\end{equation}
While the second duality matrix clearly is not of the form $T^N$, one could hope that there is another choice of Smith decomposition where it does take that form. However, that can never happen (this is related to the fact that many matrix entries share common factors, e.g.~$\gcd(20,72)=4$ and $\gcd(72, 45)=9$). To see more clearly that is cannot happen, it is instructive to look at the lattice of line operators that remain genuine, which has been depicted in figure \ref{fig:tau3i/2}. Explicitly, it can be spanned as 
\begin{equation}
    \Gamma(D_{72, 5, 78}) = \langle (3,6)\rangle_{\mathbb{Z}} + \langle (0, 78) \rangle_{\mathbb{Z}}\, .
\end{equation}
Indeed, we confirm that this lattice never contains the 't Hooft operator with charge $(0,1)$. In fact, it does not even contain any Wilson--'t Hooft line operator $(q,1)$ with unit magnetic charge. The smallest magnetic charge of a genuine line operator is equal to $p=6$. So it follows that this defect can never be written as an electric gauging (with torsion) followed by an $SL(2,\mathbb{Z})$ duality wall; it needs a general $SL(2,\mathbb{Z})$ transformation before the gauging. In other words, this duality defect arises from the gauging of a genuine dyonic one-form symmetry subgroup.

\section{Conclusions}
We have investigated non-invertible duality defects in 4d $U(1)$ Maxwell theory. By using the complex multiplication symmetries of an auxiliary two-torus, we have described at which values of the coupling these defects arise, and how they act on the Wilson--'t Hooft line operators and Gukov--Witten symmetry surfaces of the theory. Through the Smith decomposition of their $SL(2, \mathbb{Q})$ matrix representation, we have presented a general construction of these defects as two $SL(2,\mathbb{Z})$ duality walls separated by a $\mathbb{Z}_k \subset U(1)^{(1)}_e$ gauging wall. We have also set up the explicit 3d Chern--Simons theory of these defects, using continued fractions to describe the worldvolume theory of $SL(2,\mathbb{Z})$ duality walls. 
Crucially, the Chern--Simons descriptions allows us to explicitly demonstrate that our definition of the physical defect does not suffer from the algebraic ambiguity of Smith decomposition.
Finally, we have described the fusion of these defects, including decoupled TQFTs and condensation defects that may appear, in two complementary ways: through lattice intersections of Gukov--Witten symmetry surfaces that can end on the defects, and through the worldvolume Chern--Simons theory of the fusion product; the two approaches yield the same orders for the condensation and decoupled TQFT sectors in all the examples we have studied.

This investigation inspires several directions for future work. Our approach has given a concrete handle on the duality defects of 4d $U(1)$ Maxwell theory and their fusion products, but does not completely identify the general fusion algebra that underlies these Diophantine families. In particular, our lattice analysis gives a direct way to compute the orders of condensation and decoupled TQFT sectors, but does not determine which TQFT arises. So far, we have used the underlying Chern--Simons description to fix this in examples, but it would be valuable to have a general rule.

Another pressing direction is to extend the arithmetic methods developed here to other classes of theories, such as non-Abelian gauge theories or supergravities. 
For instance, frameworks such as class ${\cal S}$ provide a natural setting to explore the connection between duality defects and CM-like geometric symmetries on higher-genus Riemann surfaces.
In a similar vein, in supergravities the gauge couplings are no longer constants but are parametrized by moduli fields. Dualities in this setting are known as Gaillard--Zumino symmetries \cite{Gaillard:1981rj}, whose non-invertible aspects were explored in \cite{Apruzzi:2025byj}. It would be interesting to see how our approach extends to such theories obtained from compactifications of string theory, where analogous arithmetic structures should exist for the compactification geometry; see \cite{Moore:1998pn, Moore:1998zu, Candelas:2019llw, Grimm:2024fip} for applications of arithmetic techniques to Calabi--Yau manifolds in the context of attractor points and flux vacua.

Complementary to 4d theories, there are also the worldsheet theories of strings with geometrical target spaces. When these target spaces possess additional symmetries, the resulting 2d CFTs are constrained: they may become rational \cite{Gukov:2002nw,Jockers:2025fgv}, or acquire non-invertible defect lines \cite{Kidambi:2022wvh, Okada:2022jnq, Cordova:2023qei, Jockers:2025fgv, Angius:2025zlm, Caldararu:2025eoj, Angius:2025ium, Angius:2026jpv}. In forthcoming work \cite{workinprogress} we apply our methodology to 2d CFTs with $T^2$ target spaces, giving a complete classification of the non-invertible $SO(2,2;\mathbb{Q})$ defect lines that can arise. Ultimately, these worldsheet symmetries may produce constraints on the target space theory such as selection rules, cf.~for some applications of such geometrical symmetries in gravitational theories \cite{Ahmed:2023cnw, Heckman:2024obe, Kaidi:2024wio, Bharadwaj:2024gpj}.

\subsubsection*{Acknowledgments}
We would like to thank Roberta Angius and Eduardo Garcia-Valdecasas for discussions. GM acknowledges support by Deutsche Forschungsgemeinschaft through the Emmy Noether program 557478919, Deutsche Forschungsgemeinschaft under Germany’s Excellence Strategy EXC 2121 Quantum Universe 390833306 and by Deutsche Forschungsgemeinschaft through the Collaborative Research Center 1624 ``Higher Structures, Moduli Spaces and Integrability.''

\appendix

\section{Orders of condensation defects and decoupled TQFTs}\label{app:orders}
In this appendix we compute the orders of the quotient groups defined above in section~\ref{sec:fusion}. This is particularly instructive from the point of view of the factor $N_{\rm cond+dec} = k_1 k_2/k_3$ that dropped out in our defect matrix multiplication; here we reproduce it from the lattice data. Let us begin with the quotient group $\mathcal{Z}_{12}$ that sees both the condensation defect and decoupled TQFT. Its order is more easily computed through by rewriting \eqref{eq:Z12} to
\begin{equation}
    \mathcal{Z}_{12} \cong  \frac{\mathbb{Z}^2}{L_1+L_2} \cong \frac{\mathbb{Z}^2}{L_1} \Big/\frac{L_1+L_2}{L_1} \, ,
\end{equation}
where we introduced the shorthands $L_1 = \mathbb{Z}^2 \cap D_1^{-1} \mathbb{Z}^2$ and $L_2 = \mathbb{Z}^2 \cap D_2 \mathbb{Z}^2$. In the first line we dualized under the Dirac pairing, while in the second step we used an isomorphism for lattice quotients. Using furthermore that $(L_1+L_2)/L_1 \cong L_2/L_1 \cap L_2$, we find for the quotient order
\begin{equation}\label{eq:ordZcondTQFT}
|\mathcal{Z}_{12}| = \frac{[\mathbb{Z}^2 : L_1 ] }{[L_2 : L_1 \cap L_2]} = \frac{[\mathbb{Z}^2 : L_1][\mathbb{Z}^2 : L_2]}{[\mathbb{Z}^2 : L_1 \cap L_2]}
 =  \frac{k_1 k_2 }{ [ \mathbb{Z}^2: \mathbb{Z}^2\cap D_1^{-1} \mathbb{Z}^2 \cap D_2\mathbb{Z}^2]  } \, ,
\end{equation}
where in the second step we used that $L_2$ and $L_1 \cap L_2$ are both sublattices of $\mathbb{Z}^2$.

Computing the order of the other lattice quotient $Z_{\rm cond}$ is a bit more involved; however, we can exploit some useful identities in order to connect with the order of $Z_{12}$. Let us rewrite the lattice quotient in the denominator of \eqref{eq:ordZcondTQFT} as
\begin{align}
[\mathbb{Z}^2: \mathbb{Z}^2\cap D_1^{-1} \mathbb{Z}^2 &\cap D_2\mathbb{Z}^2]  = [\mathbb{Z}^2 : \mathbb{Z}^2 \cap D_1^{-1} \mathbb{Z}^2][\mathbb{Z}^2 \cap D_1^{-1} \mathbb{Z}^2 : \mathbb{Z}^2\cap D_1^{-1} \mathbb{Z}^2 \cap D_2\mathbb{Z}^2] \\
&= k_1[\mathbb{Z}^2 \cap D_1 \mathbb{Z}^2 : \mathbb{Z}^2\cap D_3 \mathbb{Z}^2 \cap D_1 \mathbb{Z}^2]  = [\mathbb{Z}^2 : \mathbb{Z}^2\cap D_3 \mathbb{Z}^2 \cap D_1 \mathbb{Z}^2]  \, , \nonumber
\end{align}
where, in the second line, we rotated the second quotient by $D_1$ in the first step, and in the second step used that $k_1 =[\mathbb{Z}^2: \mathbb{Z}^2 \cap D_1 \mathbb{Z}^2]$ and combined the ratios, following the reverse logic of the first equality. The rewritten denominator can be expanded as 
\begin{align}
 [\mathbb{Z}^2 : \mathbb{Z}^2\cap D_3 \mathbb{Z}^2 \cap D_1 \mathbb{Z}^2]   &= [\mathbb{Z}^2 : \mathbb{Z}^2 \cap D_3 \mathbb{Z}^2 ] [\mathbb{Z}^2 \cap D_3 \mathbb{Z}^2 : \mathbb{Z}^2 \cap D_3 \mathbb{Z}^2 \cap D_1 \mathbb{Z}^2 ]\\
 &= k_3  [D_1^{-1} \mathbb{Z}^2 \cap D_2\mathbb{Z}^2 : \mathbb{Z}^2 \cap D_1^{-1} \mathbb{Z}^2 \cap D_2 \mathbb{Z}^2 ]\, . \nonumber
\end{align}
We recognize the second factor as the order of $\mathcal{Z}_{\rm cond}$ through the isomorphism
\begin{equation}
    \mathcal{Z}_{\rm cond} = \frac{\mathbb{Z}^2 + D_1^{-1}\mathbb{Z}^2 \cap D_2 \mathbb{Z}^2}{\mathbb{Z}^2} \cong \frac{D_1^{-1}\mathbb{Z}^2 \cap D_2 \mathbb{Z}^2}{\mathbb{Z}^2 \cap D_1^{-1}\mathbb{Z}^2 \cap D_2 \mathbb{Z}^2}\, .
\end{equation}
Using these relations we can write the order of the condensation subgroup as
\begin{equation}
|\mathcal{Z}_{\rm cond}| = [ D_1^{-1} \mathbb{Z}^2 \cap D_2 \mathbb{Z}^2 : \mathbb{Z}^2 \cap D_1^{-1} \mathbb{Z}^2 \cap D_2 \mathbb{Z}^2]=  \frac{[ \mathbb{Z}^2: \mathbb{Z}^2\cap D_1^{-1} \mathbb{Z}^2 \cap D_2\mathbb{Z}^2]  }{k_3}\, .
\end{equation}
Multiplying these two quotient group orders $|\mathcal{Z}_{12}|, |\mathcal{Z}_{\rm cond}|$, we find that they indeed combine into the common divisor \eqref{eq:Nextra} that canceled out in the defect matrix multiplication
\begin{equation}\label{eq:Nextra2}
N_{\rm extra}=\frac{k_1 k_2}{k_3} = |\mathcal{Z}_{12}| |\mathcal{Z}_{\rm cond}| = |\mathcal{Z}_{\rm cond}|^2  |\mathcal{Z}_{\rm dec}|\, .
\end{equation}
where we used that $|\mathcal{Z}_{\rm dec}| = |\mathcal{Z}_{12}|/|\mathcal{Z}_{\rm cond}|$. This expansion of the common divisor $N_{\rm cond+dec}$ also makes sense from the perspective of examples of condensation defects and decoupled TQFTs: on the one hand, when a defect gauging a $\mathbb{Z}_k$ and its orientation reversal are fused together, a factor $N_{\rm cond+dec}=k^2$ drops out, where indeed $|\mathcal{Z}_{\rm cond}|=k$ and $|\mathcal{Z}_{\rm dec}|=1$; on the other hand, when two identical triality defects gauging a $\mathbb{Z}_k$ subgroup, following the construction in \cite{Choi:2022zal}, are fused, they produce yet another triality defect plus a decoupled TQFT, and indeed $N_{\rm cond+dec}=|\mathcal{Z}_{\rm dec}|=k$ is the factor that cancels out.

\section{Equivalence of Chern--Simons theories under Smith decompositions} \label{app:equiv_chern_simo_smith_deco}

Given a representation $D \in SL(2,\mathbb{Q})$ of a duality defect, this admits many Smith decompositions $D = U \text{diag}(1/k, k) V$, each of them yielding a different description \eqref{eq:Kmatfull} for its Chern--Simons theory. In order for the duality defect to be well-defined, these descriptions must all be equivalent.\footnote{This is in the same spirit as the expansions of $SL(2,\mathbb{Z})$ elements into words in $S,T$ as considered in section \ref{ssec:SL2walls}. There, we also argued that different words representing the same duality wall have equivalent Chern--Simons theories.} In this appendix we hence want to show that, given a Smith decomposition of a non-invertible defect  $D$, the resulting physics does not depend on the details of $U$ and $V$, that is to say, the resulting Chern--Simons theories are equivalent.
This proof has been developed with the help of AI, and has been separately verified by the authors.

In order to prove this, let us start by the fact that, given any Smith decomposition of the matrix representation of a defect $D$, the gauging wall will be always represented by $G={\rm diag}(1/k,\, k)$, and only the invertible, outer $SL(2,\mathbb{Z})$ walls $U,V$ can depend on the specific decomposition considered. In particular, two equivalent Smith decompositions are related as:
\begin{equation}
    U_1 \text{diag}(1/k,k) V_1 = U_2 (U_2^{-1} U_1) \text{diag}(1/k, k) (V_1 V_2^{-1}) V_2 \, .
\end{equation}
Since these are Smith decompositions of the same matrix $D = U_i \text{diag}(1/k, k) V_i$ ($i=1,2$), it follows that the transfer matrices $U_{21}=U_2^{-1} U_1$ and $V_{12}= V_1 V_2^{-1} $ obey the relation
\begin{equation}\label{eq:Smithrelation}
    U_{21} \text{diag}(1/k, k) V_{12} = \text{diag}(1/k, k)\quad  \iff \quad  V_{12} = \text{diag}(k, 1/k) U_{21}^{-1} \text{diag}(1/k, k)\, .
\end{equation}
Requiring $V_{12} \in SL(2,\mathbb{Z})$ constrains $U_{21}$ to the Hecke congruence subgroup
\begin{equation}\label{eq:U_in_Gamma_0_K2}
    U_{21} \in \Gamma_0(k^2) = \left\{\left(\begin{smallmatrix}
        a & b \\
        c & d
    \end{smallmatrix}\right) \in SL(2,\mathbb{Z}) \, | \, c = 0 \mod k^2 \right\}\, .
\end{equation}
Equivalently, requiring $U_{21}$ to be integral yields $V_{12} \in \Gamma^0(k^2)$, i.e.~with $b = 0 \mod k^2$ instead. In order to show equivalence of the two Chern--Simons descriptions, we thus have to argue that the walls satisfy the relation
\begin{equation}\label{eq:gauging_wall_equivalency}
    \mathcal{U}_{21} \mathcal{G}_{\rm el} \mathcal{V}_{12} = \mathcal{G}_{\rm el}\, , 
\end{equation}
where $\mathcal{G}_{\rm el}$ denotes the $\mathbb{Z}_k$ electric gauging wall, at the Chern--Simons theory level. The proof will be divided in 4 steps.

Before moving to the proof, in the following we will often rely on the concept of Schur complement, which was already introduced in Section~\ref{ssec:SL2walls}. Let us consider a quadratic form described by a symmetric block matrix 
\begin{equation}\label{Schur_complement_integrating_internal_fields}
    K=\begin{pmatrix}
        A&B\\B^T&C
    \end{pmatrix}, \qquad P(x,y) = (x^T,y^T) K \begin{pmatrix}
    x \\
    y
    \end{pmatrix} = x^T A x + 2 x^T B y + y^T C y\, .
\end{equation}
The Schur complement allows then to obtain the effective matrix governing the variables $x$ once we eliminate the variables $y$ by imposing $\partial_y P(x,y) = 0$. Inserting the solution $y = -C^{-1} B^T x$ into the quadratic form, we find
\begin{equation}
    P(x,y) \big|_{y = -C^{-1}B^T x} = x^T(A-BC^{-1}B^T)x\, .
\end{equation}
Notice that we have to require $C$ to be invertible. For our purposes, this procedure turns out to be useful in order to manipulate effective actions with both bulk fields and worldvolume fields. For instance, by eliminating the worldvolume fields in a Chern--Simons action such as \eqref{eq:effconstr}, we recover the effective Chern--Simons couplings for the bulk fields $A_L,\,A_R$ through the respective Schur complements. In this context, this procedure can be understood as effectively integrating out the worldvolume fields and obtaining the effective action on the external fields that couple to it.

\paragraph{Step 1: reducing the gauging wall.}Let us start from the words associated with the invertible walls in~\eqref{eq:gauging_wall_equivalency}; as we saw in Section~\ref{ssec:SL2walls}, these can always be expanded in terms of elementary duality walls $\mathcal{S}$ and $\mathcal{T}$ as:\footnote{For computational purposes, we use a slightly different convention with respect to Section~\ref{ssec:SL2walls}, which, however, is completely equivalent to the one presented above. In particular, here we use the fact that any word can always be manipulated in such a way that it starts and ends with an $\mathcal{S}$ wall; this can always be done by attaching a finite number of $\mathcal{T}^0$ factors. Moreover, we also took negative sign exponents for the $\mathcal{T}$-walls in the duality wall expansions. Again, this is just for convenience in order to avoid carrying additional signs that would make the notation heavier; it is possible to check that by changing the sign of both $p_i$ and $q_j$ in~\eqref{eq:chern_simons_initial_proof_smith}, the proof follows equivalently. Finally, some of the following manipulations are valid only for $n,m\geq2$, while they could slightly change for shorter words. However, notice that it is always possible to add an $\mathcal{S}^4=1$ wall in such a way that the word is long enough to be able to apply this procedure.}
\begin{equation}
    \mathcal{U}_{21}=\mathcal{S}\mathcal{T}^{-p_1}\mathcal{S}\mathcal{T}^{-p_2}\dots\mathcal{S}\mathcal{T}^{-p_n}\mathcal{S},\quad \mathcal{V}_{12}=\mathcal{S}\mathcal{T}^{-q_1}\mathcal{S}\mathcal{T}^{-q_2}\dots\mathcal{S}\mathcal{T}^{-q_m}\mathcal{S}.
\end{equation}
In this way, the Chern--Simons Lagrangian representing the gauging wall $\mathcal{U}_{21}\mathcal{G}_{\rm el}\mathcal{V}_{12}$ takes the form:
\begin{equation}\label{eq:chern_simons_initial_proof_smith}
\mathcal{L}_{\mathcal{D}}
    = \frac{i}{4\pi}\, \cK_{\cI\cJ}\, \mathcal{A}^\cI \wedge \dd \mathcal{A}^\cJ\,,\quad\mathcal{K}=
\scalebox{0.8}{$\left(
\begin{array}{c|cccc|ccc|cccc|c}
0 & 1 & 0 & \dots &   &   &   &   &   &   &   &   &   \\
\hline
1 & p_1 & 1 & 0 &   &   &   &   &   &   &   &   &   \\
0 & 1 & p_2 & \ddots & \ddots &   &   &   &   &   &   &   &   \\
\vdots & 0 & \ddots & \ddots & 1 &   &   &   &   &   &   &  \\
  &   & \ddots & 1 & p_n & 1 &   &   &   &   &   &   &   \\
\hline
  &   &   &   & 1 & 0 & 1 & 0 &   &   &   &   &   \\
  &   &   &   &   & 1 & 0 & -k &   &   &   &   &   \\
  &   &   &   &   & 0 & -k & 0 & 1 &   &   &   &   \\
\hline
  &   &   &   &   &   &   & 1 & q_1 & 1 &   &   &   \\
  &   &   &   &   &   &   &   & 1 & q_2 & \ddots &   &   \\
  &   &   &   &   &   &   &   &   & \ddots & \ddots & 1 &    \\
  &   &   &   &   &   &   &   &   &     & 1 & q_m & 1 \\
\hline
  &   &   &   &   &   &   &   &   &   &  & 1 & 0
\end{array}
\right)$},
\end{equation}
where $\mathcal{A}=(A_L;a_1,\dots,a_n;\phi,b,\psi;c_1,\dots,c_m;A_R)$. Due to the fact that both words for $\mathcal{U}_{21}$ and $\mathcal{V}_{12}$ start and end in $\mathcal{S}$, it follows that $A_L,\phi,\psi,A_R$ are level 0 fields, e.g., no term such as $\frac{1}{4\pi}A_LdA_L$ appears in the Chern--Simons action.

In order to prove~\eqref{eq:gauging_wall_equivalency}, let us start by noticing that the central $3\times3$ gauging block, associated with $(\phi,b,\psi)$, presents 3 consecutive 0 along the diagonal, i.e. the three fields are level-0; moreover, since $b$ couples to $\phi$ through a unitary coefficient, we can integrate it out without any gauge ambiguity. Explicitly, this takes  the form:
\begin{align}
    \mathcal{L}_{\rm gauging}&=\frac{i}{2\pi}\left(\phi\wedge da_n+b\wedge d\phi-k\,\psi\wedge db+c_1\wedge d\psi\right)\nonumber\\&=\frac{i}{2\pi}b\wedge\left(d\phi-k\,d\psi\right)+\frac{1}{2\pi}\left(\phi\wedge da_n+\psi\wedge dc_1\right)\nonumber\\
    \text{solving for }b\implies\phi=k\psi\rightarrow\;&=\frac{i}{2\pi}\psi\wedge(k\,da_n+dc_1).
\end{align}
At this point, $\psi$ itself enters just as a Lagrange multiplier, again with a unitary coefficient, imposing the condition $c_1=-ka_n$. Integrating out $\psi$ then leads to a new Lagrangian of the form:
\begin{equation}\label{eq:reduced_chern_sim_step_1}
    \mathcal{L}_{\mathcal{D}}
    = \frac{i}{4\pi}\, \cK_{\cI\cJ}\, \mathcal{A}^\cI \wedge \dd \mathcal{A}^\cJ\,,\quad\mathcal{K}=
\scalebox{0.8}{$\left(
\begin{array}{c|cccc|cccc|c}
0 & 1 & 0 & \dots &   &   &   &   &   & \\
\hline
1 & p_1 & 1 & 0 &   &   &   &   &   &\\
0 & 1 & p_2 & \ddots & \ddots &   &   &   &   &\\
\vdots & 0 & \ddots & \ddots & 1 &   &   &   &\\
  &   & \ddots & 1 & p_n+k^2q_1 & -k &   &   &   &\\
\hline
  &   &   &   & -k & q_2 & 1 &   &   &   \\
  &   &   &   &   & 1 & q_3 & \ddots &   &   \\
  &   &   &   &   &   & \ddots & \ddots & 1 &    \\
  &   &   &   &   &   &     & 1 & q_m & 1 \\
\hline
  &   &   &   &   &   &   &  & 1 & 0
\end{array}
\right)$},
\end{equation}
where the surviving fields are $\mathcal{A}=(A_L;a_1,\dots,a_n;c_2,\dots,c_m;A_R)$ and $\cI,\cJ=0,\dots,N+1$, with $N=n+m-1$. In particular, the overall effect of this manipulation is a shifting of the $a_n$ level and the direct coupling between $a_n$ and $c_2$ with coefficients $-k$, reminiscent of the gauging wall that has been integrated out.

\paragraph{Step 2: recovering the gauging order.}Now that we have reduced the $\cK$-matrix by integrating out the 0-level fields,\footnote{Note that there could be additional 0-level fields, since any $p_i,q_i\in\bbZ$ can be trivial itselves; however, these will depend on the specific example considered and do not represent a general feature of the theory.} we want to show that the reduced matrix still encodes the gauging of a $k$-electric subgroup. For computational purposes, in the following, we will consider the reduced $N\times N$ matrix $\hat{\cK}=\cK^{(\hat{0},\widehat{N+1})}$, where the notation $M^{(\hat{l})}$ implies the deletion of the $l$-th row and column from the matrix $M$, taking into account only the internal fields $\cA_i=(a_1,\dots,a_n;c_2,\dots,c_m)$ (the lower-case indexes take values $i,j=1,\dots,N$) and discarding the external fields $A_L,\,A_R$. 

The first thing to notice is that ${\rm rank}(\hat{\cK})\geq N-1$. This can be easily seen by the fact that the matrix is tri-diagonal, with  all non-vanishing links (in particular, if we denote by $s_j$ the link between $\cA_j$ and $\cA_{j+1}$, then $s_n=-k$ and $s_j=1$ for $j\neq n$; for later purposes, we also define the level of the field $\cA_j$ as $l_j$). Indeed, by considering the $(N-1)\times(N-1)$ sub-matrix obtained by deleting the first row and the last column of $\hat{\cK}$, the latter has all the links as diagonal entries, and the determinant is $-k$; hence, the full matrix $\hat{\cK}$ has ${\rm rank}(\hat{\cK})\geq N-1$. 

Now that we have seen that the $\hat{\cK}$-matrix has at most corank 1, we want to show that this is precisely the case. In order to do so, let us construct the associated null vector $v$. By definition, a null vector has to satisfy $\hat{\cK}\,v=0$; to solve this, let us exploit the tri-diagonal form of $\hat{\cK}$. In particular, the j-th row of $\hat{\cK}\,v=0$ takes the form:
\begin{equation}
    s_{j-1}\,v_{j-1}+l_j\,v_j+s_{j}\,v_{j+1}=0\implies v_{j+1}=-\frac{s_{j-1}\,v_{j-1}+l_j\,v_j}{s_{j}},\quad j=1,\dots,N.
\end{equation}
Since $A_L$ is a background field, we can impose $v_0=0$. Moreover, we can also choose $v_1=1$ as an additional initial condition for the series; indeed, $v_1=0$ would lead to the trivial vector $v=\mathbf{0}$, while any other choice of $v_1\neq0$ would lead to a vector proportional to the one considered here.\footnote{Moreover, $v_1=1$ is also necessary in order to have a primitive integer null vector, which will lead to a unimodular transformation later, as we will comment further.} With the value of $v_1$ fixed, all the following entries are then constrained by the recursive condition above. However, since $A_R$ is not part of $\cA_i$, i.e. we are working with the $N\times N$ matrix $\hat\cK$, in order for $v$ to be a well-defined null vector, we need to check that $v_{N+1}=0$. To do so, we can exploit the additional recursive formula $v_{j+1}=\frac{(-1)^jD_j}{s_1\,s_2\cdots s_j}$, where $D_j$ represents the determinant of the $j\times j$ top-left block of $\hat{\cK}$, which can be obtained from the identity $D_j=l_j\,D_{j-1}-s^2_{j-1}\,D_{j-2}$. In particular, since $s_j\neq0$ for every $j$, we immediately see that $v_{N+1}=0\iff D_N=\det(\hat{\cK})=0$, i.e. if the corank of $\hat{\cK}$ is precisely 1. The last equality then follows from the Schur complement introduced above; indeed, we have that $\det(\hat\cK)=(-1)^{n+m}\,k\,(U_{21}\,G\,V_{12})_{21}$\footnote{Notice that with this notation we mean ``the entry $(2,1)$ of the matrix $(U_{21}\,G\,V_{12})$''; in particular, whenever the subscript is outside of the parentheses, we refer to the matrix entry.} and, together with the defining property $U_{21}\,G\,V_{12}=G$, we get $\det(\hat\cK)=(G)_{21}=0$, since $G={\rm diag}(1/k,k)$. 

Now that we have shown that $v$ is a genuine null vector, let us compute its other components. In particular, since $s_j=1$ for $j\neq n$, this guaranties that $v_j\in\bbZ$ for $j< n$; this is not the case for $j=n$, however, for which we get
\begin{equation}
    v_{n+1}=\frac{v_{n-1}+p_nv_n+k^2q_1v_n}{k}.
\end{equation}
While the last factor is trivially integer, the first two require some manipulation. Their integrality can be proven using the Schur complement; indeed, by applying it to $U$, we get that $v_{n-1}+p_nv_n=-(U_{21})_{21}=-c$, and, since \eqref{eq:U_in_Gamma_0_K2} ensures that $U_{21}\in\Gamma_0(k^2)$, then $k^2|c=0$ and $v_n$ is integer. In particular, this implies that $v_j\in\bbZ$ for $j=1,\dots,N$. Moreover, it is also possible to compute the explicit value of $v_N$; again, this can be done in terms of the associated Schur complement, obtaining $v_N=-(U_{21}\,G\,V_{12})_{22}=-(G)_{22}=-k$. The null vector then takes the overall form $v=(1,\dots,-k)$, with integer entries.

Finally, before moving to the next step, let us perform one last manipulation. As we just showed, the $\hat{\cK}$ matrix is a corank 1 matrix with associated null vector $v$. In particular, we can perform the following field redefinition\footnote{Note that the field $\phi$ used here to shift $a_i$ and $c_j$ is not related to the one integrated out in Step 1.}
\begin{equation}
    a_1\mapsto\phi,\;a_2\mapsto v_2\,\phi+\tilde{a}_2,\dots,\;c_m\mapsto v_N\,\phi+\tilde{c}_m,
\end{equation}
which in matrix form becomes:
\begin{equation}
    \begin{pmatrix}
    a_1\\a_2\\\vdots
\end{pmatrix}=\hat P\begin{pmatrix}
    \phi\\\tilde{\boldsymbol a}^T
\end{pmatrix}=\left(\begin{array}{ccc}
  v_{1} & 0 & \cdots \\[6pt]
  v_{2} & \multicolumn{2}{c}{\multirow{2}{*}{$\mathbbm{1}_{N-1}$}} \\[6pt]
  \vdots & &
\end{array}\right)\begin{pmatrix}
    \phi\\\tilde{a}_2\\\vdots
\end{pmatrix}.
\end{equation}
With this field redefinition, $\hat \cK$ then takes the new form:
\begin{equation}
    \hat\cK'=\hat P^T\hat\cK \hat P=\begin{pmatrix}
        0&\boldsymbol{0}^T\\\boldsymbol0&\hat\cK^{(\hat1)}
    \end{pmatrix},
\end{equation}
where $\hat\cK^{(\hat1)}$ is the $(N-1)\times(N-1)$ matrix obtained by eliminating the first row and column of $\hat\cK$; this follows from the fact that, by definition, $\hat\cK v=0$. We can also perform a similar analysis for the full $(N+2)\times(N+2)$ $\cK$ matrix; indeed, by defining 
\begin{equation}
    P=\begin{pmatrix}
        1&\boldsymbol{0}^T&0\\ \boldsymbol{0}&\hat P&\boldsymbol{0}\\0&\boldsymbol{0}^T&1
    \end{pmatrix},
\end{equation}
we obtain:
\begin{equation}\label{eq:reduced_chern_sim_step_2}
    \cK'=P^T\cK P=\begin{pmatrix}
        0&1&\boldsymbol{0}^T&0\\1&0&\boldsymbol{0}^T&-k\\ \boldsymbol{0}&\boldsymbol{0}&\hat\cK^{(\hat1)}&e_{N-1}^T\\0&-k&e_{N-1}&0
    \end{pmatrix},
\end{equation}
where $(e_{N-1})_i=\delta^{N-1}_i$. In particular, we can immediately see that $\phi$ is a level 0 field that couples only with $A_L$ and $A_R$ with links 1 and $-k$, similarly to the $b$ field in~\eqref{eq:chern_simons_initial_proof_smith}. However, in order to prove~\eqref{eq:gauging_wall_equivalency}, we still need to understand the role of $\hat\cK^{(\hat1)}$ and the coupling between $A_R$ and $\tilde{c}_m$ due to the presence of $e_{N-1}$.

\paragraph{Step 3: unimodularity of the complement.}

Now that we have reduced and manipulated~\eqref{eq:chern_simons_initial_proof_smith} to the form~\eqref{eq:reduced_chern_sim_step_2}, we have to show that this is indeed equivalent to the gauging wall~\eqref{eq:K_matrix_gauging}. First of all, we need to show that the additional sector $\hat{\cK}^{(\hat1)}$ describes an invertible theory that does not contribute to the anyon group of the gauging wall. This is guaranteed if $|\det(\hat\cK^{(\hat1)})|=1$, and can be checked by explicit computation. To do so, let us start by splitting $\hat\cK^{(\hat1)}$ at the $-k$ link, coupling $\tilde{a}_n$ and $\tilde{c}_2$, and lets call the top-left $(n-1)\times(n-1)$ sub-matrix $L$, while the bottom-right $(m-1)\times(m-1)$ one $R$; $\hat\cK^{(\hat1)}$ then takes the form:
\begin{equation}
    \hat\cK^{(\hat1)}=\left(
\begin{array}{ccc|ccc}
  & \multicolumn{2}{c|}{\multirow{2}{*}{$L$}} & & \\
  & & & -k & & \\\hline
  & & -k & \multicolumn{2}{c}{\multirow{2}{*}{$R$}} \\
  & & & & &
\end{array}
\right).
\end{equation}
In this way, we can then split the determinant formula in the form:\footnote{Note that, since we are working with $\hat\cK^{(\hat1)}$ and not $\hat\cK$, the $(n-1)$-th position on the diagonal is the level associated with $\tilde{a}_n$.}
\begin{equation}\label{eq:splitting_det_left_right}
    \det(\hat\cK^{(\hat1)})=\det L\det R-k^2\det L^{(\widehat{n-1})}\det R^{(\hat1)}.
\end{equation}
This expression is then useful because we can now compute the different determinants explicitly using the Schur complement. 

Starting from $L$, let us call $\cK_U$ the $\cK$-matrix associated with the invertible defect $\mathcal{U}_{21}$; moreover, recall that, after the reduction in~\eqref{eq:reduced_chern_sim_step_1}, the level $l_n$ was shifted to $l_n=p_n+k^2q_1$. We can then split the determinant of $L$ as $\det(L)=\det(\cK_U^{(\hat1)})+q_1k^2\det(\cK_U^{(\hat1,\hat n)})$. Taking the different Schur complement of $\cK_U$, we then get $\det(L)=(-1)^n(a+q_1k^2b)$, where $a,b$ are the $U_{21}$ entries from~\eqref{eq:U_in_Gamma_0_K2}. Similarly, we get $\det(L^{(\widehat{n-1})})=\det(\cK_U^{(\hat1,\hat n)})=(-1)^nb$. A similar computation leads to $\det(R)=(-1)^ma'$ and $\det(R^{(\hat1)})=(-1)^m(q_1a'-c')$, where $a',c'$ are the entries of $V_{12}$. However, we can now exploit~\eqref{eq:Smithrelation}, obtaining $a'=d$ and $c'=-c/k^2$. The determinant of $\cK^{(\hat1)}$ then takes the form:
\begin{align}\label{eq:detK=1}
    \det(\hat\cK^{(\hat1)})&=(-1)^{n+m}\left[(a+q_1k^2b)a'-k^2b(q_1a'-c')\right]=(-1)^{n+m}(aa'+k^2bc')\\\nonumber&=(-1)^{n+m}(ad-bc)=(-1)^{n+m},
\end{align}
where the identity follows from $U_{21}\in\Gamma_0(k^2)$, showing that $\hat\cK^{(\hat1)}$ represents indeed an invertible theory.

\paragraph{Step 4: vanishing of the contact term.}

The last step in order to prove~\eqref{eq:gauging_wall_equivalency} is then to show that the invertible sector associated with $\tilde{\boldsymbol a}$ decouples from the external field $A_R$. To show the decoupling, we can perform a redefinition of the fields of the form $\tilde{x}\mapsto\tilde{x}'-wA_R$, where $w=(\hat\cK^{(\hat1)})^{-1}e_{N-1}$. This shift of the fields is well defined since $\det(\hat\cK^{(\hat1)})=\pm1$; hence, all the entries of $(\hat\cK^{(\hat1)})^{-1}$ are integers. In particular, the Chern--Simons action associated with $\hat\cK^{(\hat1)}$ will now take the form
\begin{equation}
    \mathcal{L}_{\hat\cK^{(\hat1)}}=\frac{i}{4\pi}(\hat\cK^{(\hat1)})_{ij}(\tilde x')^i\wedge d(\tilde x')^j-\frac{i}{2\pi}(e_{N-1})_i (\tilde x')^i\wedge dA_R+\frac{i}{4\pi}(e_{N-1}^T(\hat\cK^{(\hat1)})^{-1}e_{N-1})A_R\wedge dA_R
\end{equation}
and the second clearly cancels the coupling term between $\tilde c_m$ and $A_R$ in~\eqref{eq:reduced_chern_sim_step_2}. This proves the decoupling of the invertible sector from $A_R$. However, the shift has also produced an additional term, the last one, which would represent a non-null level $l_{A_R}$ for $A_R$, spoiling the identity~\eqref{eq:gauging_wall_equivalency}, since it would represent the presence of an additional $\mathcal{T}^{-l_{A_R}}$ wall on the right hand side of the equivalence, i.e. a shift in the $\theta$ angle associated with $A_R$. In order to prove the identity, we then need to show that this term vanishes. To do so, let us notice that $e_{N-1}^T(\hat\cK^{(\hat1)})^{-1}e_{N-1}=((\hat\cK^{(\hat1)})^{-1})_{\tilde c_m,\tilde c_m}=\frac{\det(\hat\cK^{(\hat1,\hat N)})}{\det(\hat\cK^{(\hat1)})}$. Since the denominator is different from 0, we then need the numerator to vanish. This can be computed with a similar trick to~\eqref{eq:splitting_det_left_right}, obtaining:
\begin{equation}
    \det(\hat\cK^{(\hat1,\hat N)})=(-1)^{n+m}\left[-k^2b(a+q_1k^2b)+k^2b(a+q_1k^2b)\right]=0,
\end{equation}
proving that no additional $\theta$ term is generated for $A_R$.

\paragraph{End result.} The final Chern--Simons action obtained after the four steps just presented then reads:
\begin{equation}
    \mathcal{L}_\mathcal{D}=\underbrace{\frac{i}{2\pi}\phi\wedge d(A_L-kA_R)}_{\bbZ_k-{\rm gauging}\;\mathcal{G}_{\rm el}}\;+\;\underbrace{\frac{i}{4\pi}\hat\cK^{(\hat1)}_{ij}(\tilde x')^i\wedge d(\tilde x')^j}_{\text{decoupled, invertible sector}}.
\end{equation}
This shows that~\eqref{eq:gauging_wall_equivalency} holds in general and any two Smith decompositions of the same defect $\mathcal{D}$ differing for $U,U'$ and $V,V'$ lead indeed to the same Chern--Simons action, up to a decoupled, invertible sector.

\bibliographystyle{JHEP}
\inputencoding{latin2}
\bibliography{bibliography}
\inputencoding{utf8}
\end{document}